\documentclass[aps,twocolumn,amsfonts,amssymb,amsmath,nofootinbib,preprintnumbers,eqsecnum,superscriptaddress]{revtex4-1}

\pdfoutput=1

\usepackage{epsfig}
\usepackage[latin1]{inputenc}
\usepackage{float}
\usepackage{graphicx}
\usepackage{cancel}
\usepackage{mathrsfs}
\usepackage{amssymb}
\usepackage{amsfonts}
\usepackage{amsmath}
\usepackage{slashed}
\usepackage{hyperref}
\usepackage[usenames,dvipsnames]{color}

\newcommand{\beq}{\begin{equation}}
\newcommand{\eeq}{\end{equation}}
\renewcommand\r[1]{(\ref{#1})}
\def\({\left(}  \def\){\right)}  \def\[{\left[}  \def\]{\right]}  \def\<{\langle}  \def\>{\rangle} 
\newcommand{\bew}{\begin{widetext}}
\newcommand{\ew}{\end{widetext}}

\def\kr#1{{\color{red}{ #1}}}

\begin{document}

\title{Heavy quark energy loss far from equilibrium in a strongly coupled collision}

\author{Paul~M.~Chesler}
\affiliation{Center for Theoretical Physics, Massachusetts Institute of Technology, Cambridge, MA 02139}

\author{Mindaugas~Lekaveckas}
\affiliation{Center for Theoretical Physics, Massachusetts Institute of Technology, Cambridge, MA 02139}
\affiliation{Physics Department, Theory Unit, CERN, CH-1211 Gen\`eve 23, Switzerland}

\author{Krishna~Rajagopal}
\affiliation{Center for Theoretical Physics, Massachusetts Institute of Technology, Cambridge, MA 02139}
\affiliation{Physics Department, Theory Unit, CERN, CH-1211 Gen\`eve 23, Switzerland}

\preprint{MIT-CTP 4464, CERN-PH-TH/2013-109}

\begin{abstract}
We compute and study the drag force acting
on a heavy quark propagating through the matter
produced in the collision of two sheets of energy
in a strongly coupled gauge theory that can be
analyzed holographically.  Although this matter is
initially far from equilibrium, we find that the equilibrium
expression for heavy quark energy loss in a homogeneous strongly coupled
plasma with the same instantaneous energy density or pressure as that
at the location of the quark
describes many qualitative features of our results. 
One interesting exception is that there is a time delay
after the initial collision before the heavy quark energy loss becomes significant.
At later times,
once a liquid plasma described by viscous hydrodynamics has formed,
expressions based upon assuming instantaneous homogeneity and 
equilibrium provide a semi-quantitative description of our results --- as long
as the rapidity of the  heavy quark is not too large. For a heavy quark
with large rapidity,
the gradients in the velocity of the hydrodynamic fluid result in qualitative
consequences for the `drag' force acting on the quark. In certain circumstances,
the force required to drag the quark through the plasma can point
opposite to the velocity of the quark, meaning that the
force that the plasma exerts on a quark moving through it acts in the {\it same} direction
as its velocity.
And, generically, the force includes a component perpendicular to the
direction of motion of the quark.  Our results support a straightforward
approach to modeling the drag on, and energy loss of, heavy quarks with modest rapidity
in heavy ion collisions, both before and after the quark-gluon plasma hydrodynamizes,
and provide cautionary lessons at higher rapidity. 
\end{abstract}

\date{\today}

\maketitle

\section{Introduction}

The discovery that strongly coupled quark-gluon plasma (QGP) is produced in ultrarelativistic
heavy ion collisions has prompted much interest in the real-time dynamics of strongly 
coupled non-Abelian plasmas. For example, heavy quark energy loss has received
substantial attention. If one shoots a heavy quark through a non-Abelian plasma, how
much energy does it lose as it propagates?  Equivalently, what is the drag force
required to pull a heavy quark
through such a plasma at a specified velocity?  This question has been
answered~\cite{Herzog:2006gh,Gubser:2006bz,CasalderreySolana:2006rq} for homogeneous plasma in thermal equilibrium in strongly
coupled ${\cal N}=4$ supersymmetric Yang-Mills (SYM) theory in the large number
of colors ($N_c$) limit, where holography permits a semiclassical description of
energy loss in terms of string dynamics in asymptotically AdS$_5$ spacetime~\cite{Maldacena:1997re,Witten:1998qj,Karch:2002sh,Herzog:2006gh,Gubser:2006bz,CasalderreySolana:2006rq}.
One challenge (not the only one)
in using these results to glean qualitative insights into 
heavy quark 
energy loss in heavy ion collisions is that a heavy quark produced during the initial collision event
must first propagate through the initially far-from-equilibrium
matter produced in the collision before it later plows through the
expanding, cooling, hydrodynamic fluid of strongly coupled QGP. 
In this paper we shall describe calculations that provide some
qualitative guidance for how to meet this challenge.
For a review of many other ways in which holographic calculations have
yielded qualitative insights into properties of strongly coupled QGP and dynamics
in heavy ion collisions, see Ref.~\cite{CasalderreySolana:2011us}.

We want a toy model  in which we can reliably calculate 
how the energy loss rate
of a heavy quark moving through the far-from-equilibrium matter present just after a collision
compares to that in strongly coupled plasma close to equilibrium.
We 
study the energy loss of a heavy quark moving through the debris produced by the collision of
planar sheets of energy in strongly coupled SYM theory introduced in Ref.~\cite{Chesler:2010bi}
and analyzed there and in Refs.~\cite{Casalderrey-Solana:2013aba,PaulLarryToAppear}. %\kr{Add Paul+Larry recent.}  
The incident sheets of energy move at the speed of light in the $+z$ and $-z$ directions 
and collide at $z = 0$ at time $t= 0$.  They each have a Gaussian profile in the  $z$ direction and are translationally invariant in the two directions  ${\vec  x_\perp} = {x,y}$ orthogonal to $z$. Their
energy density per unit transverse area is 
$\mu^3(N_c^2/2\pi^2)$, with $\mu$ an arbitrary scale with respect to which all 
dimensionful quantities in the conformal  theory 
that we are working in can be measured.
The width $w$ of the Gaussian energy-density profile of each sheet
is chosen to be $w = 1/(2 \mu)$.  We shall describe this setup, and its holographic
description, in Section~\ref{sec:CollidingSheets}.
Although there is no single right way to compare the widths of these translationally
invariant sheets of energy with Gaussian profiles to the widths of a nucleus that
has been Lorentz-contracted by a factor of 107 (RHIC) or 1470 (LHC), reasonable
estimates suggest that our choice of $w\mu$ corresponds to sheets with a thickness somewhere
between the thickness of the incident nuclei at RHIC and the LHC~\cite{Chesler:2010bi}.  
The recent investigations of Refs.~\cite{Casalderrey-Solana:2013aba,PaulLarryToAppear} %\kr{Add Paul+Larry recent.} 
suggest
that it would be interesting to repeat our analyses for varying values of $w\mu$, but we leave
this for future work since here we shall only be seeking to draw qualitative lessons.

The principal lesson that has been learned to date from analyses
of the collisions of strongly coupled sheets of energy as 
in Refs.~\cite{Chesler:2010bi,Casalderrey-Solana:2013aba,PaulLarryToAppear}
%\kr{Add Paul+Larry recent} 
and from many other analyses of how strongly coupled
plasma forms from a large number of
widely varied far-from-equilibrium strongly coupled initial collisions 
(for example, see Refs.~\cite{Janik:2005zt,Chesler:2008hg,Chesler:2009cy,Booth:2009ct,Heller:2011ju,Heller:2012km,vanderSchee:2012qj,Heller:2013oxa}) 
is that the fluid {\it hydrodynamizes}, i.e.~comes
to be described well by viscous hydrodynamics, after a time $t_{\rm hydro}$ that is at most 
around $(0.7-1)/T_{\rm hydro}$, where
$T_{\rm hydro}$ is the effective temperature (for example, defined from the fourth root
of the energy density)
at the hydrodynamization time $t_{\rm hydro}$.
At $t_{\rm hydro}$, the fluid can still have sufficiently large velocity gradients and
pressure anisotropies that the dissipative effects of viscosity are significant.  
In the context of hydrodynamic modeling of heavy ion collisions at RHIC (for a recent example
see Ref.~\cite{Shen:2010uy})
$t_{\rm hydro}\sim 0.7/T_{\rm hydro}$ corresponds to a time $\sim 0.3$~fm$/c$ when
$T\sim 500$~MeV.  This is about a factor of two earlier in time than the upper
bounds on the hydrodynamization times inferred from hydrodynamic 
modeling of RHIC collisions~\cite{Kolb:2003dz,Heinz:2004pj,Shen:2010uy}.
Because QCD is asymptotically free, the dominant dynamics at the earliest
moments of a sufficiently energetic heavy ion collision are expected to
be weakly coupled, with the relevant (weak) coupling being $\alpha_{\rm QCD}$
evaluated at the (short) distance scale corresponding to the mean spacing between
gluons in the transverse plane at the moment when the two highly Lorentz-contracted
nuclei collide. So, it would be inappropriate to take the estimates obtained in a context
in which the colliding sheets of 
matter are strongly coupled from beginning to end as estimates for the
hydrodynamization times of heavy ion collisions per se. The impact of these estimates
is that they teach us that the $\sim 10$ year old result~\cite{Kolb:2003dz,Heinz:2004pj} 
that the matter produced
in RHIC collisions takes at most 0.6-1~fm$/c$ to hydrodynamize should not
be seen as `rapid thermalization' since this timescale is comfortably longer
than what we now know to expect if the physics of heavy ion collisions were
strongly coupled from the start.
After we have calculated the drag force on a heavy quark
that finds itself in the midst of the colliding sheets of
energy density, we shall seek similarly qualitative lessons to those
that have been drawn from the analyses of the collisions themselves.

We compute heavy quark energy loss by
inserting a heavy quark moving at constant velocity $\vec \beta$ between the colliding sheets
before the collision and calculating 
the force needed to keep its velocity constant throughout the collision.
% then gives the energy and momentum loss rates of the quark.  
%The energy density of the colliding sheets as well as two sample trajectories of a quark getting sandwiched between them are
%shown in the left panels of Figs.~\ref{z0} and \ref{znon0}.
Via holography, the colliding planar sheets of energy in SYM theory map into colliding planar gravitational 
waves in asymptotically AdS$_5$ spacetime~\cite{Chesler:2010bi}.   The addition of a heavy quark moving at constant velocity $\vec \beta$ amounts to including a classical 
string attached to the boundary of the geometry~\cite{Karch:2002sh} and dragging 
the string endpoint  at constant velocity $\vec \beta$, pulling the string
through the colliding gravitational wave geometry.  We show how to compute
the profile of the string in this dynamical background in Section~\ref{sec:StringDynamics}.
The force needed to maintain the velocity of the string endpoint, which
we compute in Section~\ref{sec:ExtractingTheForce}, yields the
energy loss rate of the heavy quark~\cite{Herzog:2006gh,Gubser:2006bz}. 

We describe our results in Section III, beginning in Section~\ref{sec:ZeroRapidity} with the case in which
the heavy quark  is moving with $\vec\beta$ perpendicular to the $z$-direction, meaning that it has zero rapidity.
We compare the drag force that we calculate to what it would be in a homogeneous plasma
in thermal equilibrium that has the same energy density or transverse pressure or longitudinal
pressure as the matter that the quark finds itself in at a given instant in time.
We find that the peak value of the drag force, which occurs at a time when the
matter produced in the collision is still far from equilibrium, is comparable to the peak value
of the drag force
in a static plasma with the same instantaneous energy density or pressure.
However, we find that both the initial rise in the drag force and its peak are
delayed in time relative to what they would be in a static plasma with the
same instantaneous energy density or pressure.  In Appendix B we provide
some evidence that this time delay is of order $1/\pi T_{\rm hydro}$
at low $\beta$ and increases slowly as $\gamma\equiv 1/\sqrt{1-\beta^2}$ increases.
All these results are robust, in particular in the sense that we see them again when
we consider a heavy quark moving with some nonzero rapidity in 
Sections~\ref{sec:ZeroTransverseMomentum}
and \ref{sec:NonzeroBoth}.  

The message from our results at early times is that there is no sign of any 
enhancement in the energy loss experienced by a heavy quark
by virtue of the matter that it finds itself moving through
%produced in the collision   
being far from equilibrium. In broad
terms, the energy loss is comparable to what it would be in 
an equilibrium plasma with the same energy density; when
looked at in more 
detail, it can be significantly less by virtue of the initial delay in its 
rise.  This is quite different than at weak coupling, where instabilities
in the far-from-equilibrium matter can arise and can result in substantially
enhanced rates of heavy-quark energy loss~\cite{Carrington:2013tz}. 
There are, however, no signs of any instabilities
in the debris produced in the collisions of the sheets
of strongly coupled matter that we analyze~\cite{Chesler:2010bi,Casalderrey-Solana:2013aba} or, for
that matter, in any analyses of far-from-equilibrium
strongly coupled matter to date.

When we look at the drag force at late times, after the strongly coupled fluid has
hydrodynamized, we find different results depending on whether the heavy quark
has small or large rapidity. At small rapidity, the drag force that we calculate
is described semiquantitatively by assuming a homogeneous plasma in equilibrium
with an appropriate time-dependent temperature.  At large rapidity, however,
this approximation misses qualitative effects that, we show in 
Sections~\ref{sec:ZeroTransverseMomentum}
and \ref{sec:NonzeroBoth}, can be attributed to the presence of gradients
in the fluid velocity.  We find that a velocity gradient in the fluid has the greatest effect
on the energy loss of the heavy quark
when the direction of motion of the heavy quark is most closely aligned
with the velocity gradient.  As a consequence, effects of velocity gradients
are larger at larger rapidity.  We find generically that the force that must
be exerted on the quark in order to move it along its trajectory includes
a component perpendicular to the direction of motion of the quark, a component
that can be substantial in magnitude.
In certain cases we also find that, as a consequence of
gradients in the fluid velocity, the $z$-component of the force
that is required to move the quark in the positive-$z$ direction points
toward negative $z$!    We conclude in Section IV with a look at the lessons
on how best to model the drag on heavy quarks produced in heavy ion collisions
that can be drawn from our results.

\section{Holographic Description}
\label{sec:Methods}

In a strongly coupled conformal gauge theory with a dual gravitational description, a heavy quark moving through 
out-of-equilibrium matter that is 
on its way to becoming strongly coupled plasma 
consists of a string moving in some
non-equilibrium, but asymptotically AdS$_5$, black brane spacetime.
For an infinitely massive quark, the endpoint of the string is attached to the four-dimensional boundary 
of the geometry.  The geometry of the boundary is that of Minkowski space and the trajectory
of the string endpoint on the boundary coincides with the trajectory of the quark.  

We shall focus on the case in which the gauge theory is ${\cal N}=4$ SYM 
theory with $N_c$ colors, although our results can immediately
be generalized to any conformal gauge theory with a gravity dual upon making a suitable
modification to the relationship between the 't Hooft coupling 
$\lambda\equiv g^2 N_c$
of the gauge theory with $g$ the gauge coupling constant,
and corresponding quantities in the gravity dual.
In the limit of large $N_{c}$ and large $\lambda$,
the evolution of the black brane geometry is governed by Einstein's equations 
\begin{equation}
R_{MN} - {\textstyle \frac{1}{2}} G_{MN} \left (R - 2 \Lambda \right) = 0,
\end{equation}
with cosmological constant $\Lambda = -6$. In this limit,  the back reaction of the string on the 
geometry is negligible meaning that we can solve Einstein's equations first,
independently of the string equations, and then subsequently determine the shape
of the string in a background given by the solution to Einstein's equations.

\subsection{Gravitational description of colliding sheets of energy}
\label{sec:CollidingSheets}

\begin{figure*}[t]
\includegraphics[scale=0.8]{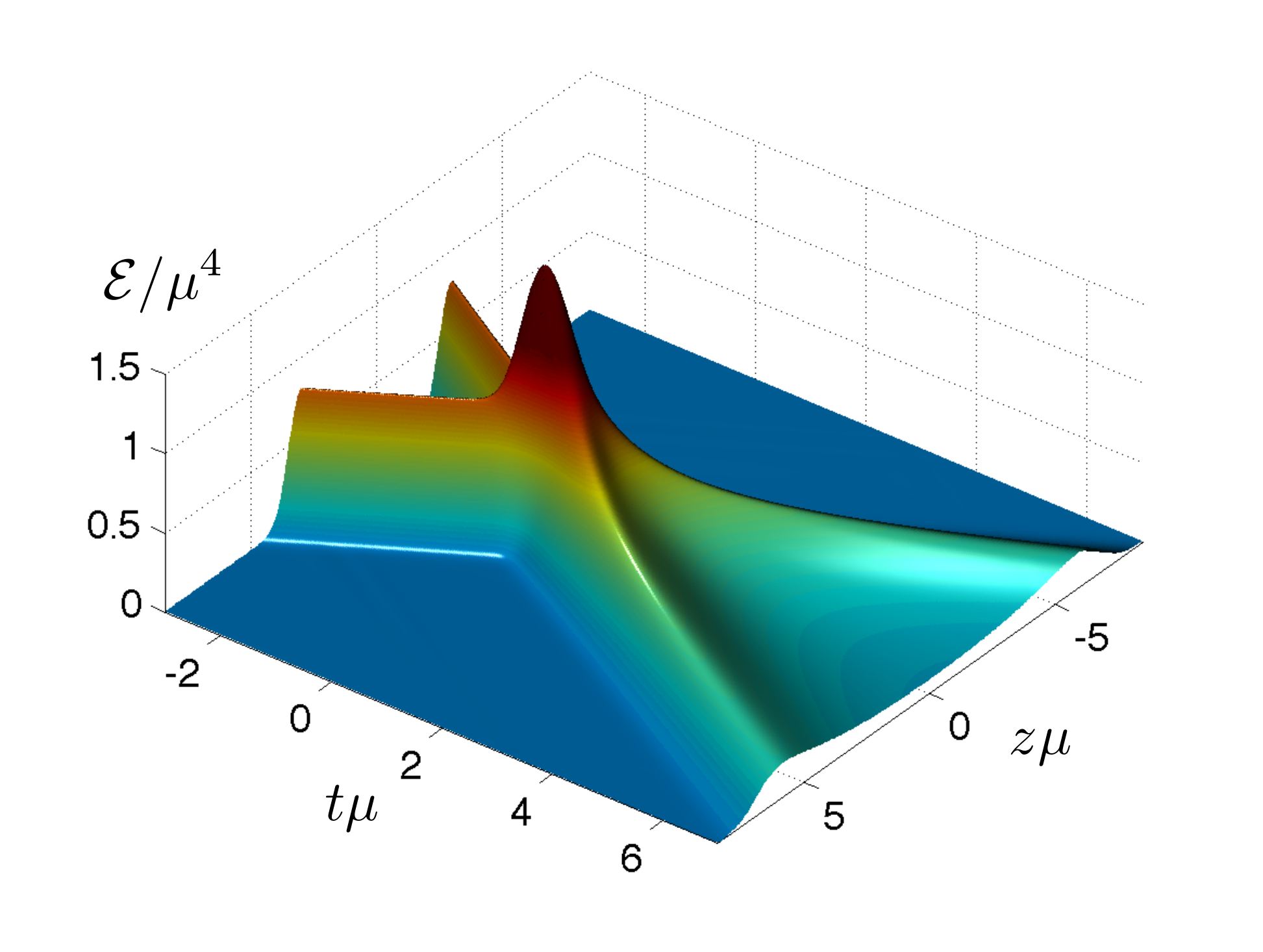}
\caption
  {
  A plot of the boundary energy density as a function of time $t$ and position along the `beam' direction
  $z$ for the colliding sheets of energy.
  Before $t = 0$, the energy density is that of two sheets with 
  Gaussian profiles moving toward each other at the speed of light.
  The sheets collide around time $t = 0$ and leave debris in the forward light-cone that subsequently hydrodynamizes
  as it expands, becoming strongly coupled, liquid, plasma.
}
\label{en_dens}
\end{figure*}

The geometry we choose to study is that of the colliding
gravitational shockwaves 
studied in Ref.~\cite{Chesler:2010bi}.  In the dual gauge theory living on the
boundary this geometry corresponds to 
colliding planar sheets of energy density.
In Fefferman-Graham coordinates 
the pre-collision metric is given by~\cite{Grumiller:2008va}
\begin{equation}
\label{fginitialdata}
ds^2 = \frac{-dx_+ dx_- + d \vec x_{\perp}^2 + u^4 \left [h_+ dx_+^2 + h_- dx_-^2\right ]+ du^2}{u^2},
\end{equation} 
where $x_{\pm} = t \pm z$ and $h_\pm \equiv h(x_{\pm})$ for some function $h(x_{\pm})$ that
specifies the profile of the incident gravitational shockwaves
and hence the incident sheets of energy in the boundary theory.  The boundary 
of the geometry is located at AdS radial coordinate $u = 0$.
The gravitational shockwaves move in the $\pm z$ direction at the speed of light.  
If $h(x_{\pm})$ has compact support then
$h(x_+)$ and $h(x_-)$ do not overlap in the distant past and the metric (\ref{fginitialdata})
is an exact pre-collision solution to Einstein's equations. 
Following Ref.~\cite{Chesler:2010bi}, we choose Gaussian profiles
\begin{equation}
h(x_{\pm}) = \mu^3 (2 \pi w^2)^{-1/2} e^{-\frac{1}{2} x_{\pm}^2/w^2},
\end{equation}
where $\mu$ defines the energy scale, meaning that we shall measure
all other dimensionful quantities in units of $\mu$.

For a given solution to Einstein's equations, the near-boundary behavior of the metric encodes 
the boundary stress tensor $ T^{\mu \nu} $ \cite{de Haro:2000xn,Skenderis:2000in}. 
In the boundary gauge theory the energy density of each shock is proportional $h(x_{\pm})$ \cite{Grumiller:2008va}.
The metric that we have described therefore corresponds in the boundary theory
to two sheets of energy density (infinite in extent, and translation-invariant,
in the transverse dimensions) that are moving towards each other in the $\pm z$
directions at the speed
of light. The energy density profiles of the sheets are Gaussians with
widths $w$, and the incident sheets each have an energy per unit area
that is given by $N_c^2 \mu^3/(2\pi^2)$~\cite{Chesler:2010bi}.
We choose the width of each shock to be $w = 0.5/ \mu$, meaning that
we shall be probing the collision of sheets of energy that are thinner by 
a factor of $2/3$  than those
in Ref.~\cite{Chesler:2010bi}.

Near the collision time, when the functions $h(x_\pm)$ begin to overlap, the metric (\ref{fginitialdata}) ceases to be a solution to Einstein's equations.
Using (\ref{fginitialdata}) as initial data in the distant past, one must therefore compute the future evolution of the geometry
numerically.  Our numerical scheme for solving Einstein's equations can be found in 
Refs.~\cite{Chesler:2010bi,PaulLarryToAppear}. 
%\kr{[[plus paul's upcoming paper with larry]]}.
In what follows we simply state some of the salient features.  A useful choice of coordinates for the numerical evolution is that 
of infalling Eddington-Finkelstein coordinates where the metric takes the form
\begin{align}
ds^2  = \frac{- A dt^2  + \Sigma^2 \[ e^B d \vec x_\perp^2 + e^{-2B} dz^2\] +2 dt \left ( F dz - du \right )}{u^2} 
%&=- A dv^2 + \Sigma^2 \[ e^B d \vec x_\perp^2 + e^{-2B} dz^2 \] + 2dv (dr + F dz) \notag \\
\label{ansatz}
\end{align}
where $A,B,\Sigma$ and $F$ are functions of $t$, $z$ and $u$.  Lines 
of constant time $t$ (and spatial coordinates $\vec x_\perp$ and $z$) are infalling null geodesics.%
\footnote
  {
  To the best of our knowledge there does not exist a closed form coordinate transformation
  taking the pre-collision metric (\ref{fginitialdata}) onto the metric (\ref{ansatz}) used for numerical 
  evolution.  Therefore, for initial data we compute the coordinate transformation
  required to put the initial metric (\ref{fginitialdata}) in the form of (\ref{ansatz}) numerically.  For details see
 Refs.~\cite{Chesler:2010bi,PaulLarryToAppear}. %\kr{[[plus paul's upcoming paper with larry]]}.
  }

An important practical matter when solving Einstein's equations is fixing the computational domain in $u$.  
The geometry we study in this paper contains a black brane with planar topology.   Moreover, the 
event horizon exists
in the infinite past, even before the collision takes place on the boundary~\cite{Chesler:2008hg,Chesler:2010bi}.
Therefore,  a natural choice is 
to excise the geometry inside the horizon, as this region is causally disconnected from the outside geometry.  To perform the excision
we identify the location of an apparent horizon (which always lies inside the event horizon) 
and choose to stop integrating Einstein's equations
any further into the black brane at its location.  Our choice of coordinates makes this procedure particularly simple.  The 
metric ansatz (\ref{ansatz}) is invariant under the residual diffeomorphism 
\begin{equation}
\label{shiftdiff}
\frac{1}{u} \to \frac{1}{u} + \xi(t, z),
\end{equation}
where $\xi$ does not
depend on the radial coordinate $u$ but is an arbitrary function of the boundary spacetime coordinates $t$ and $z$.  
One may fix $\xi$ by demanding that the location 
of the apparent horizon be at $u = 1$.  With this choice of coordinates,
the boundaries of the computational domain are static: $0<u<1$. 
% and an inverse radial coordinate $z \equiv 1/r$ can be used in the numerics with $z \in [0,1]$.
The numerical procedure for determining $\xi$ and solving Einstein's equations 
is described in Refs.~\cite{Chesler:2010bi,PaulLarryToAppear} %\kr{Add Paul+Larry recent} 
and we shall not review it here.
Following this procedure yields $\xi$ as well as the bulk metric functions $A$, $B$, $\Sigma$ and $F$,
whose asymptotic near-boundary behavior determines the 
stress-energy tensor $T^{\mu\nu}$ of the colliding sheets of energy in
the boundary gauge theory, as described in Ref.~\cite{Chesler:2010bi}.

In the distant past the apparent horizon lies very deep in the bulk.  This presents a computational problem for solving Einstein's equations numerically
as the coefficient functions in the metric  (\ref{fginitialdata}) diverge deep in the bulk.  As described in 
Ref.~\cite{Chesler:2010bi},
to regulate this problem we choose to study
study shocks which propagate and collide in a low temperature background
plasma of temperature $T_{\rm background}$.  In the distant past the effect of the background 
temperature is to push 
the apparent horizon up towards the boundary and thereby control the size 
of the metric coefficient functions deep in the bulk.
Our choice of background temperature is  $T_{\rm background}=0.085\,\mu$ which
corresponds to an initial background energy density 213 times smaller than the energy density
at the center of the sheets of energy and 450 times smaller than the energy density at $t=z=0$,
during the collision.

Fig.~\ref{en_dens} shows 
a plot of the rescaled energy density 
\beq
\mathcal E \equiv \frac{2 \pi^2}{N_{\rm c}^2}  T^{00}  
\label{RescaledEps}
\eeq
for our colliding sheets 
as a function of position and time.  Before $t = 0$, the energy density is that of two sheets of
energy with Gaussian profiles
moving toward each other at the speed of light.
The sheets collide around time $t = 0$ and leave debris in the forward light-cone.
We shall calculate the rate of energy loss of (i.e.~the `drag force' acting on) a heavy
quark moving through the far-from-equilibrium matter right near $t=z=0$, during
the collision when the energy density is largest.  We shall then follow the quark forward
in time and see how it loses energy in the expanding, cooling, fluid that forms in the
forward lightcone.
In gravitational terms, we wish to study the dynamics of a string moving in the
black brane geometry during and after the collision. 

After the collision, the fluid in the forward lightcone is expanding meaning that at
$z\neq 0$ it has a velocity that points in the direction of increasing $|z|$.
The fluid velocity is defined to be the 
future-directed time-like eigenvector of $T^{\mu}_{\ \nu}$,
normalized such that $u_\mu u^\mu = -1$: 
\begin{equation}
T^{\mu}_{\ \nu} u^\nu = - \frac{N_c^2}{2\pi^2}\,\varepsilon\, u^\mu,
\label{umuDefinition}
\end{equation}
with $- \varepsilon$ the eigenvalue, $\varepsilon$ being the
proper energy density rescaled as in (\ref{RescaledEps}).
At any spacetime point, we can find the local fluid rest frame by first using (\ref{umuDefinition})
to determine $u^{\mu}_{\rm lab~frame}$ and then boosting to a frame in which, at this
spacetime point, $u^\mu$ is given by  $u^{\mu}_{\rm rest~frame}=(1,0,0,0)$.  In this local fluid rest frame,  
the stress-energy tensor at the spacetime point of interest is diagonal and can be written as
\beq
T^{\mu\nu}_{\rm rest~frame} =\frac{N_c^2}{2\pi^2}\, {\rm diag}(\varepsilon,{\cal P}_\perp,{\cal P}_\perp,{\cal P}_\parallel)\,.
\label{RestFrameTmunu}
\eeq
Note that at $z=0$ the velocity of the fluid vanishes by symmetry, 
meaning that at $z=0$ in (\ref{umuDefinition}) we
have simply $u^\mu_{\rm lab~frame}=u^{\mu}_{\rm rest~frame}=(1,0,0,0)$ 
and $\varepsilon={\cal E}= T^{00}(2\pi^2)/N_c^2$.

\begin{figure}
\includegraphics[scale=0.45]{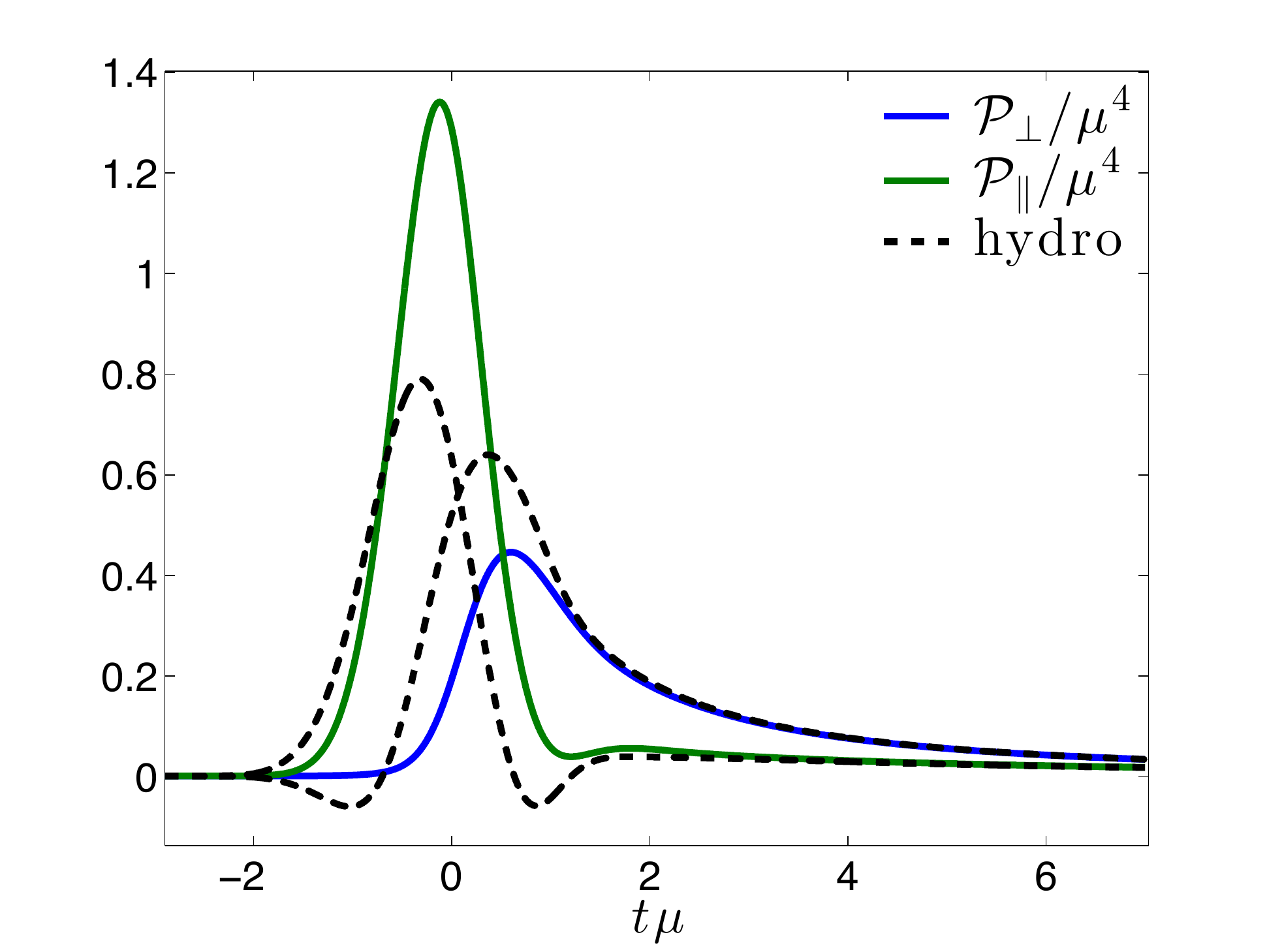}
\caption{Pressure, both parallel to the direction of motion of  the colliding
sheets (${\cal P}_\parallel$) and in the transverse directions (${\cal P}_\perp$), 
as a function of time at $z=0$ during the collision
of the sheets of energy illustrated in Fig.~\ref{en_dens}. 
We compare the pressures in the collision, shown as solid colored curves,
to what they would be if the hydrodynamic
constitutive relations were satisfied, shown as dashed curves. We see
that the strongly coupled fluid produced in the collision hydrodynamizes
to a reasonable degree 
before $t=2/\mu$, although with the more strict operational criterion defined in the text, the hydrodynamization
time is $t_{\rm hydro}=2.8/\mu$.
}
\label{pressures}
\end{figure}

%So, at $z=0$ as in Fig.~\ref{pressures}, the fluid is at rest in the  `lab frame', meaning that 
%the pressures plotted in that figure can be obtained directly from the stress-energy tensor
%in the lab frame, without the need for a boost.

Before we turn to describing the dynamics of the string and then to calculating the rate at which
the heavy quark loses energy, we should justify referring to the matter that is formed in the
forward light-cone as a fluid.  In Fig.~\ref{pressures} we compare the pressures in the
`beam' direction and perpendicular to it at $z=0$ to what those pressures would be if the matter
were a fluid described by the equations of viscous hydrodynamics for the ${\cal N}=4$ SYM
plasma, to first order in a derivative expansion.  In particular, the hydrodynamic
constitutive relations allow us to determine ${\cal P}_{\parallel}$ and ${\cal P}_\perp$
%from the energy density and the fluid velocity.  
from the 
proper energy density $\varepsilon$ and the fluid velocity $u^\mu$, determined
via (\ref{umuDefinition}).  (Because we are making this comparison at $z=0$, where the
fluid is at rest in the lab frame, in this instance (\ref{umuDefinition}) is 
trivial as described above.)
We plot the pressures determined 
by applying the first order hydrodynamic
constitutive relations to $\varepsilon$ and $u^{\mu}$ 
%determined in this way
as dashed curves in Fig.~\ref{pressures}. 
We see that from a time that is clearly  
before $2/\mu$ onwards the flow of the fluid is described very well by hydrodynamics.
We can use the same operational definition of the hydrodynamization time $t_{\rm hydro}$
as in Ref.~\cite{Chesler:2010bi}, namely the time after which the actual pressures within the 
matter produced in the
initially far-from-equilibrium collision are both within 15\% of the values derived
from the hydrodynamic constitutive relations.
With this definition, in Fig.~\ref{pressures} hydrodynamization occurs
at $t=t_{\rm hydro}=2.8/\mu$.  It is clear from the Figure that one could choose a reasonable
but less strict criterion for hydrodynamization with respect to which hydrodynamization occurs
well before $2/\mu$. The reason that with the strict criterion $t_{\rm hydro}$ 
is a little later is that ${\cal P}_\parallel$ is so small that a 15\% relative
deviation corresponds to a very small absolute deviation.

%Note that $t_{\rm hydro}\mu$ is smaller here than in Ref.~\cite{Chesler:2010bi} by
%a factor of $XX$, corresponding to the fact that the sheets of energy whose collisions
%we are probing are thinner by a factor of $2/3$ than those in Ref.~\cite{Chesler:2010bi}.

We shall calculate the drag force
on the heavy quark as it plows through the matter produced in the collision both
before the hydrodynamization time $t_{\rm hydro}$, when the matter is far from equilibrium,
and after $t_{\rm hydro}$ when we have an expanding, cooling, hydrodynamic liquid.
As in Ref.~\cite{Chesler:2010bi}, we see from Fig.~\ref{pressures} that at the
hydrodynamization time $t_{\rm hydro}$ the parallel and transverse pressures
are very different.  The fluid is very anisotropic at the time that it  hydrodynamizes
and becomes locally isotropic only at a much later time, beyond that shown in the Figure.
We shall therefore be studying the drag force on a heavy quark when it is first
in the far-from-equilibrium matter at $t\sim 0$, well before hydrodynamization,
and then later in an expanding cooling  hydrodynamic fluid that is far from isotropic.

\subsection{String dynamics}
\label{sec:StringDynamics}

For a given solution to Einstein's equations, for example the one above
that describes
the dynamics of a black brane in the bulk and the collision, hydrodynamization,
and hydrodynamic flow of the strongly coupled matter in the boundary
theory, we can add an infinitely heavy quark moving through the boundary theory
matter by adding a string
in the bulk geometry whose endpoint follows the trajectory of the quark along
the boundary.
The dynamics of the string are governed by the Polyakov action
\beq
S_{\rm P} = - \frac{T_0}{2} \int d^2 \sigma \sqrt{-\eta} \eta^{ab} G_{MN} \partial_a X^M \partial_b X^N,
\label{Polyakov}
\eeq
where $\sigma_1  \equiv \tau$ is a temporal worldsheet coordinate, $\sigma_2 \equiv \sigma$ is a spatial worldsheet coordinate, 
$X^M(\tau,\sigma) = \left (t(\tau,\sigma),\vec x(\tau,\sigma), u(\tau,\sigma) \right)$ are the string embedding functions
that describe where the string is located within the spacetime metric that we are interested
in probing, 
$\eta_{ab}$ is the worldsheet metric and $T_0 = \frac{\sqrt{\lambda}}{2\pi}$ is the string tension.  
We choose worldsheet coordinates such that the worldsheet metric takes the form
\beq
\eta_{ab} = 
\left(
\begin{array}{cc}
  -\alpha(\tau,\sigma) & -1     \\
 -1 &      0
\end{array}
\right)
\label{eta}
\eeq
with $\alpha(\tau,\sigma)$ being an arbitrary function.  Just as we did
in the spacetime metric (\ref{ansatz}), we have chosen
worldsheet coordinates that are infalling Eddington-Finkelstein coordinates.  Lines of constant 
$\tau$ are infalling null worldsheet geodesics.  As we describe further below, 
we fix the function $\alpha(\tau,\sigma)$
by demanding $\sigma = u$.

Varying the Polyakov action with respect to the embedding functions $X^M$ yields dynamical equations of motion.
Likewise, varying the Polyakov action with respect to the worldsheet metric yields a system of constraint equations.
With our choice of infalling Eddington-Finkelstein bulk and worldsheet coordinates, the dynamical equations of motion
take the simple form
\beq
\dot X'^N + \Gamma_{AB}^N \dot X^A X'^{B} = 0 
\label{eoms}
\eeq
where $\Gamma_{AB}^N$ are Christoffel symbols associated with the metric (\ref{ansatz}) for the bulk
spacetime and 
\begin{subequations}
\begin{eqnarray}
\dot X^M &\equiv& \partial_\tau X^M - \frac{1}{2} \alpha X'^M\ ,\label{dotdef}\\
X'^M &\equiv& \partial_\sigma X^M\ .\label{primedef}
\end{eqnarray}
\end{subequations}
The constraint equations are even simpler:
\begin{subequations}
\label{cons}
\begin{eqnarray}
\label{temporalconstraint}
X'^2 &=& 0, 
\\
\dot X^2 &=& 0.
\label{boundaryconstraint}
\end{eqnarray}
\end{subequations}
The constraint equation (\ref{temporalconstraint}) is a temporal constraint.  If (\ref{temporalconstraint}) 
is satisfied every where in space at one time, the dynamical equations (\ref{eoms}) imply that it will be satisfied at all subsequent 
times.
The constraint equation (\ref{boundaryconstraint}) is a boundary constraint.  If (\ref{boundaryconstraint})
is satisfied at all times at one value of $\sigma$ the dynamical equations (\ref{eoms}) imply it will be satisfied at all 
$\sigma$.

Given $X^M$ for all $\sigma$ at some %(early) 
value of 
$\tau$, the equations of motion (\ref{eoms}) constitute a linear system of 
ordinary differential equations for $\dot X^M$.  To solve these equations,
in addition to specifying the initial conditions for $X^M$
one must specify five boundary conditions at all $\tau$ 
at the AdS boundary, where the string endpoint is located.  
Three
boundary conditions are simply that the string endpoint moves on a given trajectory $\vec x_o(t)$
\begin{equation}
\lim_{\sigma \to 0} \vec x(\tau,\sigma) = \vec x_o(t).
\end{equation}
We choose the trajectory
\begin{equation}
\label{xbc}
\vec x_o(t) = \vec \beta t,
\end{equation}
for some constant velocity $\vec \beta$.  
These boundary conditions correspond to choosing to study
a heavy quark, whose location after all coincides with the location
of the endpoint of the string, that is moving with constant velocity $\vec \beta$
and that finds itself at $z=0$ at time $t=0$, meaning that this 
heavy quark finds itself right in the center of the collision that we
wish to probe.  We shall present results for several different values of $\vec \beta$.

%For these boundary conditions the quark, whose location coincides 
%with the location of the string endpoint, passes through the collision point $z = 0$ of the shocks at time $t = 0$.

Two more boundary conditions are required.  For convenience, we choose 
\begin{equation}
\label{tbc}
\lim_{\sigma \to 0} t(\tau,\sigma) = \tau,
\end{equation}
so that the worldsheet time $\tau$ corresponds to the coordinate time $t$ at the boundary.
The remaining boundary condition comes from 
demanding that the boundary constraint 
(\ref{boundaryconstraint}) is satisfied at $\sigma = 0$. The boundary conditions (\ref{xbc})
and (\ref{tbc}) and the boundary constraint (\ref{boundaryconstraint}) are satisfied at $\sigma = 0$
provided that %
\footnote
  {
  To see that this is true one can solve the string equations of motion (\ref{eoms}) and the constraint
  equations (\ref{cons}) with a power series expansion in $\sigma$ near $\sigma = 0$.  In doing so one can 
  directly incorporate the boundary conditions (\ref{xbc}) and (\ref{tbc}) into the series expansions.  With the expansion known
  to order $\sigma$, one can easily see that the boundary limit of $\dot X^M$ takes the form (\ref{dotboundaryconditions}).
  }
\begin{align}
\label{dotboundaryconditions}
\lim_{\sigma \to 0} \dot t =  \frac{1}{2} \left [ 1 + \frac{1}{\gamma} \right ], \
\lim_{\sigma \to 0} \dot {\vec x} =  \frac{\vec \beta}{2}  , \
\lim_{\sigma \to 0} \dot u = -  \frac{1}{2\gamma}\ ,
\end{align}
where $\gamma\equiv 1/\sqrt{1-\vec \beta^2}$ as usual.

We can now see how to evolve the string forward in $\tau$, given
an initial string profile specified by  $X^M$ at some initial $\tau$.
The algorithm has two steps.  First we need to
obtain $\dot X^M= \left(\dot t, \dot {\vec x}, \dot u \right)$ at the initial $\tau$.  
We do this by observing that if we think of $\dot X^M$ as the dynamical
variables,
the equations of motion (\ref{eoms})
are first order differential equations for $\dot X^M$ with the 
independent variable being $\sigma$.  
We can solve these equations for $\dot X^M$ as functions of $\sigma$ at the initial $\tau$ as long as we
know $X'^{M}$ --- which we do since we have been given $X^M$ for all $\sigma$ at the initial $\tau$ --- and
as long as we have boundary conditions for $\dot X^M$ at $\sigma=0$ --- which
we have in (\ref{dotboundaryconditions}).  So, we solve (\ref{eoms}) starting from $\sigma=0$
and obtain $\dot X^M$ at all $\sigma$ at the initial $\tau$.
In the second step of the algorithm, 
we use (\ref{dotdef}), 
 rewritten as $\partial_\tau X^M = \dot X^M + \frac{1}{2} \alpha X'^M$, to compute
 the field velocities $\partial_{\tau} X^M$.
 To do this we need the function $\alpha$.
 %can by computed via (\ref{dotdef}), $\partial_\tau X^M = \dot X^M + \frac{1}{2} \alpha X'^M$.  
With the gauge choice $\sigma = u$, it is given 
simply by $\alpha = -2 \dot u$.  With $\partial_\tau X^M$ known, we can then
determine $X^M$ at the new $\tau$, thus completing the evolution of the system from the initial $\tau$
to a $\tau$ one time-step later.  We then repeat.

All that we still need to specify is our choice of initial conditions for $X^M$
at an initial time $t$ that we shall typically take to be $-3/\mu$, when the centers of the incident sheets
of energy are $6/\mu=12w$ apart.
In the distant past, well before the sheets of
energy collide, the near-boundary geometry between the sheets of energy, which
is where the heavy quark is located, is that of an equilibrium black
brane with a small temperature $T_{\rm background}$. 
This is so provided that we do not choose $|\vec \beta|$ so
large that at our initial time the heavy quark is within one of the incident sheets.
We shall make sure not to do this, which is to say that we shall make sure that
at our initial time the heavy quark has not yet felt the sheets of energy that
are soon going to hit it to any significant degree.
We therefore choose initial conditions 
such that the %near-boundary 
string profile coincides with the trailing string 
solution of Refs.~\cite{Herzog:2006gh,Gubser:2006bz} moving at 
velocity $\vec \beta$, given in Appendix~\ref{app:ts}.  
Such initial conditions satisfy the temporal constraint equation (\ref{temporalconstraint}) near 
the AdS boundary.\footnote{
  Our choice of initial conditions is slightly complicated by our choice of infalling Eddington-Finkelstein coordinates.
  In infalling Eddington-Finkelstein coordinates the bulk is the causal future of the boundary.  At any given Eddington-Finkelstein time 
  $t$, even in the distant past before the sheets of energy collide on the boundary, 
  the gravitational shocks  are colliding somewhere
  deep in the bulk.  As a consequence of this, choosing the trailing string of Refs.~\cite{Herzog:2006gh,Gubser:2006bz} will lead to a small violation of the temporal constraint equation
  (\ref{temporalconstraint}) deep in the bulk.  However, as 
  time progresses the violation of the constraint decreases in magnitude and then as time progress further
  towards $t = 0$ when the sheets collide on the boundary, the portion of the 
  string that violates the temporal constraint equation is rapidly enveloped by the event horizon of the black brane.
  Because of this, the initial violation of the temporal constraint equation is causally disconnected from physics near the boundary and hence is of no concern. We discuss this further in Appendix~\ref{app:ts}.
  }

Constructing initial string profiles that are equilibrium solutions to the pre-collision geometry ensures that the 
future non-trivial evolution of the string is entirely due to the change in the bulk geometry 
associated with the collision event, not due to transients that would come along as artifacts of any
other choice of initial conditions.  However, as we shall describe below, although 
perturbing 
the initial string profile 
does result in an early-time transient before the collision happens, the change that results in the rate of energy
loss of the heavy quark
\textit{during and after} the collision event is negligible.
Because in an actual heavy ion collision a heavy quark is produced
during the collision, we are not particularly interested in any aspect of the motion
of our heavy quark before the collision of our sheets of energy. This, together
with the insensitivity of the results during
and after the collision that we are interested in to initial conditions and associated
transients, means that our study could be repeated with other choices
of initial conditions for the string without changes to our conclusions.
%Note that before the collision happens, the quark should not be considered physical in the sense that in reality quark should be created during collision as the consequence of hard interaction. In the framework that we are considering, setting up the quark to have the string profile as in equilibrium ensures that string reacts to the change of the bulk geometry properly and constraint equations are satisfied.
%If perturbing the initial string profile, then constraint equations are not satisfied for the time corresponding to several sheet thickness, but once constraint equations are satisfied, the drag force relaxes to the value as if the string was set with equilibrium profile. 
%At late time, once the hydrodynamics start working, and the drag force as obtained in section III is independent of the initial conditions of the string.

We discretize the $\sigma$ coordinate using
pseudo-spectral methods.  Specifically, we decompose the $\sigma$ dependence of all functions in terms of 
the first 20-35 Chebyshev polynomials.   We then time evolve the string profile, typically from $t=-3/\mu$
to $t=+6/\mu$,
according
to the algorithm that we have described above using 
a fourth order Runge-Kutta ordinary differential equation solver.

\subsection{Extracting the drag force acting on the quark from the string profile}
\label{sec:ExtractingTheForce}

Because the quark is a point-like source
which is being pushed through the surrounding medium at 
a constant speed by some external agent, it must be
transferring energy and momentum to the surrounding medium meaning that the 
boundary stress tensor is not conserved:
\begin{equation}
\label{boundarycons}
\partial_\mu T^{\mu \nu}(x)  = F^\nu(x) = - f^\nu \delta^3 ( \vec x - \vec \beta t ),
\end{equation}
where $f^\nu$ is the four momentum lost by the quark per unit time.
Note that, as always for a force, $f^\nu$ does not
transform as a four-vector under Lorentz transformations.  
It is conventional to write
\beq
\frac{d p^\nu}{dt}=f^\nu\,,
\eeq  
with $p^\nu$ the four momentum of the quark, even though
in the setup we are analyzing the four momentum of the quark
does not actually change.  The external agent does work in
order to move the quark at constant speed and, in so doing,
transfers energy and momentum to the medium surrounding the
quark; the quark does not slow down.
If the external agent
dragging the quark is
a classical background electric field, the total stress tensor 
\begin{equation}
T^{\mu \nu}_{\rm tot} = T^{\mu \nu} + T^{\mu \nu}_{\rm EM},
\end{equation} 
is conserved
\begin{equation}
 \partial_\mu T^{\mu \nu}_{\rm tot} = 0\ .
 \end{equation}  
 Via the quark, energy and momentum flow from the electric field into the surrounding medium.   
 The quantity $f^\nu$ is thus given by 
\begin{equation}
 f^\nu(t) = - \int d^3 x\, \partial_\mu T^{\mu \nu}(x)
=  \int d^3 x\, \partial_\mu T^{\mu \nu}_{\rm em}(x)\ .
\end{equation}

Just as the near-boundary behavior of the metric encodes the expectation value of the boundary stress tensor, the near
boundary behavior of the string profile encodes $f^\nu$.  Specifically, $f^\nu$ can be identified as the 
flux of four momentum down the string~\cite{Herzog:2006se}.  The flux down the string can be extracted by 
noting that the total action for the holographic system, 
\begin{equation}
S_{\rm tot} = S_{\rm P} + S_{\rm EM},
\end{equation}  
must be diffeomorphism invariant.  The electromagnetic action $S_{\rm EM}$ only has support at the boundary and couples the string endpoint to the classical background
electric field used to drag the quark (which is to say the string endpoint) at constant velocity.

Under an infinitesimal diffeomorphism $X^M \to X^M + \chi^M$ the variation in the Polyakov action
(\ref{Polyakov}) only has support at the boundary and reads
\begin{equation}
\delta S_{\rm P} =  \lim_{ u \to 0} \int d^4 x \, n_N\, \chi^M \sqrt{-G}\, \mathcal T^{N}_{\ M},
\end{equation}
where $D_M$ is the covariant derivative with
respect to the bulk spacetime metric (\ref{ansatz}), $n_N$ is the normal to the boundary at $u = 0$, and 
\beq
\mathcal T^{MN}(Y) = \frac{-T_0}{\sqrt{-G}} \int d^2\sigma \sqrt{-\eta}\, \eta^{ab}\, \partial_a X^M \,\partial_b X^N \delta^5(Y-X),
\label{TMN}
\eeq
is the string stress tensor.  
Likewise, the variation in the electromagnetic action is
\begin{equation}
\delta S_{\rm EM} = \int d^4 x \chi^\mu \partial_\nu T^{\nu}_{\mu\,{\rm EM}} = -\int d^4 x \chi^\mu F_\mu,
\end{equation}
where in the last line we used $\partial_\nu T^{\nu}_{\mu\,{\rm EM}}  = - F_\mu$ with $F_\mu$ defined in (\ref{boundarycons}).
Upon demanding that the variation in the action vanish for all $\chi^\mu$, we conclude that
\begin{equation}
\label{boundaryflux1}
F_\mu = \lim_{\rm u \to 0} n_N \sqrt{-G} \mathcal T^N_{\ \mu},
\end{equation}
from which the drag $f^\mu$ can easily be extracted via (\ref{boundarycons}).

Setting the normal to the boundary $n_M = \delta_{M5}$, one might naively 
conclude\footnote{As we did, in a preliminary version of this study~\cite{Chesler:2012pw}.
The drag forces calculated in Ref.~\cite{Chesler:2012pw} are not correct but, comparing the figures
from that preliminary report to those in Section~\ref{sec:dis} where we present our results, 
the differences are not large.}
 that
$F_\mu = \lim_{\rm u \to 0} \sqrt{-G} \mathcal T^5_{\ \mu}$.  However, this cannot be correct. 
Near the boundary some components of the string stress tensor diverge like $1/u^2$.
As a consequence of this divergence $\lim_{\rm u \to 0} \sqrt{-G} \mathcal T^5_{\ \mu}$ 
transforms nontrivially under the {\em radial} diffeomorphism (\ref{shiftdiff}).  Simply put, the expression
$\lim_{\rm u \to 0} \sqrt{-G} \mathcal T^5_{\ \mu}$ depends on one's choice of the gauge parameter $\xi$ used
in solving Einstein's equations.

The unique remedy to the above problem is to set $n_M = \delta_{M5} +u^2 \partial_M \xi$.  This choice is simply the near-boundary limit of $\delta_{5N}$, transformed 
by (\ref{shiftdiff}).  In other words, $n_M=\delta_{M5}$ would have been correct if we had
used $\xi=0$ but is not correct once we make the transformation (\ref{shiftdiff}) that
we needed to make in order to solve Einstein's equations.
With the correct choice of $n_M$, $F_\mu$ in (\ref{boundaryflux1}) is invariant 
under all infinitesimal diffeomorphisms, including those arising from Eq.~(\ref{shiftdiff}).  We therefore conclude 
that
\beq
F_\mu = \lim_{u\to0} \[\sqrt{-G} \( \mathcal T^5_{~\mu} + u^2 \partial_\alpha \xi \(\mathcal T^\alpha_{~\mu} - \delta^\alpha_\mu \mathcal T^5_{~5} \) \)\].
\label{F1}
\eeq

We note that $f^\mu$ can also be expressed in terms of the canonical worldsheet fluxes 
\beq
\pi_M^a \equiv \frac{\delta S_{P}}{\delta \( \partial_a X^M \)} = - T_0\,\sqrt{-\eta}\, G_{MN}\, \eta^{ab}\, \partial_b X^N.
\label{pidef}
\eeq
A straightforward exercise using the string stress (\ref{TMN}) and (\ref{pidef}) shows that
\beq
f_\mu= \lim_{u\to0}\[ \pi^\sigma_\mu + u^2\, \pi^\sigma_5\, \partial_\mu \xi - u^2\, \pi^a_\mu\, \partial_\alpha \xi \,\partial_a X^\alpha  \]\ .
\label{f2}
\eeq
We see that 
for the choice $\xi = 0$ the usual identification of force in terms of the canonical 
fluxes~\cite{Herzog:2006gh,Gubser:2006bz} is reproduced.

\section{Results}
\label{sec:dis}

In this section we present and discuss the results that we have obtained 
for the drag force on a heavy quark being dragged through the colliding sheets
of Fig.~\ref{en_dens}.  We shall initially choose the quark to be moving with
a velocity $\vec \beta$ in the plane perpendicular to the direction of motion
of the colliding sheets.  Later we shall consider the more general case
where the velocity of the quark has components both parallel to and perpendicular
to the `beam' direction.  We shall compare our results to expectations based
upon comparing  the drag force that we calculated to what
it would have been in a static homogeneous plasma with the same instantaneous
energy density or parallel pressure or transverse pressure as that at the
position of the quark.   To further explore the effect of a time-dependent
background on the drag force, in Appendix~\ref{app:step} we consider a toy example in which the temperature
increases with time but the plasma remains homogeneous and isotropic and does not undergo any expansion.

\subsection{Heavy quark with zero rapidity}
\label{sec:ZeroRapidity}

\begin{figure}
\includegraphics[scale=0.65]{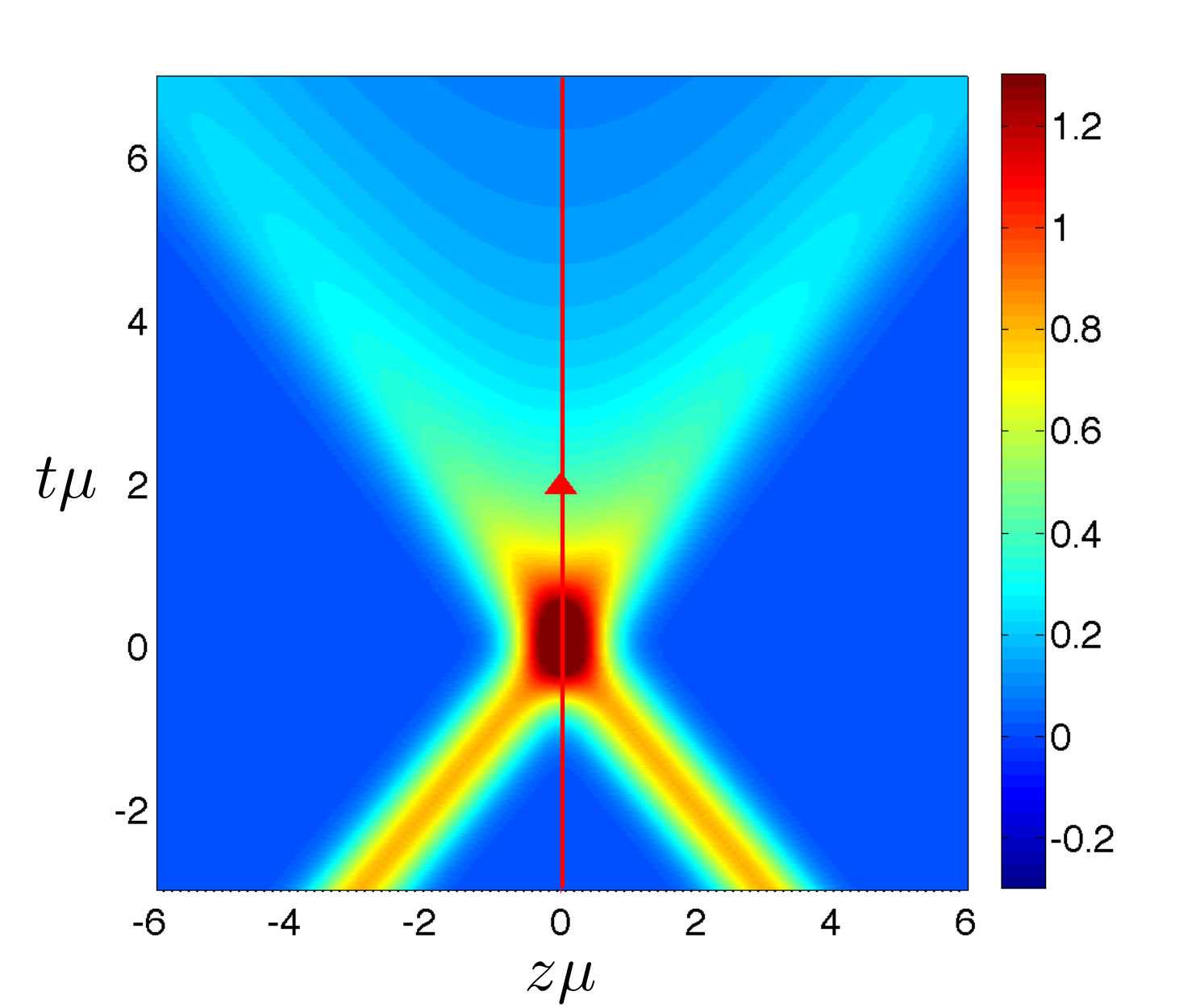}
\caption{The path in the $(z,t)$-plane
of a quark moving in the $z=0$ direction, i.e.~with zero rapidity,
is shown as the red line superimposed on a color-plot of the rescaled energy density $\mathcal E$, see
Eq.~(\ref{RescaledEps}), in units of $\mu^4$.   Thinking of the $x$-axis as perpendicular to
the page, the quark is moving out of the page with velocity $\beta_x$.
 }
\label{z0path}
\end{figure}

\begin{figure}
\includegraphics[scale=0.48]{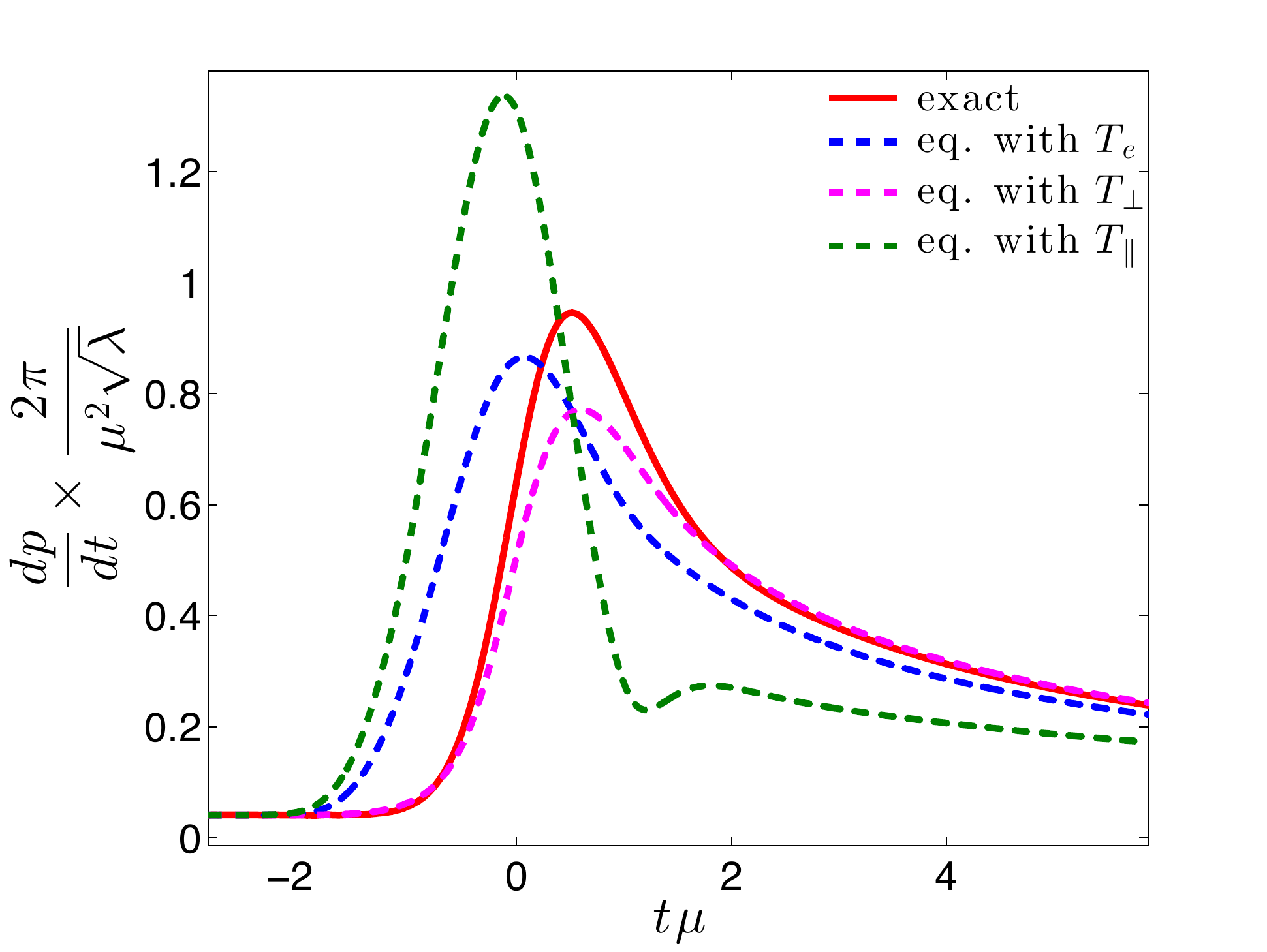}
\caption{Drag force in units of $\mu^2\sqrt{\lambda}/(2\pi)$
on a quark being pulled through the collision in the $x$-direction at $z=0$,
i.e.~at zero rapidity, with velocity $\beta_x=0.5$.
The dashed curves show the drag force that the quark would experience in an equilibrium plasma
with the same instantaneous energy density, parallel pressure or perpendicular pressure as that
at the location of the quark.  The dashed curves are described further in the text.
%equilibrium drag result with temperature $T_e$ associated with energy density , temperature $T_\perp$ associated with transverse pressure and temperature $T_\parallel$ associated with the parallel pressure.   
}
\label{betax05}
\end{figure}

\begin{figure}
\includegraphics[scale=0.48]{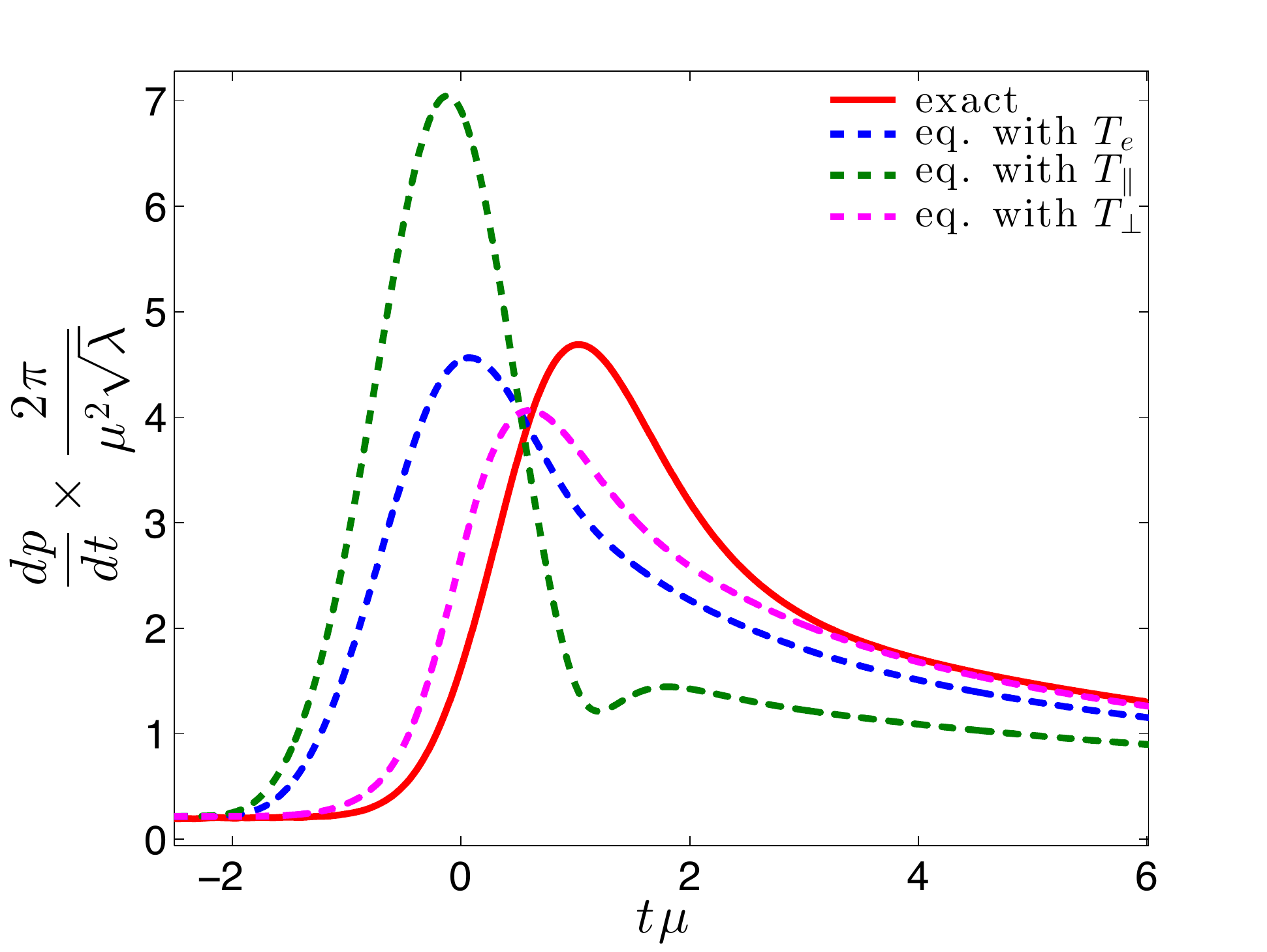}
\caption{Same as Fig.~\ref{betax05} but for a quark moving at zero rapidity
with a velocity $\beta_x=0.95$.
%Drag force on a quark being pulled through the collision at $z=0$ along the $x$-direction,
%i.e. at zero rapidity, with velocity $\beta_x=0.95$.
%The dashed curves show the drag force that the quark would experience in an equilibrium plasma
%with the same instantaneous energy density, parallel pressure or perpendicular pressure as that
%at the location of the quark.  The dashed curves are described further in the text.
}
\label{betax095}
\end{figure}

We begin with the case where the heavy quark is moving perpendicular to the
`beam' direction, which is to say perpendicular to the $z$-direction along which
the sheets of energy collide.  Although our heavy quark was present before the
collision, from a phenomenological perspective this calculation may inform how
we think about a heavy quark that is produced 
at $t=0$ in a heavy ion collision with zero rapidity 
moving with some perpendicular velocity $\vec \beta$.
The $z=0$ path along which the heavy quark is moving is illustrated in
Fig.~\ref{z0path}, as is the energy density through which it moves.
The parallel and perpendicular pressures of the material in which it
finds itself are those that we plotted in Fig.~\ref{pressures}.  
%
%
%When quark is moving along $z=0$, as shown in the Figure \ref{z0path} where the red line represents the path of the quark superimposed on the energy density color map, the drag force is directed along the drag velocity vector.
%
%The string is initialized using initial background temperature $T = T_{\rm bkg}$ and evolved in the time period $-3 < t\mu < 7$. 
The case of a quark moving at zero rapidity is simpler than the more general case
that we will turn to next 
for two reasons:
(i) a quark moving along the $z=0$ plane is always moving through fluid at rest,
with $u^\mu=(1,0,0,0)$, meaning that in this case the local fluid rest frame is the
same as the lab frame;
(ii) the drag force acting on the quark with zero rapidity
is directed antiparallel to the velocity vector, with no component of the force perpendicular
to the direction of motion of the quark.  
We have calculated the drag force as described in Section~\ref{sec:Methods}. Our results
for quarks moving at zero rapidity with speed $\beta_x=0.5$ and $\beta_x=0.95$ 
are shown as the solid red curves in 
Figs.~\ref{betax05} and \ref{betax095}, respectively.

We are interested in comparing our results to expectations 
based upon the classic result for the drag force
required to move a heavy quark with constant velocity $\vec \beta$
through the {\it equilibrium} plasma in strongly coupled SYM 
theory, namely~\cite{Herzog:2006gh,Gubser:2006bz}
\beq
\left.\frac{d \vec{p}}{dt}\right|_{\rm eq} =  \frac{\sqrt{\lambda}}{2\pi}\,(\pi T)^2\frac{\vec\beta}{\sqrt{1-\beta^2}}\,,
\label{EqbmDrag}
\eeq
where $T$ is the temperature of the equilibrium plasma. Out of equilibrium, the
matter does not have a well-defined temperature.  We can nevertheless
use (\ref{EqbmDrag}) to frame expectations for $d\vec p/dt$ at any point
in spacetime, as follows.  At $z=0$ the fluid is at rest, meaning that
the lab frame in which we are working is
%transform to 
the local fluid rest frame
%as described above Eq.~(\ref{RestFrameTmunu}), obtaining 
and the stress tensor
for the fluid takes the form (\ref{RestFrameTmunu}) which we now
rewrite as
\beq
T^{\mu\nu}_{\rm rest~frame} =\frac{\pi^2 N_c^2}{8}\, {\rm diag}(3 T_e^4, T_\perp^4 , T_\perp^4 , T_\parallel^4)\,.
\label{RestFrameTmunu2}
\eeq
If the fluid were at rest in equilibrium, we would have $T_e=T_\perp=T_\parallel=T$.
In the nonequilibrium setting of interest, through (\ref{RestFrameTmunu2}) we have defined
three different ``effective temperatures'', no one of which is a true temperature since none
can be defined.  We can then use each of these three ``temperatures'' in the equilibrium
expression (\ref{EqbmDrag}), obtaining the three dashed curves in Figs.~\ref{betax05} and \ref{betax095}.
Because the stress tensor is traceless, any one of the dashed curves can
be obtained from the other two; only two are independent.  This 
also explains why wherever two of the dashed curves cross
the third must cross also.
Note that none of the three dashed curves should be seen as a ``prediction'' for the actual drag force experienced
by the heavy quark in the far-from-equilibrium conditions created by the collision of the sheets of energy.
Rather, the three dashed curves tell us what the drag force would be in a static plasma
with the same instantaneous energy density or transverse pressure or parallel pressure
as that present in the far-from-equilibrium conditions at time $t$ and position $z=0$.
The dashed curves are devices through which we use what we know about
the drag force in an equilibrium plasma to  frame expectations for the nonequilibrium 
case.\footnote{  
Some support for this strategy is provided by analyses~\cite{Giecold:2009wi,Stoffers:2011fx,Abbasi:2012qz}
of the drag force on a heavy quark moving through a fluid that is undergoing boost invariant
expansion in one dimension and is translation invariant in the other two dimensions.
In this setting, the gradient expansion of hydrodynamics becomes an expansion in
powers of $1/\tau^{2/3}$, with $\tau$ the proper time. (See, for example, Refs.~\cite{Janik:2005zt,Chesler:2009cy}.)
At leading order, which is to say at late times, the fluid can be treated as ideal meaning
that $T_e=T_\perp=T_\parallel\equiv T(\tau)$ with $T(\tau)\propto 1/\tau^{1/3}$~\cite{Bjorken:1982qr}. In this
approximation, the drag force on the heavy quark is indeed given by (\ref{EqbmDrag}) 
with $T$ replaced by $T(\tau)$~\cite{Giecold:2009wi,Stoffers:2011fx}.  
Ref.~\cite{Abbasi:2012qz} includes an investigation of the next order corrections.}

Comparing the drag force on the heavy quark, the solid red curves in Figs.~\ref{betax05} and \ref{betax095},
to the three dashed curves yields many observations.  Reading the figures from left to right, we first
see that at very early times where $T_e=T_\perp=T_\parallel=T_{\rm background}$
the drag force is indeed given by the equilibrium result (\ref{EqbmDrag}) for a static plasma
with temperature $T_{\rm background}$.  Next, we see that the increase in
the drag force due to the dramatic change in the stress tensor of the fluid corresponding
to the collision of the sheets of energy is delayed.  And, the delay in the increase of
the drag force seems to increase with increasing heavy quark velocity. We shall return to this below.
Third, we see that the peak value of the drag force 
is comparable to the expectations provided by the dashed curves, meaning
that the peak value of the drag force
in the far-from-equilibrium matter produced in the collision
 is not dramatically smaller or larger
than what it would be in a static plasma with the same instantaneous
energy density.  These second and third observations suggest
that a reasonable first-cut approach to modelling the drag
on a heavy quark in a heavy ion collision would be to 
turn the drag force on roughly one ``sheet thickness'' in time after the 
collision and from then on use
the equilibrium expression (\ref{EqbmDrag}) with an effective
temperature determined by the instantaneous energy density.
Fourth, at late times when the expansion of the fluid
is described well by viscous hydrodynamics we see in Figs.~\ref{betax05} and \ref{betax095} 
that the drag force 
is best approximated by the value that it would
have in a static plasma with the same instantaneous transverse pressure.
To some degree this agreement is coincidental as we see that for $\beta=0.5$
our result is a little below (\ref{EqbmDrag}) with the effective temperature $T_\perp$
while for $\beta=0.95$ our result is a little above this benchmark.  
What seems robust is the fact that at late times the actual force 
required to drag the quark through the expanding, cooling, anisotropic
hydrodynamic fluid lies within the band of expectations spanned by 
the drag force in a static plasma with the same instantaneous energy
density or parallel pressure or transverse pressure. For a heavy quark
with zero rapidity, we
see no qualitative deviation relative to these expectations, meaning
that we see no qualitative consequences of the presence of
gradients of the fluid velocity.

\begin{figure}
\includegraphics[scale=0.48]{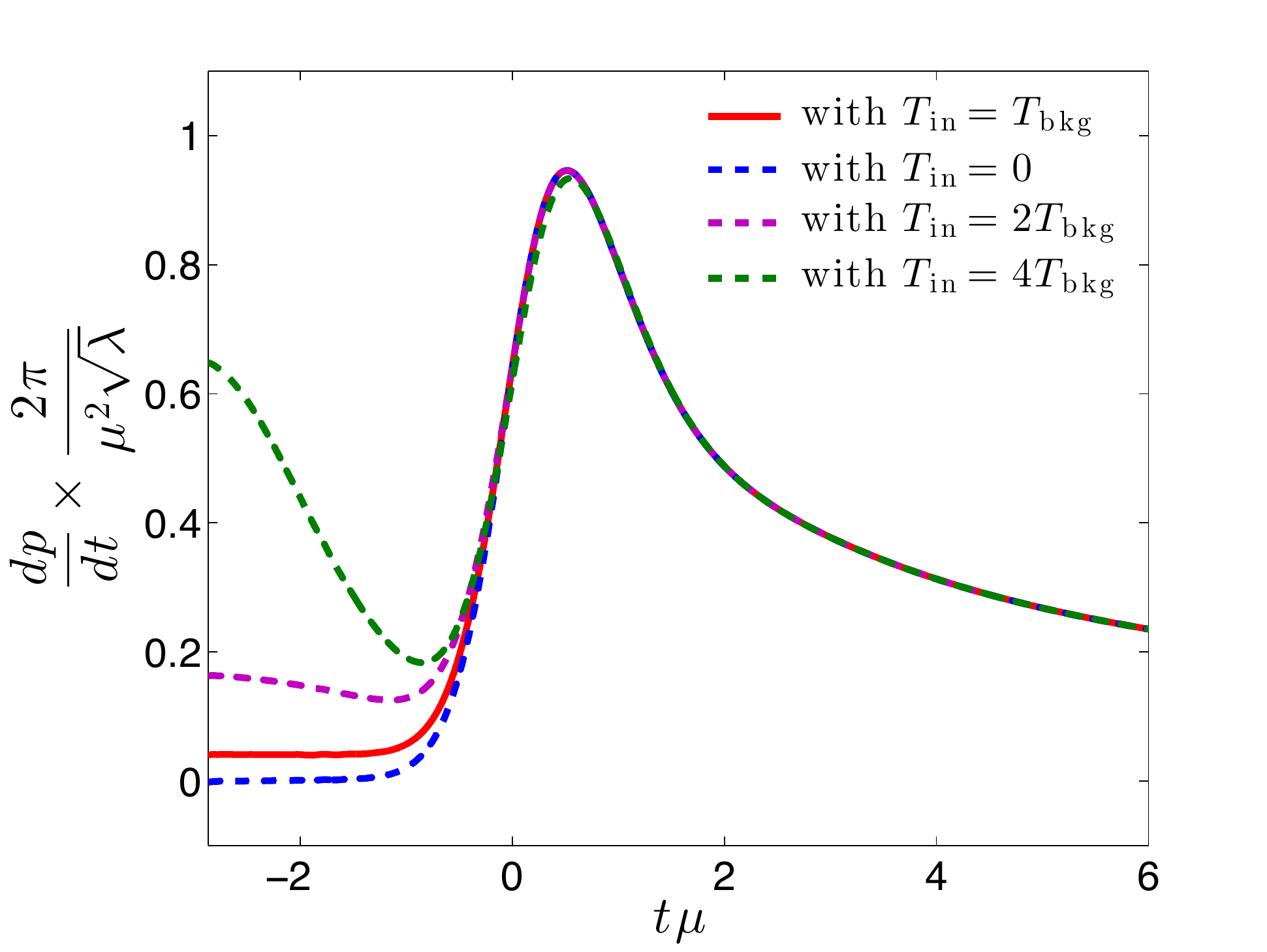}
\caption{Illustration of how the drag force depends 
on the initial string shape.   The four curves show the drag force
on a quark whose string initially has the shape of the string trailing
behind a heavy quark that is being dragged through an equilibrium
plasma with 
temperatures ranging from zero to four times the actual
temperature $T_{\rm background}$ of the low temperature
plasma present before the collision.   The solid red curve
is the same as the solid red curve in Fig.~\ref{betax05}. The other three
curves 
show that by the time the collision occurs the drag
force felt by the heavy quark is quite insensitive to the choice
of initial conditions.  Note that the initial drag force felt by the heavy
quark varies between zero and sixteen times that for the red curve; 
the perturbations to the initial shape of the string illustrated here are
not small, but their residual effects during and after the collision are.
In all cases, the heavy quark is dragged with velocity $\beta_x = 0.5$ and $\beta_z = 0$.
}
\label{initial}
\end{figure}

We are using initial conditions that correspond to a heavy quark which has been
dragged through an equilibrium plasma with the low temperature $T_{\rm background}$
for a long time before the collision.  This, of course, is not reminiscent at all 
of what actually happens
in a heavy ion collision, where the heavy quark is created during the collision.
It is therefore important to check how sensitive our results for the drag force 
on the heavy quark during and after
the collision are to the choices we are making for the initial shape of the
string well before the collision.  If our results during and after
the collision were sensitive to our choice of initial conditions 
this would be problematic, since there can be no right answer to the question of
what the initial conditions for a heavy quark before the collision should be since
in actual heavy ion collisions there are no heavy quarks present then.
Fortunately, as we illustrate in
Fig.~\ref{initial} we have found that our results of interest, namely our 
results for the drag force
on the heavy quark during and after the collision, are quite insensitive to the
choice of initial conditions.
The solid red  curve in Fig.~\ref{initial} is the same as that in Fig.~\ref{betax05}, while the 
blue, purple and green dashed curves correspond to initial string profiles chosen
such that the initial drag force is zero, four or sixteen times that for the red 
curve.
\footnote{In obtaining these results,
we have determined $t(\tau,\sigma)$ by solving the temporal constraint 
equation \r{temporalconstraint} numerically; see Appendix~\ref{app:ts} for a discussion.}
%As can be seen from Fig.~\ref{initial}, 
These perturbations to the initial shape of the string,
which are in no way small, do have transient effects at early times
but their effects on
the drag force felt by the heavy quark during and after the collision
are negligible.
%initial conditions become irrelevant shortly after the collision and the drag force has unique profile at $t \gtrsim 0$, confirming the claim made in the Section \ref{sec:Methods} that drag force is independent of the initial conditions shortly after the sheets of energy collide. 

\begin{figure}[t]
\includegraphics[scale=0.48]{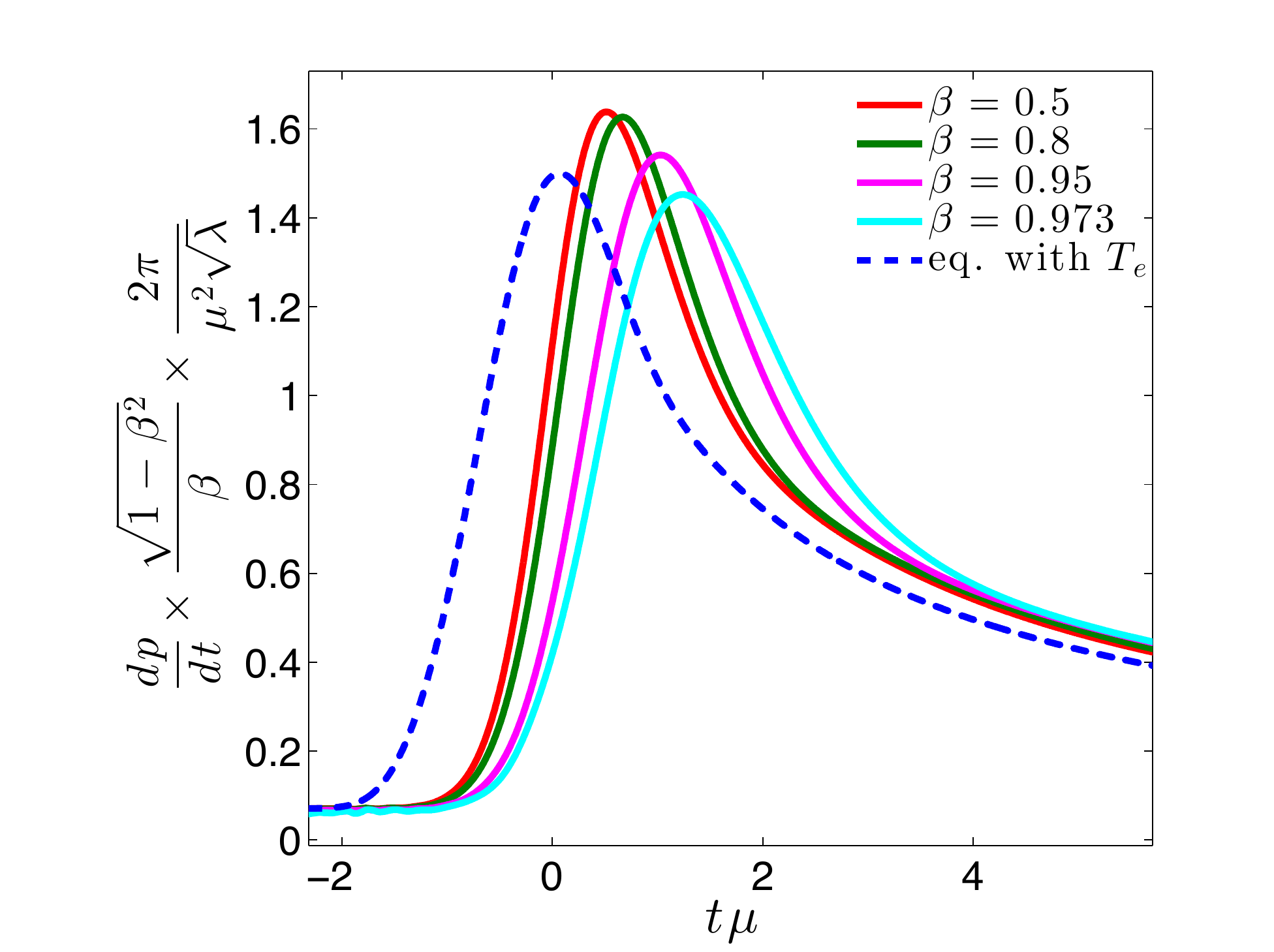}
\caption{Rescaled drag force for four different quark 
velocities $\beta$, all in the $x$-direction as in Figs.~\ref{betax05} and \ref{betax095}.
We see that the largest $\beta$-dependence of our results 
in Figs.~\ref{betax05} and \ref{betax095} can be understood
by assuming that the drag force scales roughly with $\beta\gamma$.
The remaining $\beta$-dependence seen here illustrates the
fact that the delay in the onset of, and subsequent peak in, the
drag force increases with $\beta$.
The dashed curve shows the rescaled drag force 
that the quark would experience in an equilibrium plasma
with the same instantaneous energy density.  From (\ref{EqbmDrag})
we see that, once we have rescaled by $\beta\gamma$,
 the dashed curve is $\beta$-independent.}
\label{dpxdtcomp}
\end{figure}

\begin{figure}
\includegraphics[scale=0.48]{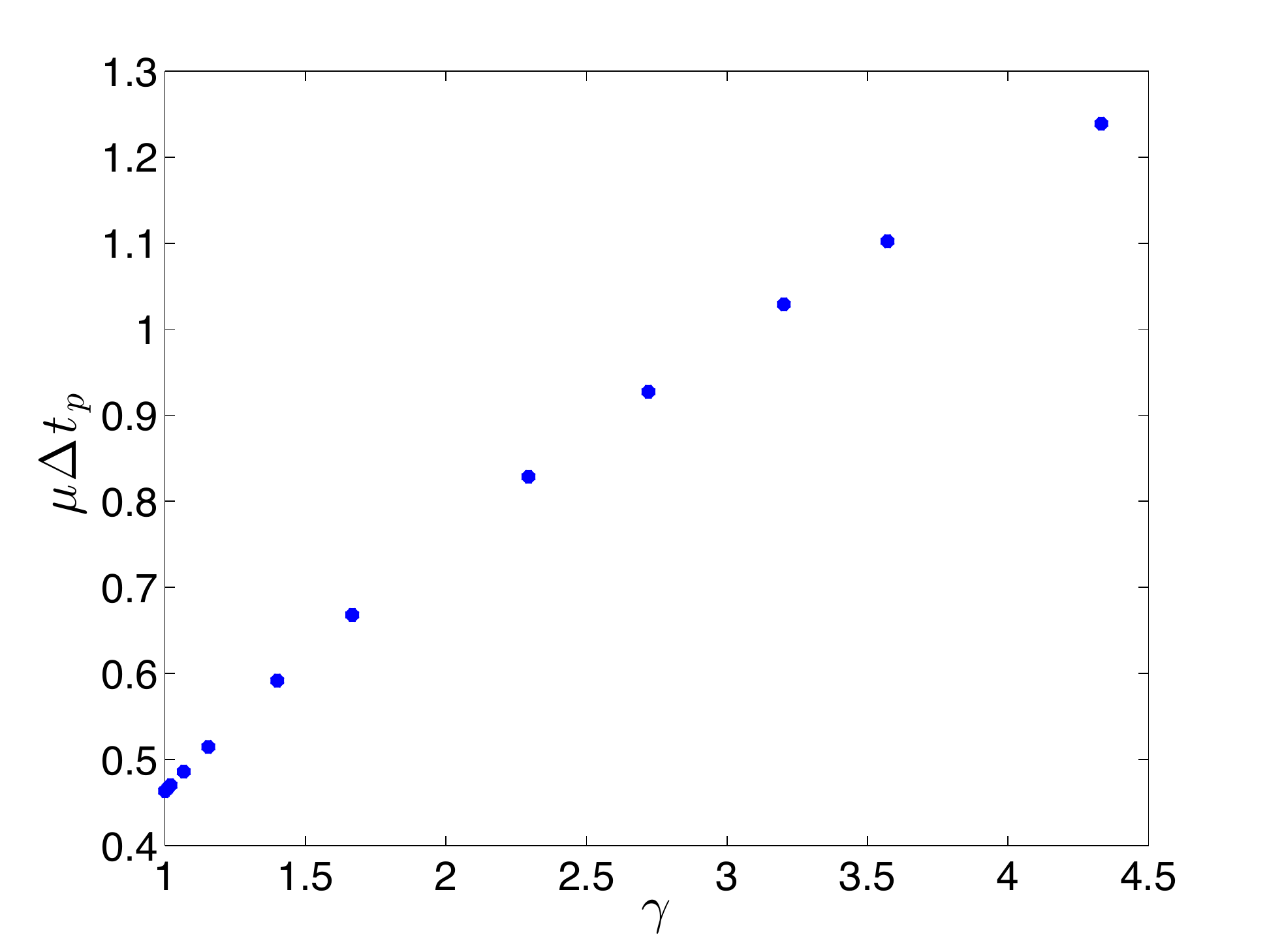}
\caption{The time delay $\Delta t_p$ between the time
at which the drag force on a heavy quark moving with velocity $\beta$
at zero rapidity peaks, i.e.~the peak in the solid curve in Fig.~\ref{dpxdtcomp},
and the time at which the dashed curve in Fig.~\ref{dpxdtcomp} peaks.
We plot $\Delta t_p$ vs. $\gamma$ for velocities ranging from $\beta=0.05$ to $\beta=0.973$.
The increase in $\Delta t_p$ clearly has a component that is linear
in $\gamma$, but the dependence is not completely linear.
The value of $\Delta t_p$ at $\gamma\rightarrow 1$, i.e.~at $\beta\rightarrow 0$,
is similar to the delay between the peaks in the purple and blue dashed
curves in Figs.~\ref{betax05} and \ref{betax095}, which is $0.53/\mu$. 
%Drag force seems to follow linear scaling for large drag velocities. 
%The corresponding velocities $\beta$ for each point are: $0.5, 0.7, 0.8, 0.95, 0.973$. 
}
\label{peak_d}
\end{figure}

Let us now return to the question of how the drag force changes with
quark velocity $\beta$.  Looking at Figs.~\ref{betax05} and \ref{betax095},
the biggest effect of increasing $\beta$ is certainly the increase in the overall
magnitude of the drag force.   Since we have seen that our
results are not dramatically different from the equilibrium expectations 
provided by the dashed curves in these figures, it is natural to look
at the equilibrium formula (\ref{EqbmDrag}) and ask to what degree
the $\beta$-dependence in our results is described by assuming that
the drag force scales with $\beta\gamma$.  Fig.~\ref{dpxdtcomp}
provides the answer.  Almost all of the change in the overall magnitude
of the drag force, e.g. the change in the height at which it peaks seen by comparing
Figs.~\ref{betax05} and \ref{betax095}, can be understood as scaling
with $\beta\gamma$.  The interesting $\beta$-dependent effect that remains
in Fig.~\ref{dpxdtcomp} is the time delay in the onset and peaking of the drag force.
This time delay increases with increasing $\beta$.  We investigate
the $\beta$-dependence of the time delay in Fig.~\ref{peak_d}.
From this figure we see that a reasonable
characterization of our results for the time at which the drag force peaks is
that at low velocity it peaks 
about one sheet thickness after the time when (around the same time as) the
drag force in an equilibrium plasma with the same 
with the same instantaneous energy density (transverse pressure) would peak while
at higher velocities its peak is delayed by a time that increases approximately
linearly with $\gamma$. 
In Appendix~\ref{app:step} we investigate
this time delay further by watching how
the drag force responds  in a background in which the temperature of the
plasma is homogeneous in space but over some
narrow range of time increases from a low value to a high value.
This investigation indicates that the fact that the time delay increases
roughly linearly with increasing $\gamma$ may be generic while the 
fact that (for the colliding sheets of energy above) we find that at low velocity the 
drag force peaks close to when the transverse pressure peaks is likely a coincidence.
It also indicates that at low velocity the drag force may generically peak
at a time that is about $1/(\pi T_{\rm hydro})$ later than the time when the
energy density peaks, where $T_{\rm hydro}$ is the temperature defined
from the fourth root  of
the energy density of the fluid at the time when the fluid hydrodynamizes. For the
collision we are analyzing, this corresponds to a time delay of around one sheet thickness.

The analysis in Appendix B supports a gravitationally intuitive picture
in which a time delay of order $1/(\pi T_{\rm hydro})$ corresponds to the
time it takes information from deep inside the bulk, near the black hole
horizon, to propagate to the boundary.  The information that the geometry near the horizon
has changed cannot propagate in the bulk faster than the speed of light, meaning
that the time it takes for such a change to affect the string near the boundary --- which is what
determines the drag force on the heavy quark --- is at least $\sim 1/(\pi T_{\rm hydro})$.
%With the intuition strengthened by Appendix B, the propagation of information from deep inside of the bulk is not instantaneous, as it cannot propagate faster than the speed of light, and the changes in the shape of the string which is caused by geometry perturbation close to the horizon takes time ~$1/T$ to propagate to the boundary. 
Furthermore,  for a quark that is moving with a large $\gamma$ 
the length of the string that stretches from the quark at the boundary down to the
near-horizon region is proportional to $\gamma$, suggesting that the time delay
for a fast quark should include a contribution that is proportional to $\gamma/(\pi T_{\rm hydro})$.
These considerations are somewhat heuristic, however, since the changes in the bulk geometry
to which the string dragging behind the heavy quark responds do not occur only in the near-horizon
region.
%from the horizon up to the boundary is proportional to $\gamma$, so for fast moving quark it takes time of ~$\gamma/T$ for the information about the updated position of the horizon to propagate to the boundary and manifest as the delayed drag force.

A time delay between a change in the stress tensor of the plasma and the
resulting change in the drag force that is proportional to $\gamma$
can be understood qualitatively in terms of the boundary gauge theory as follows.
Clearly, the drag force depends not only on the instantaneous stress tensor of the plasma
in which the quark finds itself and the velocity of the quark but also on the history of the quark.
In particular, our results suggest that the drag force takes on its ``correct'' value --- i.e.~the value it would
have in an equilibrium plasma whose stress tensor is similar to the instantaneous stress tensor
of the fluid around it --- only if the fields that dress the moving quark are configured appropriately.  
Perhaps as the energy density of the plasma increases the longer wavelength
fields that dress the moving quark must be stripped off leaving only those on length scales
of order $1/T_e$ and smaller.  Perhaps when the energy density  decreases those longer wavelength fields
need to grow back.  Whether or not such speculations are correct in detail, our results
indicate that the drag force on the quark responds to changes in the conditions around the quark only after
the fields carried along with the moving quark rearrange themselves in a way that takes some time.  If the time
this takes were constant in the rest frame of the quark, the time delay that we evaluate would
be proportional to $\gamma$.\footnote{A time delay between some change in the
environment in which a heavy quark finds itself and the resulting change in the drag force on the heavy quark
has been seen in other contexts, see for example Refs.~\cite{Peigne:2005rk,Guijosa:2011hf}, in which it has also been
attributed to the time it takes for the gluon fields around the moving quark to rearrange.}

\begin{figure}
\includegraphics[scale=0.6]{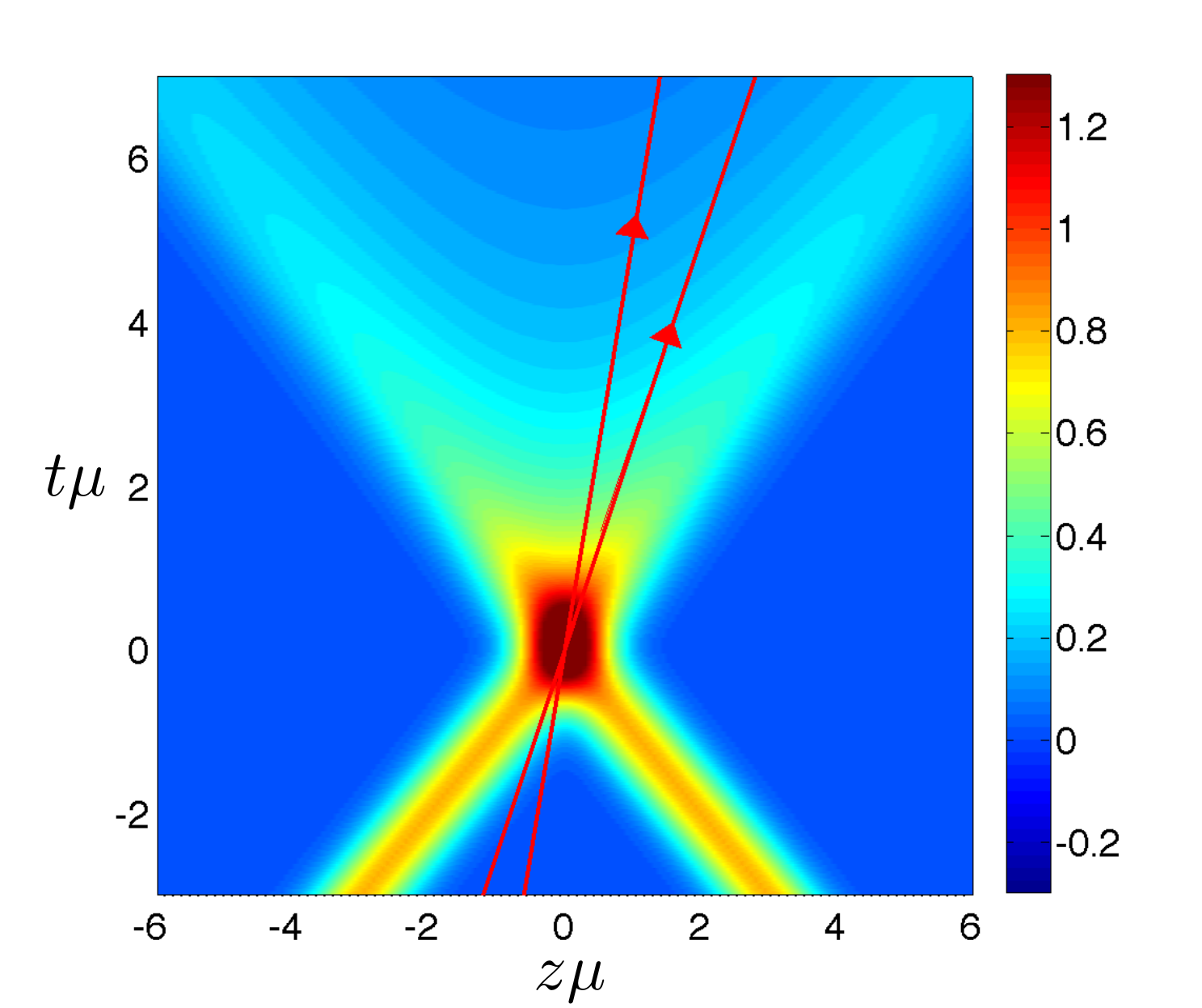}
\caption{The red arrows show the trajectories in the $(z,t)$-plane of two quarks with nonzero rapidity, 
with  $\beta_z=0.2$ and $\beta_z=0.4$.
% in case of non-zero rapidity. Two red lines with arrows correspond to transverse velocities $\beta_z = 0.2$ and $\beta_z = 0.4$ 
The trajectories are 
superimposed on the plot of the rescaled energy density ${\cal E}$ as in Fig.~\ref{z0path}.
We shall consider cases in which the quark has zero or nonzero transverse velocity $\beta_x$.
In all cases that we shall consider, the quark is pulled with constant
velocity $\vec \beta$ through the energy density produced
by the colliding sheets.
}
\label{path2}
\end{figure}

\subsection{Heavy quark with zero transverse momentum}
\label{sec:ZeroTransverseMomentum}

In this section and the next we shall consider more general cases in which
the quark has some nonzero velocity $\beta_z$ in the `beam' direction parallel
to the direction of motion of the colliding sheets, which is to say that we shall
allow the quark to have nonzero rapidity.  As illustrated in Fig.~\ref{path2}, we shall
only consider trajectories in which the velocity $\vec \beta$ of the quark
is constant and in which the initial position of the quark has been chosen
such that at $t=0$ the quark is at $z=0$, meaning that the quark passes
through the spacetime point at which the sheets of energy collide.
Although in our setup the quark has existed for all time, we are of course
interested in gaining qualitative insights into
circumstances in which a heavy quark is produced
at $z=t=0$ with some velocity $\vec \beta$.  
We shall not consider values of $\beta_z$ that are greater than 0.4
because we want to ensure that at the time $t=-3/\mu$ at which we choose
our initial conditions the quark has not yet felt the sheet
of energy that is about to catch up with it in any significant away.  

In order to obtain the greatest contrast with the case in which the
quark has zero rapidity, analyzed above, in this section we shall consider
the case in which the quark has a nonzero $\beta_z$ but has no
transverse velocity,  $\beta_x=0$.  We shall allow both $\beta_x$ and $\beta_z$ to be nonzero
in the next section.

\begin{figure}
\includegraphics[scale=0.48]{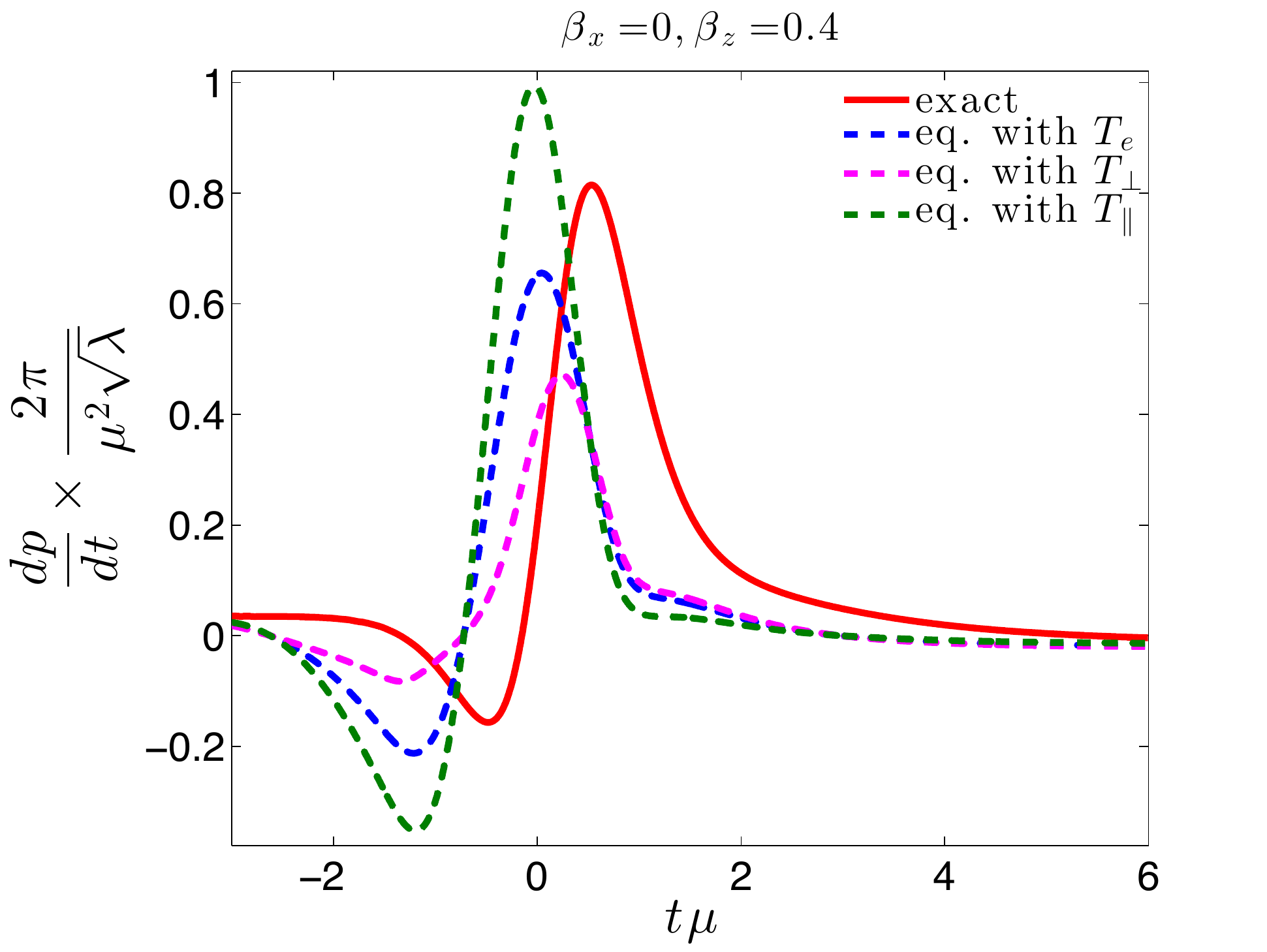}
\caption{The `drag' force (solid red curve) acting on a quark moving through the
matter produced in the collision along a trajectory with $\beta_z=0.4$ and $\beta_x=0$.
%
%along the motion of the quark in the Laboratory frame. The quark is dragged with velocity $\beta_x = 0, \beta_z = 0.4$. 
The force is shown in the lab frame.
The dashed curves show what the force would be if in the local
fluid rest frame it were given by the drag force that the quark
would experience in an equilibrium plasma with the same instantaneous
energy density, parallel pressure or perpendicular pressure as that 
in the local fluid rest frame at the location of the quark.  The dashed
curves are described further in the text.
%correspond to the equilibrium expectation of drag force with $T_e$ (blue dashed line), $T_\perp$ (magenta dashed line) and $T_\|$ (green dashed line).
}
\label{betax0betaz04}
\end{figure}

\begin{figure}
\includegraphics[scale=0.48]{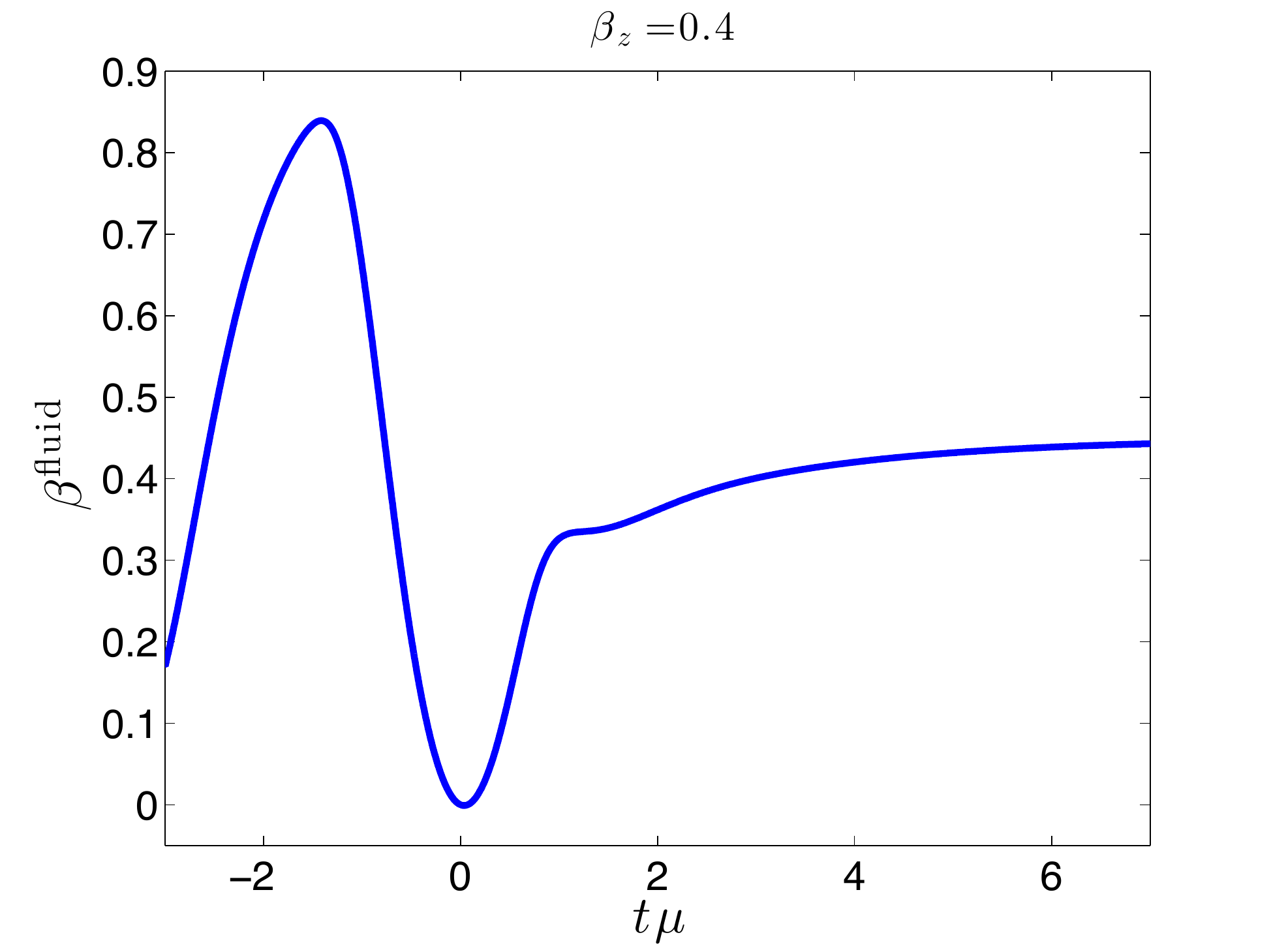}
\caption{The fluid velocity $\beta^{\rm fluid}$, defined in the text, at the location
of a quark moving along the trajectory in Fig.~\ref{path2} with 
$\beta_z=0.4$. If the hydrodynamic fluid at late times
were boost invariant, $\beta^{\rm fluid}$ would be given by $\beta_z=0.4$.
Note that at late times $\beta^{\rm fluid}>0.4$, meaning that the fluid
is moving faster than the quark. 
}
\label{betax0betaz04fluidvelocity}
\end{figure}

The solid red curve in Fig.~\ref{betax0betaz04} shows the
force on a quark with $\beta_z=0.4$ and $\beta_x=0$. 
To interpret
the force shown in Fig.~\ref{betax0betaz04}, we first need to
ask what the velocity of the fluid is
at the location of the quark as a function of time, since the drag on the quark
should depend on its velocity relative to that of the moving fluid.  
The fluid velocity is always in the $z$-direction and its magnitude is given by 
\begin{equation}
\beta_z^{\rm fluid} \equiv \frac{u^z}{u^0}
\end{equation}
where $u^\mu$ is the fluid velocity four-vector 
obtained from the stress tensor in the lab frame as described by  Eq.~(\ref{umuDefinition}).
We show
$\beta^{\rm fluid}$ in Fig.~\ref{betax0betaz04fluidvelocity}.  Reading this
figure from left to right while referring to Fig.~\ref{path2}: 
(i) We see that at the time $t=-3/\mu$ at which
we choose our initial conditions the quark is already feeling
some effects of the fast-approaching sheet of energy.
The quark is initially immersed in the background plasma, which
is at rest, meaning that the fact that $\beta^{\rm fluid}\neq 0$ can
be attributed to the Gaussian tail of the sheet of energy which is
approaching with velocity 1.  
(ii) Then, the sheet of energy incident from the left
catches up to the quark and as the energy
of the sheet overwhelms that of the background plasma $\beta^{\rm fluid}$
rises toward the speed of light.  Note that the quantity $\beta^{\rm fluid}$ can
be computed as we have described even when the matter is far from
equilibrium, as long as its energy density is positive~\cite{Casalderrey-Solana:2013aba}.  At these early times
$\beta^{\rm fluid}$ should not really be thought of as the velocity of a fluid,
although for convenience we shall refer to it as such. 
(iii) Next, the left-moving sheet
of energy slams into the right-moving sheet and, briefly, $\beta^{\rm fluid}$
is near zero. By symmetry, $\beta^{\rm fluid}=0$ at all times at $z=0$, meaning
that $\beta^{\rm fluid}=0$ at the location of the quark at $t=0$.
(iv) Finally, at late times after the fluid
has hydrodynamized, $\beta^{\rm fluid}$ is close to $\beta_z=0.4$.  Note that
if the hydrodynamic expansion at late times were boost invariant, the velocity
of the fluid at late times on a trajectory with constant rapidity like the one
that the quark is following would be given by the velocity of the quark itself.
In other words, if the expanding fluid at late times were boost invariant, the curve 
in Fig.~\ref{betax0betaz04fluidvelocity} would have $\beta^{\rm fluid}=0.4$ at late times. 

With an understanding of the velocity of the matter through which the heavy
quark is moving now in hand, we can return to Fig.~\ref{betax0betaz04} 
and compare the drag force there with expectations based upon the force (\ref{EqbmDrag}) on a heavy quark 
that is being dragged through plasma in equilibrium,
as we did in Fig.~\ref{betax05}.
The expression (\ref{EqbmDrag}) describes the force on a quark moving
through matter at rest, so to use it at any given time $t$ we must first boost
to a frame in which the matter through which the quark in Fig.~\ref{betax0betaz04}
is moving is at rest at that time. That is, 
we boost by a velocity $\beta^{\rm fluid}(t)$, plotted in Fig.~\ref{betax0betaz04fluidvelocity},
to the local fluid rest frame.  In the local fluid rest frame, the quark is moving
in the $z$-direction with a velocity 
\beq
\beta_{z,{\rm RF}}=\frac{\beta_z - \beta^{\rm fluid}}{1-\beta_z\,\beta^{\rm fluid}}
\label{betazRF}
\eeq
that can be positive or negative depending on whether $\beta^{\rm fluid}$ is
smaller or larger than the velocity $\beta_z$ of the quark in the lab frame.
Next, we compute the stress tensor in the local fluid rest frame,
where it takes the form (\ref{RestFrameTmunu2}), and use the temperatures $T_e$, $T_\perp$
and $T_\parallel$ so obtained as well as the velocity $\beta_{z,{\rm RF}}$ from (\ref{betazRF})
in the expression (\ref{EqbmDrag}). This gives us 
the drag force that the quark would experience
if, in the local fluid rest frame, it were moving with velocity $\beta_{z,{\rm RF}}$ through an 
equilibrium plasma with the same
instantaneous energy density, perpendicular pressure or parallel pressure as 
that of the matter at its location.  
Finally, we boost the three forces computed in this way
back to the lab frame, using the appropriate Lorentz transformation for forces,
given in the present context by\footnote{It is a worthwhile check
of the formalism for extracting the drag force from the bulk gravitational
quantites that we
have set out in Section~\ref{sec:ExtractingTheForce} to confirm
that upon boosting the bulk black brane metric and the trailing string profile
by a velocity $\beta^{\rm fluid}$ and evaluating (\ref{f2}) one obtains (\ref{ForceTransformation}).}
\beq
\begin{split}
f_{\rm eq,\ lab\ frame}^\mu &= \frac{1}{u^0 + u^z \beta_{z,{\rm RF}}}  
 \left(
\begin{array}{c}
u^0 f_{\rm eq,RF}^{0} + u^z f_{\rm eq,RF}^{z}\\
 f_{\rm eq,RF}^x \\
 f_{\rm eq,RF}^y \\
 u^0 f_{\rm eq,RF}^{z} + u^z f_{\rm eq,RF}^{0}
\end{array}
\right)\,,
\end{split}\label{ForceTransformation}
\eeq
with $u^\mu$  the fluid four-velocity from Eq.~(\ref{umuDefinition}).
We plot the three lab-frame forces computed as we have just described
as the three dashed curves in Fig.~\ref{betax0betaz04}.
If we apply this algorithm in the more general case in which
in the lab frame the quark is moving with both $\beta_z$ and 
$\beta_x$ nonzero, the result is 
\beq
\begin{split}
%f_{\rm eq}^v &= \frac{\sqrt{\lambda}}{2\pi} \gamma (\pi T)^2  \(u^0 (\beta)^2 - u^z \beta_z\) \\
f_{\rm eq,\ lab\ frame}^x &= \frac{\sqrt{\lambda}}{2\pi} (\pi T)^2\gamma \(u^0\beta_x - u^z \beta_x \beta_z\) \\
f_{\rm eq,\ lab\ frame}^z &= \frac{\sqrt{\lambda}}{2\pi}  (\pi T)^2 \gamma \( u^0 \beta_z  + u^z (\beta_x)^2 - u^z \) \,,
\label{forceBoosted}
\end{split}
\eeq
which can be written as
\beq
f_{\rm eq,\ lab\ frame}^\mu = -\frac{\sqrt{\lambda}}{2\pi}(\pi T)^2 \frac{1}{ \gamma}
\(u_\nu u_q^\nu u_q^\mu + u^\mu\)\ ,
\label{forceBoosted2}
\eeq
where we have defined $u^\mu_q\equiv \gamma(1,\vec \beta)$ and have allowed
for a generic choice of coordinate axes.
The $T$ used in the expressions (\ref{forceBoosted}) or (\ref{forceBoosted2})
can be either the $T_e$ or the $T_\perp$ or the $T_\parallel$ obtained from
the stress tensor
in the local fluid rest frame.

We can now compare the actual drag force on a quark
moving in the $z$-direction, the solid red curve in Fig.~\ref{betax0betaz04},
to the expectations 
%obtained from (\ref{EqbmDrag}) 
for an equilibrium plasma moving with velocity $\beta^{\rm fluid}$ with
the same instantaneous energy density or pressure, shown
as the dashed curves.  In many respects, this comparison is as we found in Fig.~\ref{betax05}
for a quark moving in the $x$-direction.  As in Fig.~\ref{betax05}, we 
see in Fig.~\ref{betax0betaz04} that the maximum value of the drag force
in the far-from-equilibrium matter produced during the collision is within the
expectations for the maximum drag force spanned by the three dashed curves.
And, as in Fig.~\ref{betax05}, we see a time delay in the rise in the magnitude of the drag force.
Here, though, the drag force first goes negative as the heavy quark is hit from the left
by the sheet of energy going in the same direction and only then goes positive as the 
second sheet of energy slams into the first.

\begin{figure}
\includegraphics[scale=0.48]{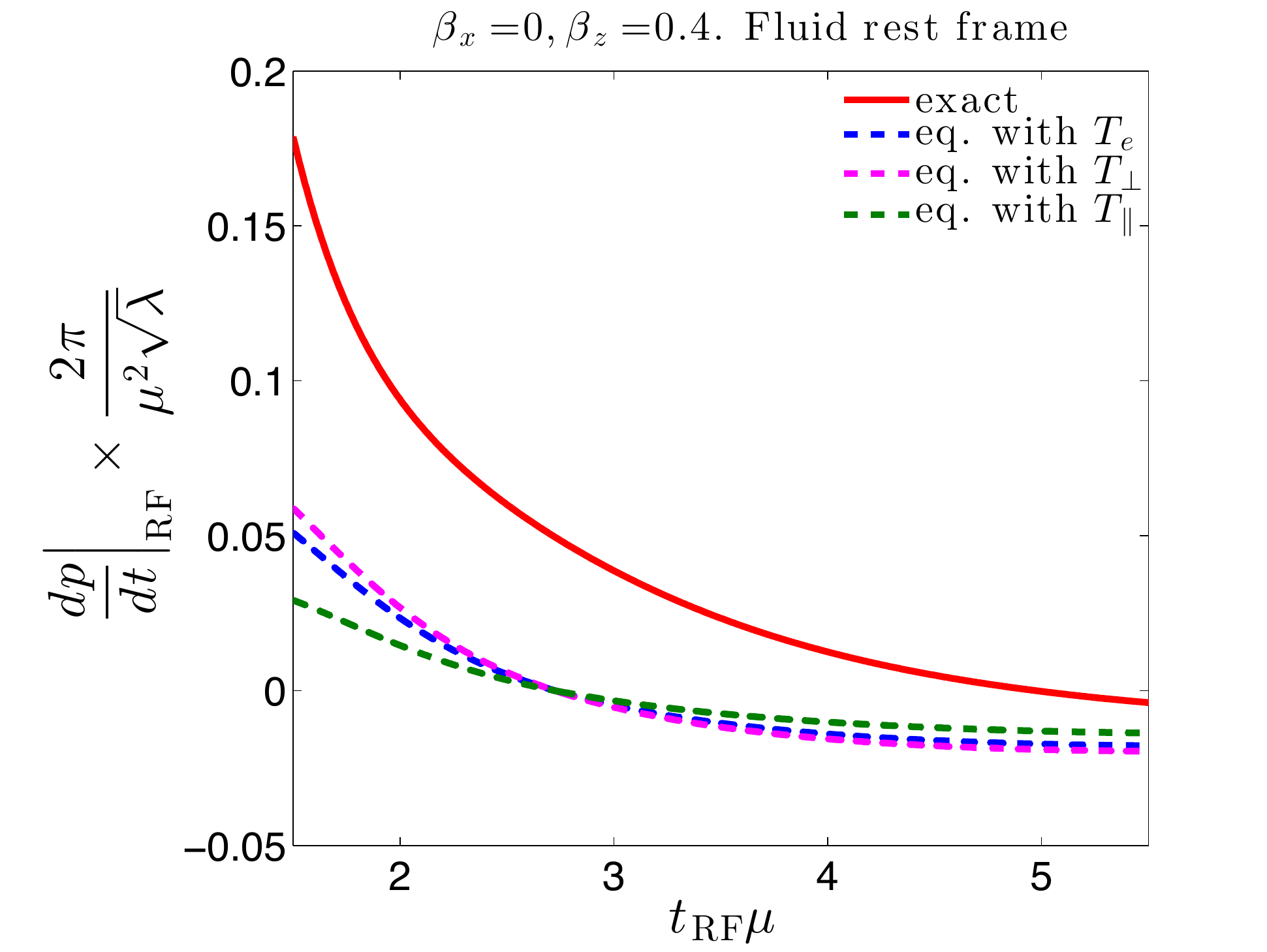}
\caption{Forces as in Fig.~\ref{betax0betaz04},
plotted here in the local fluid rest frame and
focusing on times $t_{\rm RF}>1.5/\mu$ to better illustrate 
the behavior after hydrodynamization.
After $t_{\rm RF}=2.8/\mu$, 
the quark is moving to
the left in the local fluid rest frame (in the lab frame,
the fluid is moving
faster than the quark) and the
dashed curves accordingly lead us to expect
that the force needed to drag the quark leftward
in the local fluid rest frame should be a force acting toward the left.
Instead, we see that between $t_{\rm RF}=2.8/\mu$ and $t_{\rm RF}=5.0/\mu$, 
the `drag' force that must be exerted to maintain
the leftward motion acts toward the right!
}
\label{betax0betaz04RF}
\end{figure}

The most  interesting differences between 
Figs.~\ref{betax0betaz04}
and \ref{betax05} arise at late times.  First, we see that the magnitude of the drag force at late
times is much smaller in Fig.~\ref{betax0betaz04} than in Fig.~\ref{betax05}.  This
reflects the fact that at late times the quark and the fluid are moving at comparable
velocities, see Fig.~\ref{betax0betaz04fluidvelocity}, meaning that their relative
velocity is small.  In fact, at $t\simeq 2.8/\mu$ the fluid velocity and the quark
velocity are equal --- the dashed curves in Fig.~\ref{betax0betaz04} therefore
cross zero there.  Interestingly, we see that the solid curve stays positive
for quite a long time afterwards until $t=5.0/\mu$ meaning that there is an extended period
of time when: (i) The fluid has hydrodynamized. (ii)~The fluid is moving
faster than the quark, which would suggest that the `drag' force 
on the quark needed to keep it moving at constant $\beta_z$
should be a force pulling backward on it, toward negative $z$,
pulling against the push from the fluid that is moving faster than the quark.
That is, we expect that
$dp/dt$ should be negative, as is indeed the case for the dashed curves in Fig.~\ref{betax0betaz04}.
(iii)~Instead, the quark is still being dragged forward, toward positive $z$, with $dp/dt$ positive.
This means that in the local fluid rest frame the force
that the external agent must exert in order to move the quark towards the
left acts toward the right.   The quark
is moving towards the left in this frame but the force exerted on it by the liquid
through which it is moving is {\it also} acting toward the left, pushing it in the
direction of its motion rather than dragging it in the opposite direction!
We illustrate this in Fig.~\ref{betax0betaz04RF} by plotting
the force exerted on the quark in the local fluid rest frame.
The reason that a result as surprising as this is possible
is that in formulating our expectations, as shown
via the dashed curves, we are completely neglecting
the effects of spatial gradients in the fluid velocity.\footnote{We have also
neglected spatial gradients in the energy density of the fluid.
In the following, we will show that after hydrodynamization the effects of spatial gradients 
on the drag force that we compute
%see in our results after hydrodynamization
correlate well with the behavior of the gradients of the
fluid velocity.  We have checked that the spatial gradients of
the energy density are small at these late times, as are their
effects on the force that we compute.
}
Unlike in Fig.~\ref{betax05}, in Fig.~\ref{betax0betaz04} these gradients are 
%in the fluid velocity is 
aligned parallel to the direction in which the
quark is moving; the lesson we learn is that in this 
circumstance the gradients in the fluid velocity
can have qualitative
effects on the `drag' force that must be exerted on the quark.
Qualitative effects of the gradient in the fluid
velocity arise in Figs.~\ref{betax0betaz04} and \ref{betax0betaz04RF} but not
in Fig.~\ref{betax05} 
both because the effects of the gradient
on the `drag' force are larger in absolute magnitude 
%in  Fig.~\ref{betax0betaz04}, 
when the gradient is aligned
with the motion of the quark and because in this case the drag force
in the absence of gradients would be very small since the
relative velocity of the quark and the fluid is so small.

\begin{figure}
\includegraphics[scale=0.48]{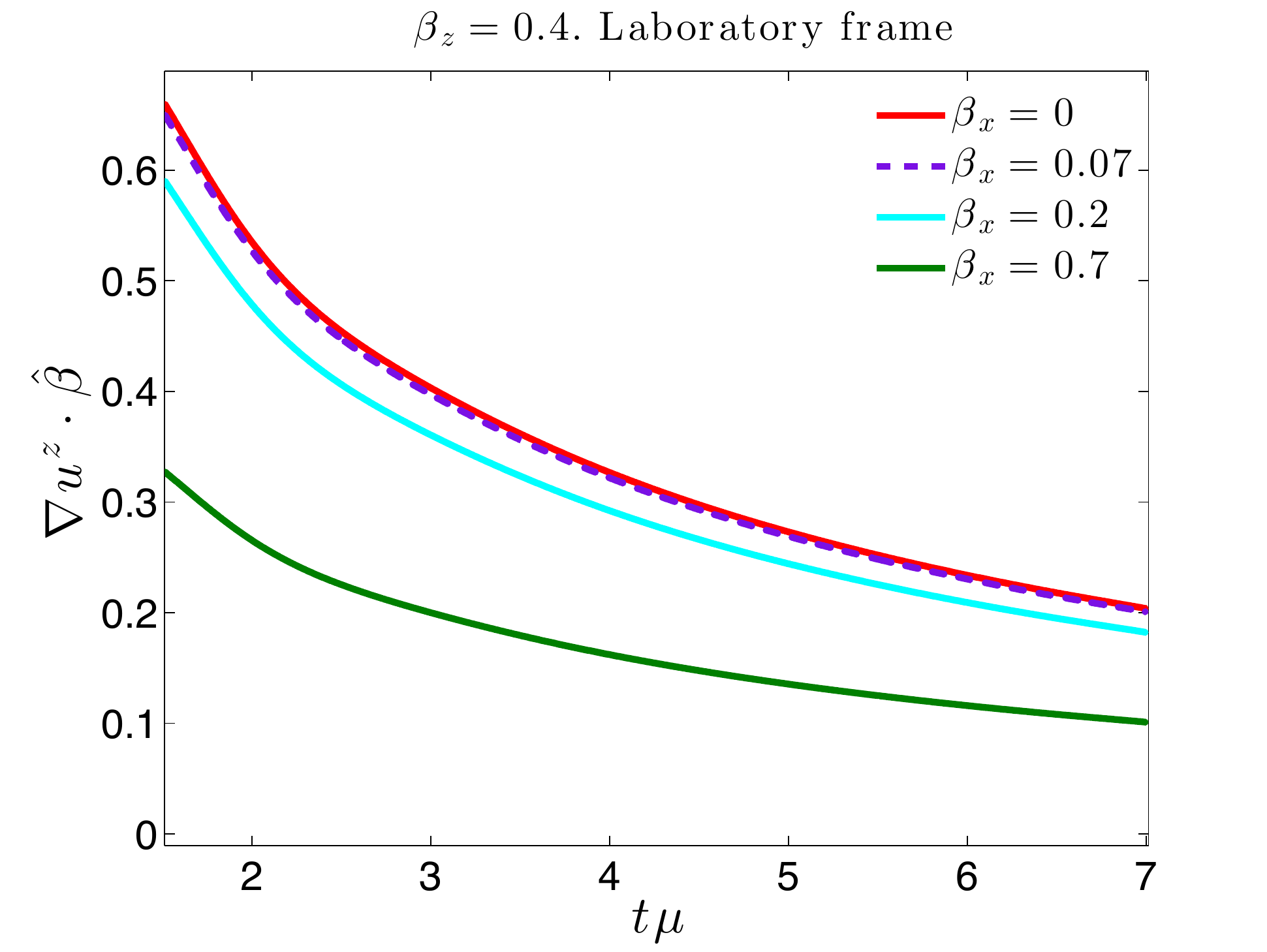}
\includegraphics[scale=0.48]{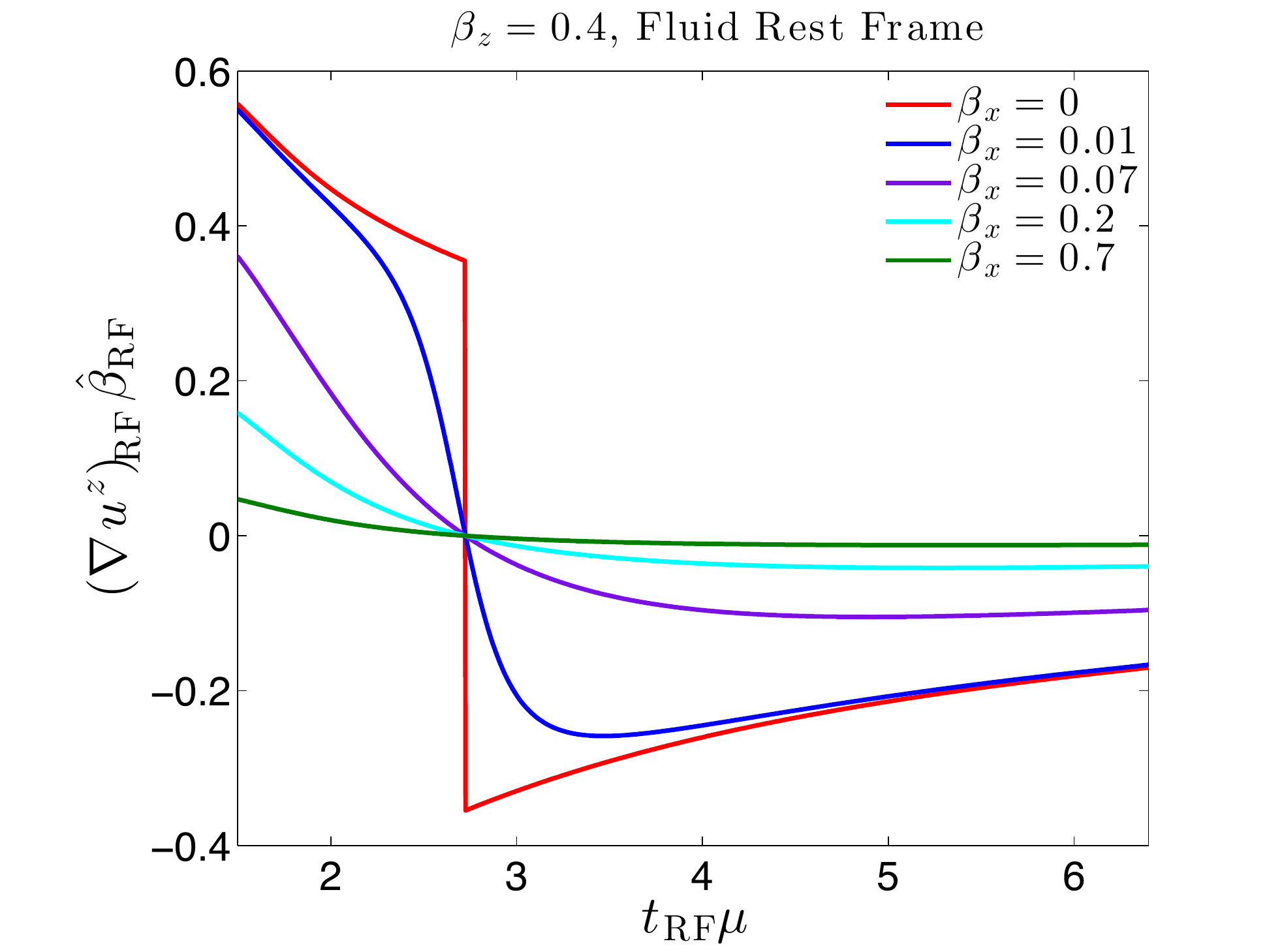}
\caption{Component of the gradient of the fluid velocity in the direction
of motion of the quark at the
location of the quark  in the 
lab frame (upper panel) and in the 
local fluid
rest frame (lower panel).  The differences between the two panels arises
because $\beta_{z,{\rm RF}}\ll \beta_z$.
In all curves in both panels,
the trajectory of the quark has $\beta_z=0.4$. 
%as in Fig.~\ref{betax0betaz04}. 
The different curves
are for trajectories with varying values of $\beta_x$.
At $t_{\rm RF}=2.8/\mu$, the sign of $\beta_{z,{\rm RF}}$ flips, which
is to say that in the lab frame the velocity of the fluid exceeds
the velocity of the quark. 
We see in the lower panel that adding a $\beta_x$ that is quite small
compared to $\beta_z$ in the lab frame, but is comparable to $\beta_{z,{\rm RF}}$,
significantly reduces 
 $(\nabla u^z)_{\rm RF} \cdot \hat \beta_{\rm RF}$.    
 }
\label{Par_grad}
\end{figure}

%In order to get more qualitative understanding of such large corrections, it is suggestive to investigate how big is the fluid velocity $\beta_{\rm fluid}$ gradient along the quark path.
%Since the spatial gradient is non-zero only along the $z-$direction, in the Fluid rest frame the spatial gradients are given by ({\bf Could show the derivation in the Appendix})
%\beq
%\left. \frac{\partial u^x}{\partial z} \right|_{RF}  = 0 \quad \text{and} \quad 
%\left. \frac{\partial u^z}{\partial z} \right|_{RF}  = \partial_z u^z + \frac{u^z}{u^0} \partial_t u^z
%\eeq
%while the quark velocity in the fluid rest frame are
%\beq
%\beta_x^{RF} = \frac{\beta_x}{u^0} \frac{1}{1 - \beta_z \beta_{\rm fluid}}
%\quad \text{and} \quad 
%\beta_z^{RF} =  \frac{\beta_z - \beta_{\rm fluid}}{1 - \beta_z \beta_{\rm fluid}}
%\eeq
%so that 

The red curves in Fig.~\ref{Par_grad} show
the gradient of the fluid velocity in the 
lab frame and in the
local fluid rest frame, in each case 
projected onto the quark velocity in that frame, namely
\beq
(\nabla u^z) \cdot \hat \beta  \equiv  \frac{\partial u^z}{\partial z}\, \frac{\beta_{z}}{\beta}
\eeq
and
\beq
(\nabla u^z)_{\rm RF} \cdot \hat \beta_{\rm RF} \equiv \left. \frac{\partial u^z}{\partial z} \right|_{\rm RF} \frac{\beta_{z,{\rm RF}}}{\beta_{\rm RF}}
\eeq
for the trajectory with $\beta_z=0.4$ and $\beta_x=0$. For trajectories
with $\beta_z=0$ and $\beta_x>0$, both quantities vanish identically.
Fig.~\ref{Par_grad} shows that if we start with $\beta_z=0.4$ and  add 
a nonzero $\beta_x$ that is small compared to $\beta_z$ in the lab
frame but that is large compared to $|\beta_{z,{\rm RF}}|$, for example
$\beta_x=0.2$,
 the quantity
$(\nabla u^z)_{\rm RF} \cdot \hat \beta_{\rm RF}$
is substantially reduced while in the lab frame
$(\nabla u^z) \cdot \hat \beta$ is not much changed.
This suggests that if we analyze the drag force on a quark that follows
a trajectory with $\beta_z=0.4$ and $\beta_x=0.2$, 
we should find results that 
in the local fluid rest frame are more similar to those in
Fig.~\ref{betax05} but in the lab frame
are more similar to those in Fig.~\ref{betax0betaz04}.  We
shall confirm this expectation in the next section, but in so doing we shall discover
a second surprise.

%In the Figure \ref{Par_grad}, the parallel component of the fluid velocity gradient parallel to the quark trajectory with $\beta_z = 0.4$ and several values of $\beta_x$ is shown. For $\beta_x = 0$, the projection becomes equal to
%\beq
%\left. \frac{\partial u^z}{\partial z} \right|_{RF} \text{sgn}(\beta_z - \beta_{\rm fluid})
%\eeq
%so that at around $t\mu = 2.8$, where $\beta_{\rm fluid}$ becomes bigger than $\beta_z$, the parallel gradient becomes a step function. When $\beta_x$ is increased from zero even slightly, $\beta_x = 0.01$ case in the plot, the step function is smoothed out. The important message is that with increasing $\beta_x$, the fluid velocity gradient along the path becomes smaller. The same statement holds for any non-zero $\beta_z$, while for $\beta_z = 0$ the parallel gradient vanishes identically since $\beta_{\rm fluid} = 0$ along $z=0$. 
%Our statement is that the difference between the parallel component of the drag force and the equilibrium expectations is proportional to fluid velocity gradient along the path of the quark. 

\subsection{Heavy quark with nonzero rapidity and transverse momentum}
\label{sec:NonzeroBoth}

In this section we analyze the force that must be exerted in order
to move a heavy quark through the colliding sheets of energy along
a trajectory with nonzero $\beta_x$ and $\beta_z$, which is to say
with both transverse momentum and rapidity nonzero.  We
start %in Figs.~\ref{betax02betaz04_parallel_lab} and \ref{betax02betaz04_perp_lab}
by considering a trajectory 
with $\beta_z=0.4$ as in Figs.~\ref{betax0betaz04} and \ref{betax0betaz04fluidvelocity} but
now with $\beta_x=0.2$.  Because the velocity of the quark now has a component
perpendicular to the velocity of the fluid at the location of the quark (this
was not the case in both previous sections) we now expect and find
that the force that must be exerted in order to move the quark along this
trajectory has a component perpendicular to the direction of motion
of the quark.  

%
%To illustrate this statement, we check the case of $\beta_x = 0.7, \beta_z = 0.4$ in the Figure \ref{betaz04_par}, where 

\begin{figure}
\includegraphics[scale=0.48]{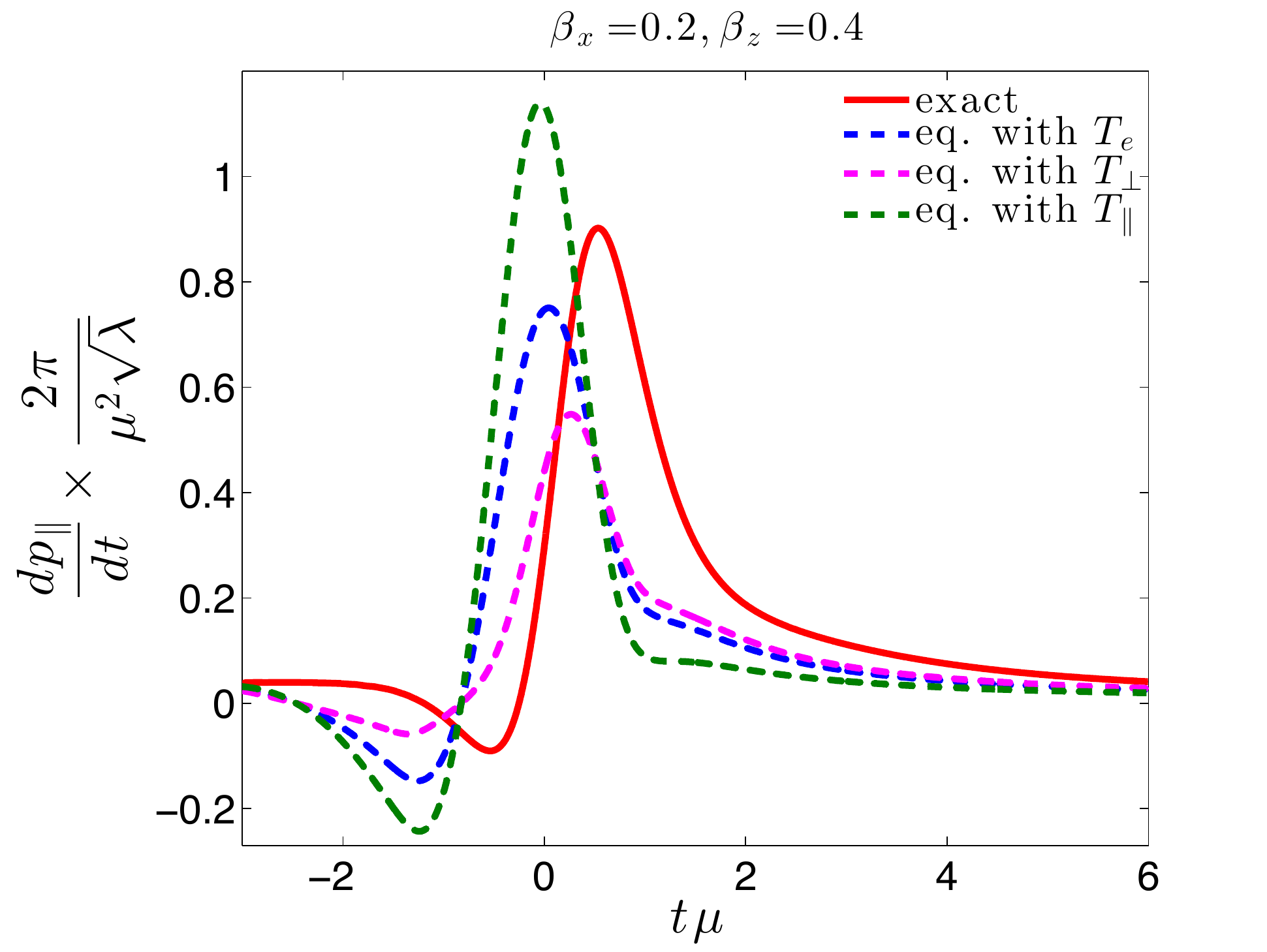}
\caption{
The force (solid red line) 
in the lab frame 
that must be exerted in the direction of motion of the quark (in the lab frame)
in order to drag it with a velocity $\beta_x=0.2$ and $\beta_z=0.4$, a trajectory 
with nonzero transverse
momentum and rapidity.
The dashed curves, computed as described in the previous section, 
show what this force would be if in the local fluid rest frame the quark were being dragged through 
a plasma with the same instantaneous energy density, perpendicular pressure
or parallel pressure as that at the location of the quark.
%perpendicular to the motion of quark in the Laboratory frame. The quark is dragged with velocity $\beta_x = 0.2, \beta_z = 0.4$. The dashed curves correspond to the equilibrium expectation of drag force with $T_e$ (blue dashed line), $T_\perp$ (magenta dashed line) and $T_\|$ (green dashed line). 
}
\label{betax02betaz04_parallel_lab}
\end{figure}

\begin{figure}
\includegraphics[scale=0.48]{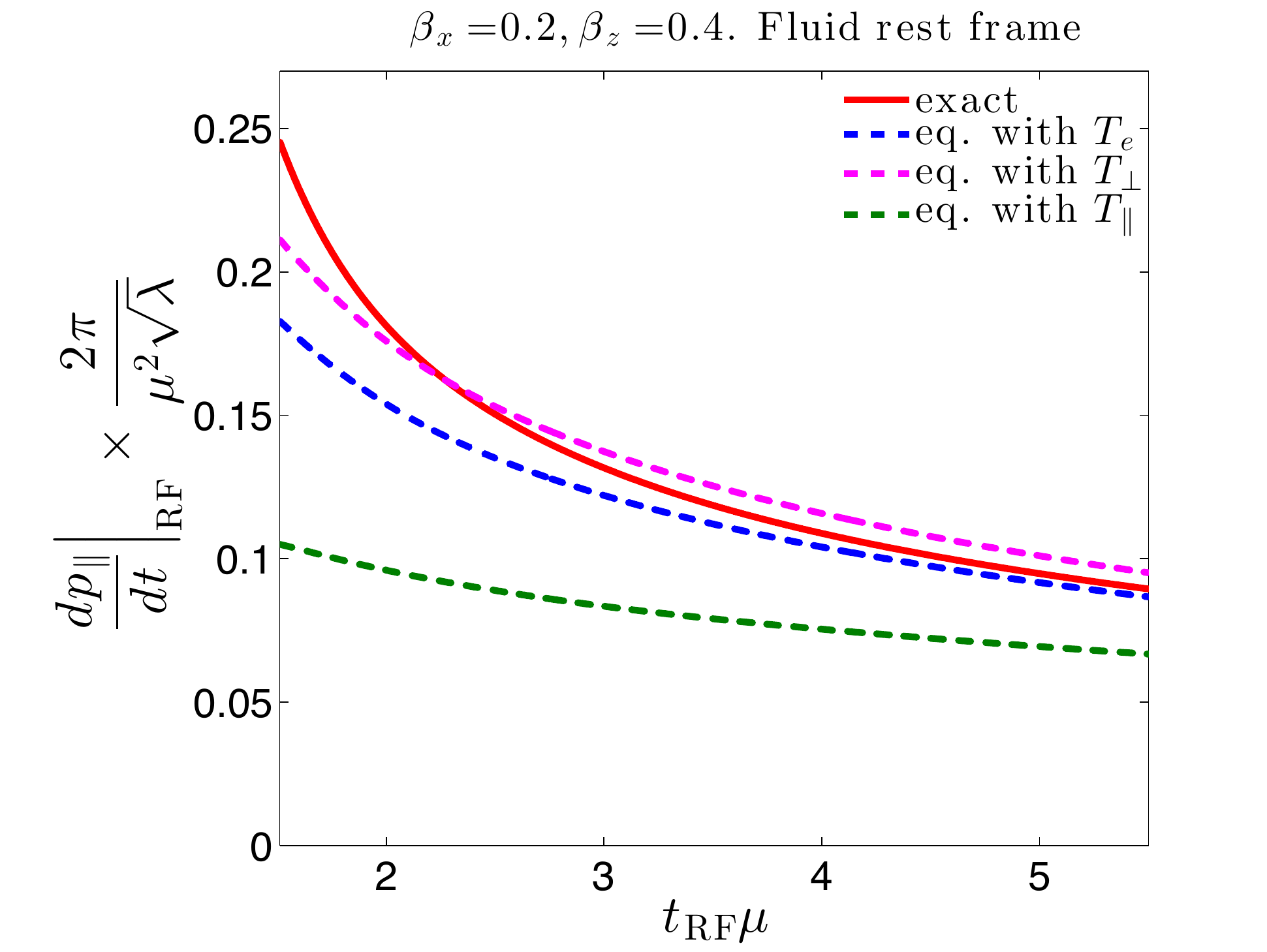}
\caption{Forces as in Fig.~\ref{betax02betaz04_parallel_lab}
plotted here in the local fluid rest frame and focusing on times $t_{\rm RF}>1.5/\mu$
to better illustrate the behavior after hydrodynamization.
We saw in Fig.~\ref{betax0betaz04RF} that when $\beta_z=0.4$ and $\beta_x=0$
the dashed curves that show what the drag force in the
local fluid rest frame would be in a spatially homogeneous
plasma with the same instantaneous energy density or transverse pressure or
parallel pressure failed to describe the actual force on the quark, because they
neglected the effects of gradients in the fluid velocity on the force.  Here,
with $\beta_x=0.2$, we
see that after hydrodynamization the actual force on the quark falls within
the range of expectations spanned by the dashed curves.
}
\label{betax02betaz04_parallel_RF}
\end{figure}

Let us first look at the drag force parallel to the direction of motion the quark in the lab frame,
%shown (red solid line) and is defined by
\beq
\frac{dp_\|}{dt} \equiv \frac{\vec f \cdot \vec \beta}{\beta}\,,
\label{parf}
\eeq
which is shown as the solid red curve in Fig.~\ref{betax02betaz04_parallel_lab}.
%When quark is dragged with both non-zero components of velocity, the applied force is not necessarily aligned with the path of the quark and both components of eq.~(\ref{forceBoosted}) are present. In the Figure \ref{betax02betaz04_perp_lab} 
(Note that $d p_\parallel /dt$ and, below, $d p_\perp/dt$ refer to the force parallel and
perpendicular to the direction of motion of the quark; in contrast, $T_\parallel$ and $T_\perp$ are defined via
the pressures in the fluid that act in the directions parallel to and perpendicular to
the direction of motion of the colliding sheets and hence of the fluid, i.e.~the $z$-direction.)
We see that, as in both Sections~\ref{sec:ZeroRapidity}
and \ref{sec:ZeroTransverseMomentum}, the dashed curves provide a reasonable guide
to the peak value of the drag force but the actual force peaks later than the dashed curves do,
a time delay that is by now familiar.   

At late times, after the fluid has hydrodynamized, we
see behavior that is more similar to that in Section~\ref{sec:ZeroTransverseMomentum} in the sense that
the drag force in the direction of motion of the quark is affected by the
gradients in the fluid velocity to such a degree that it is well outside the
expectations spanned by the three dashed curves. Unlike when $\beta_x$ was zero, though,
with $\beta_x=0.2$ at least the sign of the force is the same for the solid and dashed curves.
The dashed curves nevertheless fail to give a qualitative description of the actual
drag force after the fluid has hydrodynamized.  This is consistent with
the upper panel of Fig.~\ref{Par_grad} which shows that,
when $\beta_z=0.4$, turning on $\beta_x=0.2$ does not substantially reduce
the magnitude of the component of the gradient of the fluid velocity in the
direction of motion of the quark --- in the lab frame.
In Fig.~\ref{betax02betaz04_parallel_RF} we see that the story
is different in the local fluid rest frame.  With $\beta_x=0.2$, 
we find that the drag force in this frame does lie within the range of expectations spanned
by the three dashed curves.  This is consistent with the lower panel of Fig.~\ref{Par_grad},
where we saw that, because $|\beta_{z,{\rm RF}}| \ll \beta_z$, when we turn on
$\beta_x=0.2$ we do substantially reduce the component of the gradient
of the fluid velocity in the direction of motion of the quark --- in the local fluid
rest frame.

\begin{figure}
\includegraphics[scale=0.48]{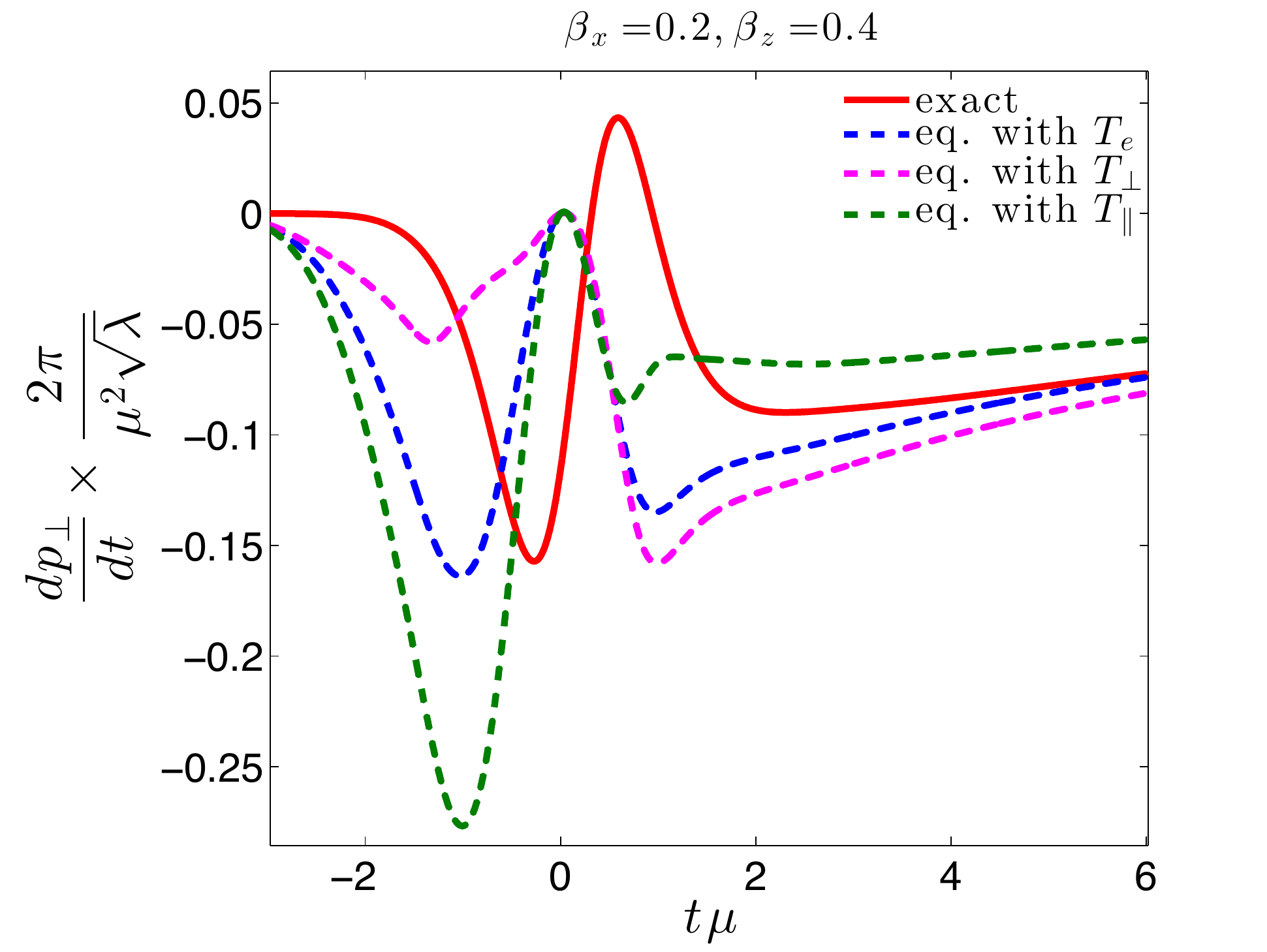}
\includegraphics[scale=0.48]{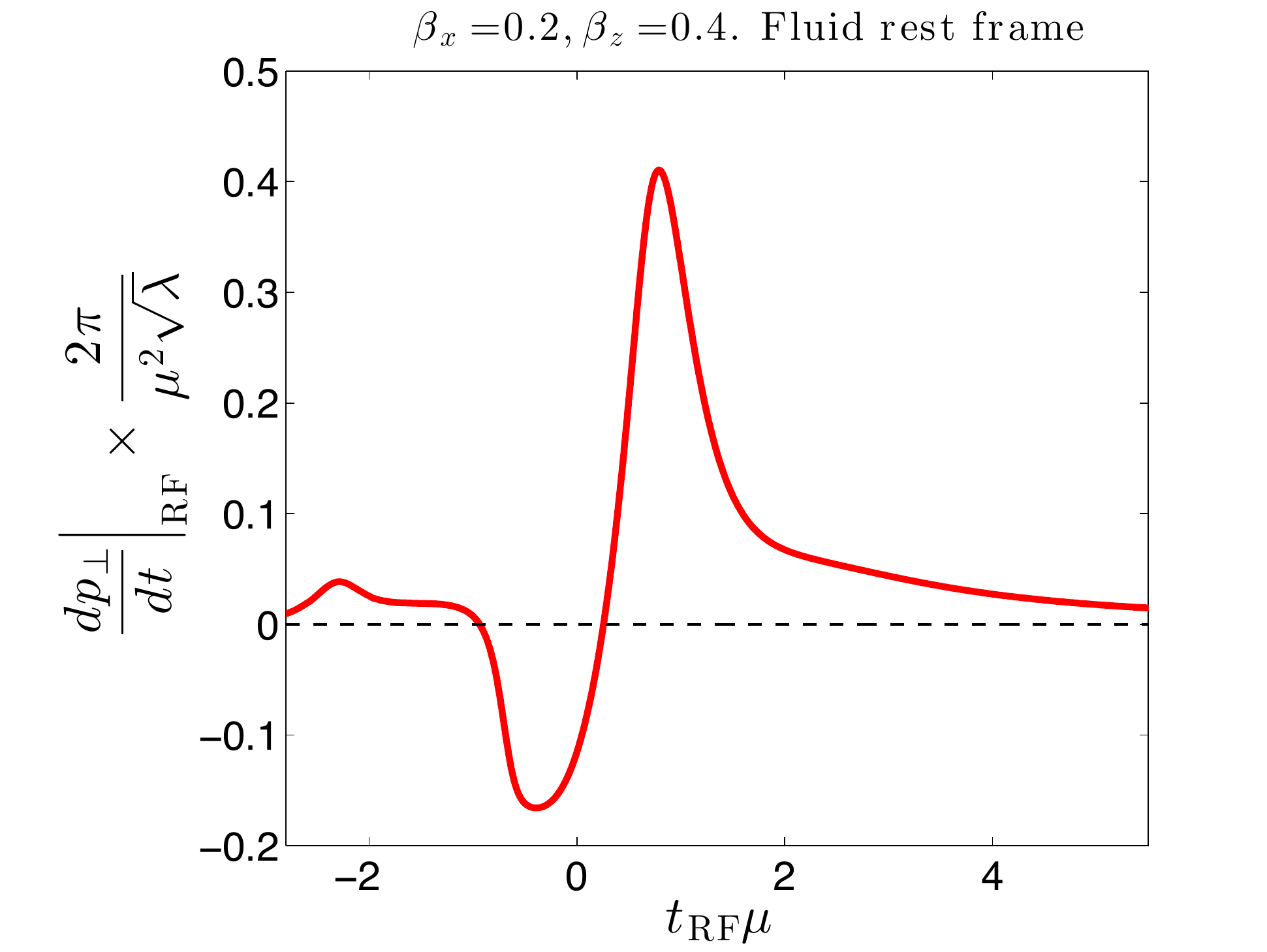}
\caption{
Upper panel: as in Fig.~\ref{betax02betaz04_parallel_lab},  
but here we plot the force that must be exerted perpendicular to the direction
of motion of the quark, in the lab frame.  Note the scale on the
vertical axis: this component of the force 
is significantly smaller than the component of the force in Fig.~\ref{betax02betaz04_parallel_lab}.
%The drag force (solid red line) perpendicular to the motion of quark in the Laboratory frame. The quark is dragged with velocity $\beta_x = 0.2, \beta_z = 0.4$. The dashed curves correspond to the equilibrium expectation of drag force with $T_e$ (blue dashed line), $T_\perp$ (magenta dashed line) and $T_\|$ (green dashed line). }
Lower panel: the force that must be exerted perpendicular
to the direction of motion of the quark in the local fluid rest frame.
%There are no dashed curves in the lower panel because 
If, in this frame, the quark were moving through a homogeneous
fluid (e.g.~the equilibrium plasma with the same instantaneous
energy density or pressure as that at the location of the quark)
the force in this frame could only
act parallel to the direction of the quark. The dashed curves therefore
all vanish in the lower panel.  The actual force does not vanish and
is in fact quite substantial at its peak:
even in the local fluid rest frame,
the force needed to drag the quark
includes a component perpendicular to its direction of motion.  
}
\label{betax02betaz04_perp_lab}
\vspace{-0.1cm}
\end{figure}

Turning now to the force acting perpendicular to the direction of motion of the
quark, we fix our sign conventions by defining this as
%the perpendicular force component to the quark movement direction is illustrated,
\beq
\frac{dp_\perp}{dt} \equiv f^z \frac{\beta_x}{\beta} - f^x \frac{\beta_z}{\beta}\ .
\label{fperp}
\eeq
It is plotted as the solid red curve in 
the upper panel of Fig.~\ref{betax02betaz04_perp_lab}.  The dashed curves show
what the force perpendicular to the direction of
motion of the quark would be in the lab frame 
if, in the local fluid rest frame, the quark were moving through an equilibrium plasma
with the same instantaneous energy density, perpendicular pressure, or parallel pressure as
that at the location of the quarks.  

The force perpendicular to the direction of motion of
the quark is a new development, present only
when both $\beta_x$ and $\beta_z$ are nonzero.  
In the lab frame, the existence of a perpendicular force is 
no surprise since in this frame of reference the 
quark is moving through a fluid whose velocity includes a component perpendicular to its own.
For this reason, in the upper panel of Fig.~\ref{betax02betaz04_perp_lab} the dashed curves show
that even if there were no velocity gradients in the fluid the force would include a component
perpendicular to the direction of motion of the quark in the lab frame.
This poses an obvious question: boost to the local fluid rest frame and ask whether
in that frame the drag force is parallel to the direction of motion of the quark. 
We show in the lower panel of Fig.~\ref{betax02betaz04_perp_lab} that the answer is no:
even in a reference frame in which at each point in time we have boosted 
to a frame in which the
quark is moving through a fluid at rest, the force required to drag the quark
along its trajectory through this fluid includes a component perpendicular to the 
direction of motion of the quark!  
In this frame, in the absence of gradients in the fluid 
there could be no 
component of the force perpendicular to the motion of the quark.
%Instead, we see in the lower panel that in order to move the quark
%along its trajectory the force required does include 
%a component perpendicular to the trajectory.
The lower panel of Fig.~\ref{betax02betaz04_perp_lab}
thus represents a stark consequence of the
presence of spatial gradients in the matter produced
in the collision.  The effect is largest at early times, when
the matter is far from equilibrium. 
At these early times, the effect is large indeed: the peak
value of the force perpendicular to the motion of the quark
is about half as large as the peak value of the drag force
that acts parallel to the velocity of the quark.
The effect is also nonvanishing
at late times, after the expanding fluid has hydrodynamized.
We find this result to be generic,
arising for any trajectory in which both $\beta_x$ and $\beta_z$ are nonzero.
A drag force that
includes a component perpendicular to the direction of motion of the quark 
through a medium
arises in other contexts in which some anisotropy in the medium has been
introduced~\cite{Chernicoff:2012iq,NataAtmaja:2010hd,Fadafan:2012qu}.
Here we see this phenomenon arising as a robust consequence of gradients in the
fluid velocity --- a form of anisotropy that must be present in heavy ion collisions.

%together with analogous curves for equilibrium expectations. At late times, the perpendicular component approaches the equilibrium expectation associated with $T_e$, indicating much smaller effect of gradient corrections as it is for the parallel component of the force, and this is true quite generally, as long as the value of $\beta_z$ is such that quark is not too close to the remnants of the shocks. 

%In this case parallel component of the force agrees well with the equilibrium expectation with $T = T_\perp$, as in the cases of $\beta_x = 0$.  

%In case the fluid were expanding as a Bjorken flow \cite{Bjorken:1982qr}, the fluid velocity would be $\beta^{BF}_{\rm fluid} = z/t$, so that if quark was moving along the path $z = \beta_z t$, fluid expansion velocity would be equal to the quark drag velocity, ensuring that the quark locally is at rest. For $\beta_x = 0$, this would imply that for the equilibrium of (parallel) drag would vanish as can be seen from eq.~(\ref{forceBoosted}). The fact that in the Figure \ref{betax0betaz04} equilibrium expectation force(s) are very close to zero is manifestation of the fact that once the hydrodynamic description of the fluid starts working, the fluid can be thought as boost-invariant.

\begin{figure}
\includegraphics[scale=0.48]{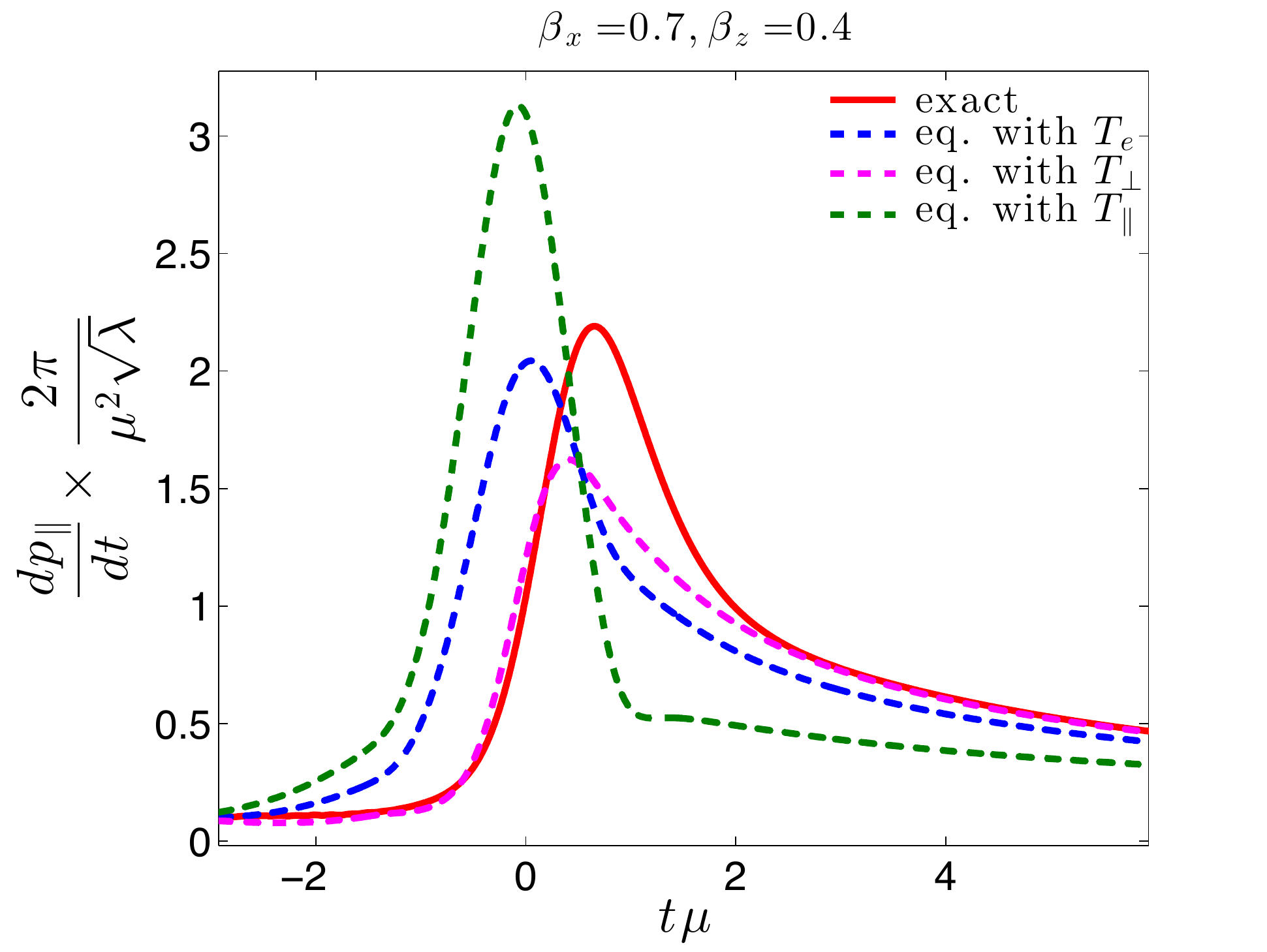}
\includegraphics[scale=0.48]{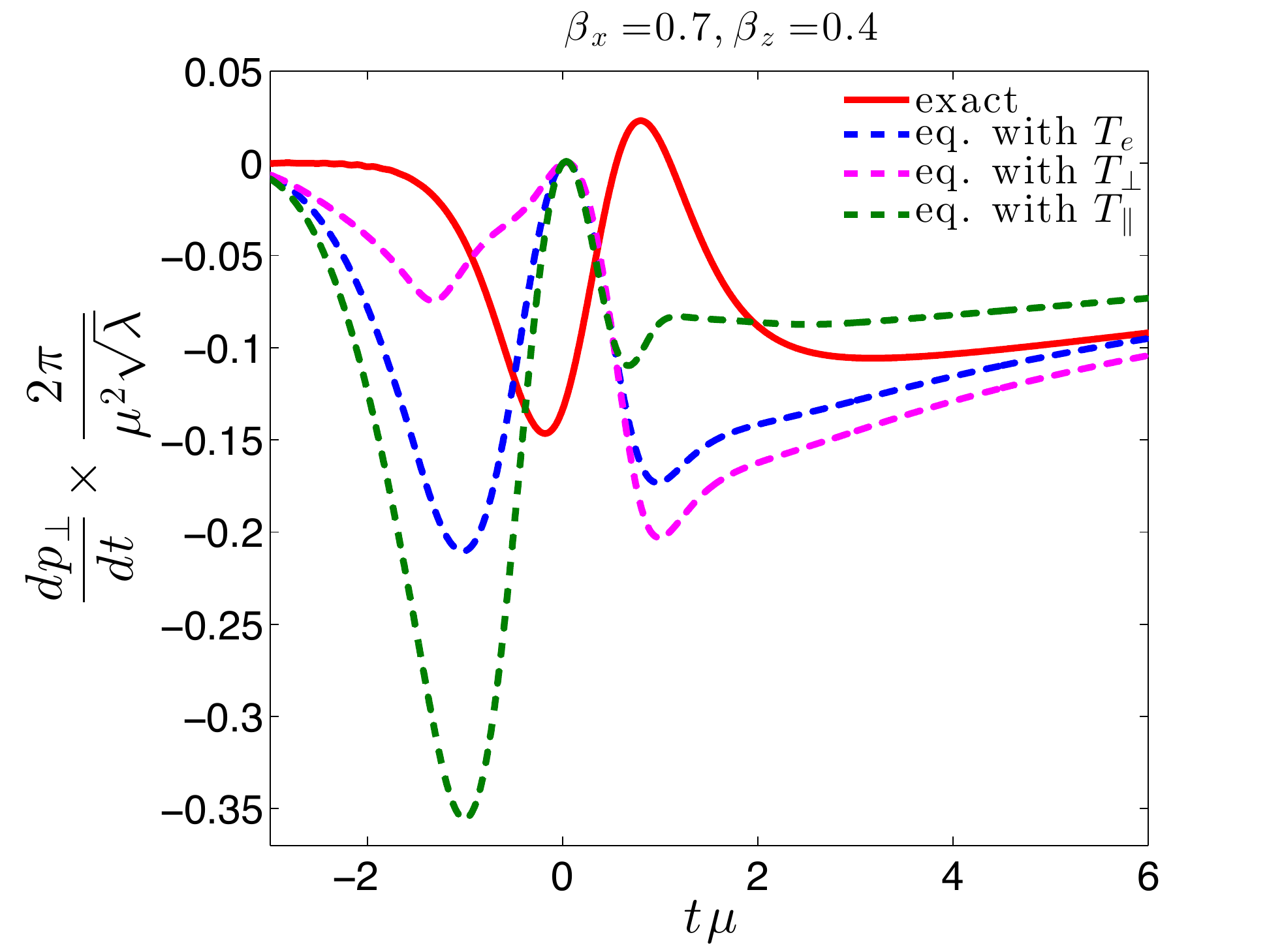}
\caption{
Drag force parallel to (upper panel) and perpendicular to (lower panel)
the direction of motion
of the quark in the lab frame, as in Fig.~\ref{betax02betaz04_parallel_lab} and 
the upper panel of Fig.~\ref{betax02betaz04_perp_lab},
 but here for 
a quark with $\beta_x=0.7$ and $\beta_z=0.4$. 
%Lower panel: drag force perpendicular to the direction
%of motion of the quark in the laboratory frame, as in Fig.~\ref{betax02betaz04_perp_lab}, but here for 
%a quark with $\beta_x=0.7$ and $\beta_z=0.4$.  
}
\label{betax07betaz04_lab}
\end{figure}

\begin{figure}
\includegraphics[scale=0.48]{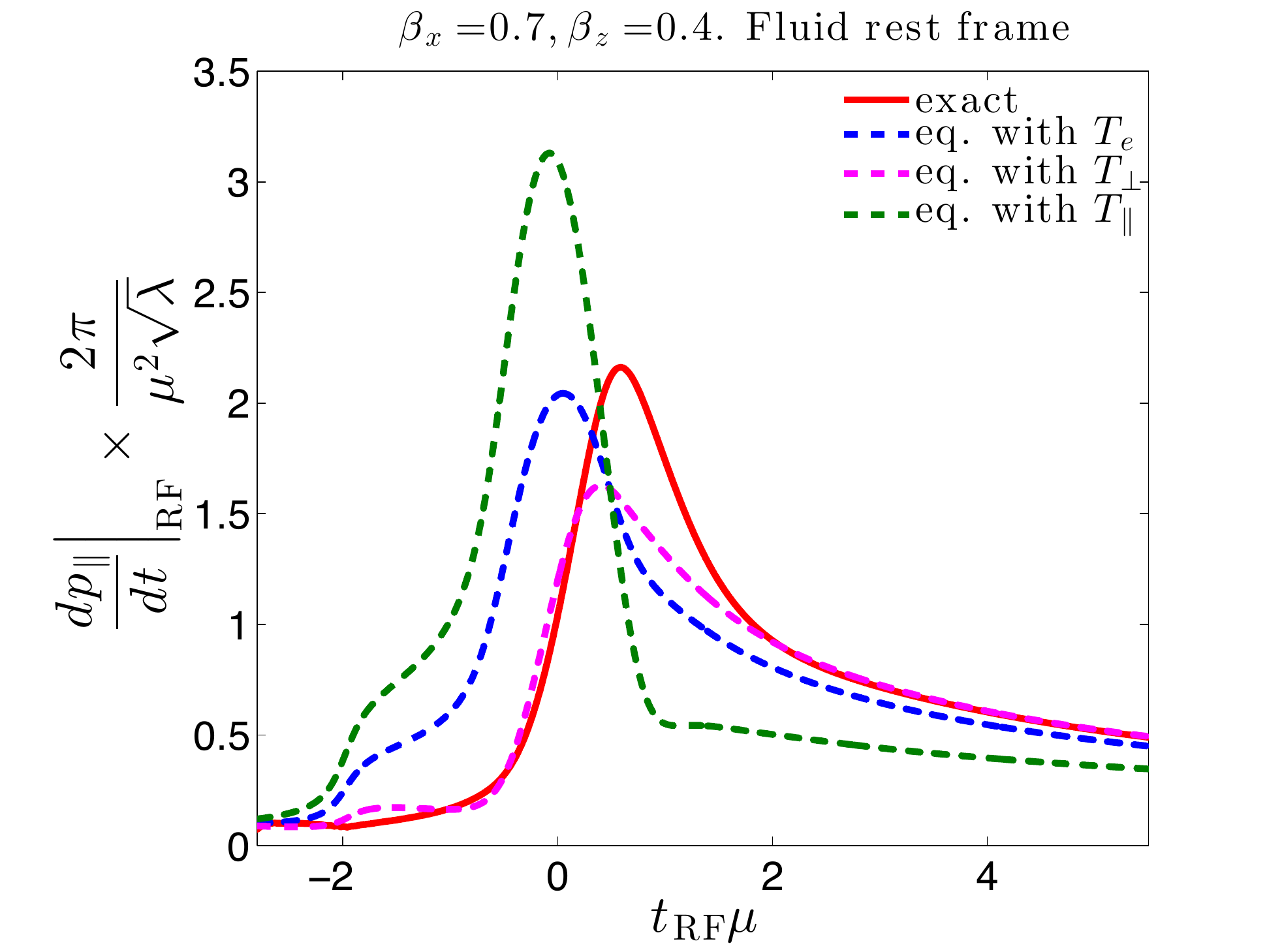}
\includegraphics[scale=0.48]{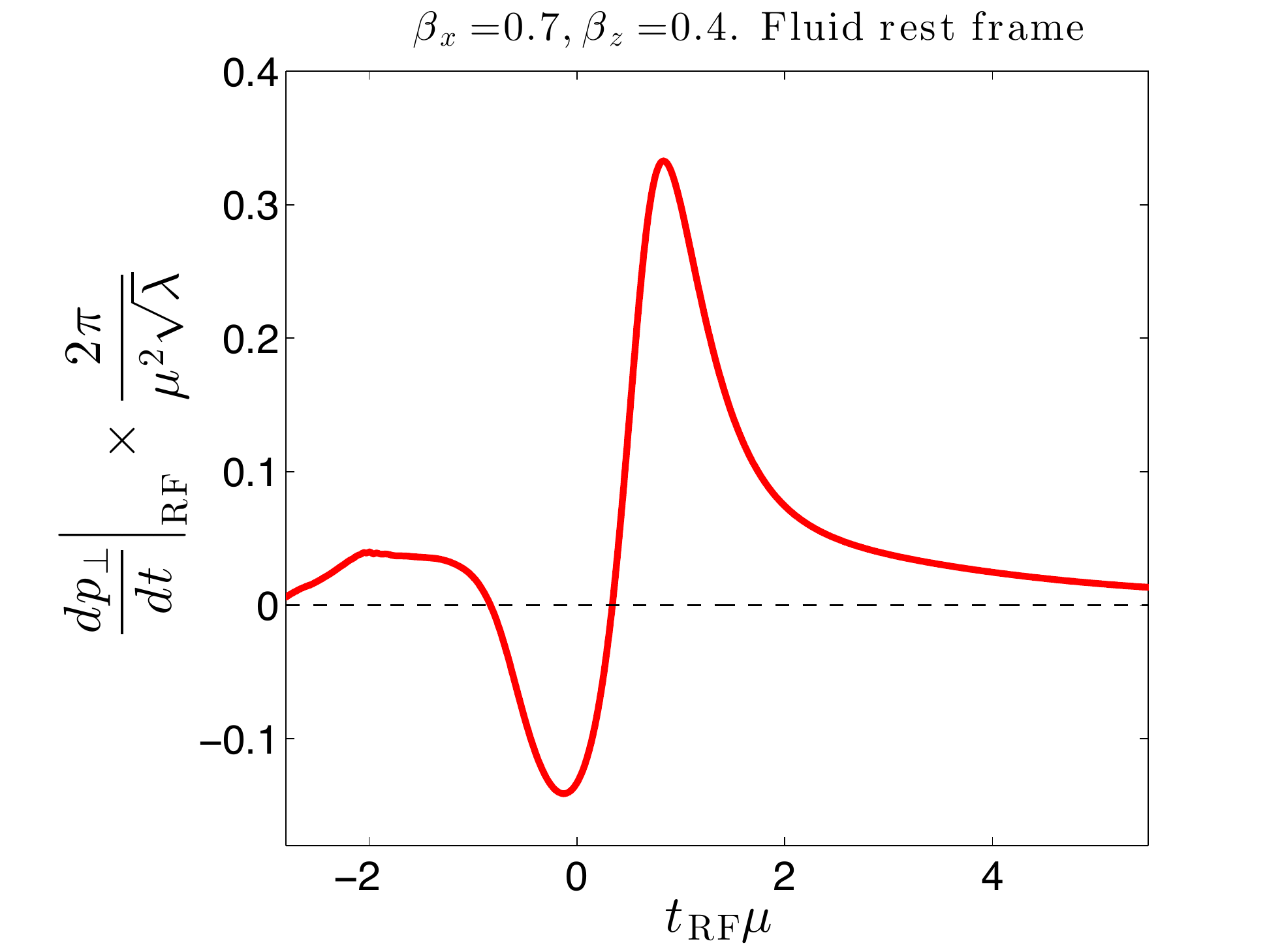}
\caption{
As in Fig.~\ref{betax07betaz04_lab}, but here in the local fluid rest frame.
%The upper panel shows that boosting to the local fluid rest frame has no qualitative effect on the
%force acting parallel to the direction of motion of the quark, or on 
%the dashed curves that describe our expectations for this quantity.
}
\label{betax07betaz04_RF}
\end{figure}

%\begin{figure}
%\includegraphics[scale=0.48]{PB1_RF_perp}
%\caption{Perpendicular component of the force in the fluid rest frame. The equilibrium expectations have no perpendicular component, since in the fluid rest frame equilibrium expectation force vector is parallel to the drag velocity. }
%\label{betaz02perpRF}
%\end{figure}

To get a sense of how generic our results are, we 
close this section by illustrating them for the case where the
heavy quark is dragged along a trajectory with $\beta_x=0.7$ and $\beta_z=0.4$.
In Fig.~\ref{betax07betaz04_lab} we first 
show our results in the lab frame, finding
results that are similar in most respects to those 
at smaller $\beta_x$ that we saw
in Figs.~\ref{betax02betaz04_parallel_lab}
and \ref{betax02betaz04_perp_lab}.
The important distinction is that
in Fig.~\ref{betax07betaz04_lab} we see that
at late times, after hydrodynamization, the drag force
that we calculate is well described by the dashed
curves meaning that, unlike 
in Fig.~\ref{betax02betaz04_parallel_lab},
here
there are no qualitative
effects of the gradients in the fluid velocity apparent.
We can understand this by noting from 
the upper panel of Fig.~\ref{Par_grad}
that by increasing $\beta_x$ from 0.2 to 0.7
we have substantially reduced the component
of the gradient of the fluid velocity in the direction
of motion of the quark in the lab frame.

In Fig.~\ref{betax07betaz04_RF}, 
we show our results for the case with $\beta_x=0.7$ and $\beta_z=0.4$ after
boosting to the local fluid rest frame. Again, we find
that even in a frame in which the quark is moving through
a fluid that is instantaneously at rest the force required to drag the quark
along its trajectory
includes a component perpendicular
 to the trajectory.
We see that this effect of gradients in the fluid velocity is somewhat smaller in absolute terms, and much smaller
relative to the drag force in the direction of motion of the quark,  here
where $\beta_x=0.7$ than it was in the lower panel of 
Fig.~\ref{betax02betaz04_perp_lab}, where $\beta_x=0.2$. This reflects
the fact that as $\beta_x$ increases at fixed $\beta_z$ the component of the
gradient of the fluid velocity in the direction of motion of the quark decreases.

\section{Conclusions and Lessons for Heavy Ion Physics}

The straightforward approach to modelling the 
rate of energy loss of heavy quarks produced in a heavy ion collision
%and, consequently,
%its rate of energy loss 
%and, via an Einstein relation, its diffusion constant 
proceeds as follows: (i) Use the equilibrium
equation of state to turn the proper energy density as a function
of space and time in the collision (for example
described via viscous hydrodynamics) into an effective
temperature as a function of space and time;  (ii) Use perturbative
QCD to calculate the distribution of the initial positions and momenta of heavy
quarks produced via hard scattering at the earliest moments of the heavy ion collision;
(iii) Use the effective temperature from (i) in
the expression (\ref{EqbmDrag}) for the drag force in a homogeneous plasma
in thermal equilibrium,
perhaps with the overall prefactor in (\ref{EqbmDrag}) turned into a parameter to be fit
to data;  (iv) Use the resulting spacetime-dependent drag force, and consequent energy loss
rate, in a Langevin equation employed to model the dynamics of heavy quarks in 
heavy ion collisions, as for example 
in Refs.~\cite{Moore:2004tg,vanHees:2005wb,vanHees:2007me,Akamatsu:2008ge,Rapp:2009my,Alberico:2011zy}.

Our results indicate that a straightforward approach along the lines above can reasonably
be applied even at very early times, before hydrodynamics
applies.  In particular, even though the peak value
of the energy loss in the matter produced in the collision of the two sheets
of energy that we have analyzed
occurs before hydrodynamization, at a time when
the matter produced in the collision is still far from equilibrium,
this peak value is nevertheless reasonably well reproduced by the straightforward
approach.  
Certainly
there is no sign of any significant  ``extra'' energy loss arising by virtue of being far from equilibrium. 
The message of our calculation seems to be  that if we want to 
use (\ref{EqbmDrag}) to learn about heavy quark energy loss in heavy ion
collisions, it is reasonable to apply it throughout the collision, even before
equilibration, defining the $T$ that appears in it through the energy density.
The error that one would make by treating the far-from-equilibrium energy loss
in this way is likely to be smaller than other uncertainties.
We anticipate that this is the most robust lesson for heavy ion physics that
we can draw from our results.

It also seems to be a generic feature of our results that there is some time delay
after the collision before the rate of energy loss of the heavy quark rises
to its peak value, even though it is during  this very earliest time that the matter in
which the quark is immersed has the very highest energy density.
Although we have not characterized the delay time in the case of
the collision of sheets of energy quantitatively, our analysis of  this
delay in a simpler setting suggests that it is of order $1/(\pi T_{\rm hydro})$
(where $T_{\rm hydro}$ is the temperature of the fluid when it
hydrodynamizes)
for a heavy quark whose velocity $\beta$ through the matter is not
relativistic, and increases slowly as $1/\sqrt{1-\beta^2}$ increases.
This would correspond to a delay of something like 0.1-0.2~fm$/c$ in
a heavy ion collision at RHIC.  This delay suggests that 
it takes a little time after the heavy quark is enveloped by
matter with a high energy density before the gluon fields
around the heavy quark respond to the presence of the matter, with
the drag and energy loss rising only after this response.
Although our strongly coupled calculation cannot provide a complete characterization
of the very earliest moments of a heavy ion collision, it is possible that
this qualitative lesson may carry over. It would be interesting to 
use a model of heavy quark dynamics in heavy ion collisions to
investigate whether a time delay along these lines has observable
consequences.

The straightforward approach to modelling the drag force on
a heavy quark in a heavy ion collision is built upon 
the result (\ref{EqbmDrag}) for a homogeneous fluid
and therefore neglects all effects of gradients in the fluid
velocity.  A third qualitative lesson that we can infer from
our results is that this neglect works reasonably well for heavy quarks
with small rapidity, whose velocities are close to perpendicular
to the gradient in the fluid velocity. 

We have found qualitative consequences for the drag force on heavy
quarks with
larger rapidity, moving closer to parallel to the gradient
in the fluid velocity, arising from the 
presence of a velocity gradient, which is to say qualitative phenomena
that are not present at all in the straightforward approach that
we have sketched above. For example, we have found that because
of the gradients in the fluid the
force exerted by the fluid on a heavy quark that has a small velocity relative
to the fluid at its location can sometimes point in the same direction
as the velocity of the quark, rather than dragging on it.   And, generically, we find that the force exerted
by the fluid on a heavy quark will have a component perpendicular to
the velocity of the quark, even as seen in the local fluid rest frame.
This perpendicular force can be large; we found instances where
at early times, before hydrodynamization,
its peak value in the local fluid rest frame was about half as large as the maximum 
drag force acting parallel to the velocity
of the quark.  The perpendicular force is nonzero at late times
too, when the quark is propagating through a liquid that is
described well by viscous hydrodynamics.
This perpendicular force can also be attributed to the presence of
gradients in the fluid velocity.  Here too, it would be interesting to
use a model of heavy quark dynamics in heavy ion collisions
to investigate the consequences of these effects.  That being
for the future, at present the fourth lesson from our results is
that the straightforward approach to modelling heavy quark dynamics
in heavy ion collisions should be used with caution for heavy quarks
at high rapidity.

From a more theoretical perspective, next steps that 
our results motivate include repeating our analysis for
the collisions of sheets of energy with varying widths
and seeking an analytical understanding of the effects
of gradients in the fluid velocity (and temperature) on
heavy quark energy loss.

\begin{acknowledgments}
We would like to thank Jorge Casalderrey-Solana, Michal Heller, Hong Liu, David Mateos, 
Stanislas Mrowczynski, Juan Pedraza, Silviu Pufu, 
Wilke van der Schee and Urs Wiedemann
for helpful discussions.
This work was supported by the U.S. Department of Energy
under cooperative research agreement DE-FG0205ER41360.
\end{acknowledgments}

\appendix

%\section{Eddington-Finkelstein coordinates}
%\label{app:EF}
%
%The Schwarzschild metric for the black hole in AdS$_5$ with radial coordinate $u$ (boundary is located at $u = 0$), is given by
%\beq
%ds^2 = \frac{1}{u^2} \( -f(u/u_h) dt^2 + d\vec x^2 + \frac{1}{f(u/u_h)} du^2 \)
%\label{stmet}
%\eeq
%where $f(u) = 1-u^4$ and $u_h = 1/(\pi T)$. Switching to the Eddington-Finkelstein coordinates corresponds to performing transformation on the time coordinate, so that $t \to v = t + h(u)$ and the term $du^2$ in the metric \r{stmet} which is divergent as $u \to u_h$ disappears. This translates to requirement on function $h(u)$, 
%\beq
%h'(u) = \pm \frac{1}{f(u/u_h)}
%\eeq
%The solution to this differential equation with positive sign corresponds to outgoing while negative sign defines the infalling light-like geodesics. We are interested in the infalling case and the solution relating $t$ and $v$ coordinates in this case is given by
%\beq
%v = t - \frac{u_h}{2} \( \tan^{-1} \frac{u}{u_h} + \tanh^{-1} \frac{u}{u_h} \)
%\eeq
%The static metric for the infalling Eddington-Finkelstein coordinates becomes
%\beq
%\begin{split}
%ds^2 &= G_{MN} dx^M dx^N \\
% &= \frac{1}{u^2} \( -f(u/u_h) dv^2  + d\vec x^2 - 2dvdu \)
%\end{split}
%\label{stBH}
%\eeq

\section{Trailing string solution in an equilibrium background}
\label{app:ts}

The trailing string solution that describes an infinitely heavy quark
being dragged in the $\vec x$-direction with velocity $\vec\beta$
through an equilibrium plasma with 
temperature $T_{\rm background}$ is~\cite{Herzog:2006gh, Gubser:2006bz}
\beq
\vec x = \vec \beta \(t + u_h \tan^{-1} \frac{u}{u_h} \)\,,
\label{tsgen}
\eeq
where we have not yet made the transformation (\ref{shiftdiff}) meaning that the event horizon
of the black brane in the bulk is located at $u_h=1/(\pi T_{\rm background})$.
With this choice of string profile, when we choose
worldsheet coordinates of the form (\ref{eta}) the temporal constraint (\ref{temporalconstraint}) 
becomes a differential equation for $t$ as a function of $\tau$ and $u$ that, in the case of
a time-independent metric corresponding to a static plasma with temperature $T_{\rm background}$,
has the solution
\beq
t + u_h \tan^{-1}\frac{u}{u_h} = \tau + u_h\sqrt{\gamma} \tan^{-1} \frac{u\sqrt{\gamma}}{u_h}
\label{temporalconstraint2}
\eeq
where $\gamma=1/\sqrt{1-\vec\beta^2}$.
After we make the transformation (\ref{shiftdiff}), the trailing string profile (\ref{tsgen})  is given by
\beq
\vec x = \vec \beta \( t + u_h \tan^{-1} \frac{u}{(1 + u\xi)u_h} \) 
\label{tstransformed}
\eeq
and for the case of a static plasma the solution to
the temporal constraint equation is now  %(\ref{temporalconstraint2}) takes the form
\begin{equation}
t + u_h \tan^{-1} \frac{u}{(1+u\xi)u_h} 
= \tau + u_h\sqrt{\gamma} \tan^{-1}\frac{u\sqrt{\gamma}}{(1+u\xi)u_h}\ .
\label{temporalconstraint3}
\end{equation}
As we have described in Section~\ref{sec:StringDynamics},  upon making the transformation (\ref{shiftdiff}) we
choose worldsheet coordinates in which $u=\sigma$.  This together with the 
expressions (\ref{tstransformed}) and (\ref{temporalconstraint3}) give us our initial
conditions for $X^M$, which is to say $u$, $t$ and $\vec x$, as functions of $\tau$ and $\sigma$.

If all we were interested in doing was dragging a heavy quark through static plasma, there
would be nothing to add.  The background of interest to us, however, is one where the heavy
quark is initially in a region of spacetime filled with static plasma but in which in the future
the quark will be slammed by two sheets of energy, incident upon it from the $+z$ and $-z$ directions.
As we mentioned in 
Sec.~\ref{sec:StringDynamics},
at any given early Eddington-Finkelstein time $t$, even well before the sheets of energy
collide on the boundary, the gravitational shocks are already colliding somewhere deep
in the bulk.  Eddington-Finkelstein coordinates on the worldsheet are advantageous from the point of
view of making it possible to solve the evolution equations but from the point of view of specifying
the initial conditions this feature is a complication.  There are two possible ways to proceed.
The route that we have followed in obtaining
all the results that we show  is to use (\ref{tstransformed}) and (\ref{temporalconstraint3})
as our initial conditions at all $\sigma$, even deep in the bulk, even though deep in the bulk
where the gravitational shocks are already colliding at the early $t$ at which we are initializing
the string profile the choice (\ref{temporalconstraint3}) does not satisfy the temporal constraint
equations.  The other option is to use (\ref{tstransformed}) and then to solve the temporal constraint
differential equation numerically, replacing (\ref{temporalconstraint3}) by a numerically determined
$t(\tau,\sigma)$.  Because we are initializing at a time $t$ when the quark at the boundary
is in a region of static plasma, these two options yield identical results close to the boundary, where (\ref{temporalconstraint3}) itself solves the temporal constraint equation.  They are inequivalent deep
in the bulk, and indeed there is no one ``right answer'' for how to initialize the string profile
deep in the bulk in Eddington-Finkelstein coordinates.  We have checked, however, that these
two options yield identical results for the drag force on the heavy quark, meaning that for
our purposes they are equivalent.  The reason that the distinction between them is irrelevant
is that it arises only deep enough within the bulk that well before the sheets collide at the
boundary the region of the string that is affected by these considerations has been
enveloped by the event horizon of the black brane.  Because of this, the initial violation
of the temporal constraint equation (assuming we pursue the first option) and the
initial arbitrariness associated with choosing the profile (\ref{tstransformed}) where
there is no reason to do so (in either the first or second option) are causally disconnected
from the boundary.  Since the drag force on the heavy quark is computed from the near-boundary
asymptotics of the string profile, nothing in this paragraph affects it.

%\beq
%t = \tau + u_h \( \tan^{-1} \frac{u_h}{u} + \sqrt{\gamma} \tan^{-1} \frac{u}{u_h\sqrt \gamma} \)
%\label{vtsP}
%\eeq
%
%Finally, if we perform transformation of radial coordinate $r \to r + \xi$ or $u \to \frac{u}{1 + u\xi}$ with space time dependent gauge parameter $\xi$, the trailing string solution for the Polyakov action with world sheet metric given in eq.~\r{eta}, takes the form 
%
%
%\beq
%t = \tau + u_h \( \tan^{-1}  \frac{(1 + u\xi)u_h}{u} + \sqrt{\gamma} \tan^{-1} \frac{u}{(1 + u\xi)u_h\sqrt \gamma}\)
%\eeq

%which are used in the section discussing the initial conditions and given by eqs.~\r{xts} and \r{vts} there.

\section{Varying Temperature}
\label{app:step}

We discovered in Section~\ref{sec:ZeroRapidity} that when the energy density
and pressures of the matter through which the heavy quark is moving change there
seems to be a time delay in the response of the drag force experienced by the
heavy quark, relative to  the way in which the drag force would change if it
were given by its value in a static plasma with the same instantaneous energy density or pressure.
In this Appendix we shall quantify this time delay in a time-dependent 
background that is much simpler than the colliding sheets of energy that
are our focus throughout the rest of this paper.  We shall consider
a background that, at all times, is a spatially homogeneous plasma
whose properties are exactly as if it is
in thermal equilibrium at temperature $T$, and in
particular which always has zero fluid velocity throughout, 
but by hand we shall make $T(t)$ 
change with time.  This of course violates energy conservation in the boundary theory.  It is therefore
no surprise that the bulk metric that provides the dual gravitational
description of this {\it ad hoc} setup is not a solution to Einstein's equations.
The setup in this Appendix is therefore not a model for anything; it is simply
a device with which to evaluate the time delay in the response of the drag
force on a heavy quark to a change in the conditions in which the heavy
quark finds itself.  Note that in this setup there are no spatial gradients of the 
fluid velocity since $u^{\mu}=(1,0,0,0)$ at all times.  Also, $T_e=T_\perp=T_\parallel=T(t)$
at all times.

%It is instructive to consider simpler problem than colliding shocks to gain understanding about the delay of the of the drag force with respect to the values based on the equilibrium expectations. To do so, we consider the static metric and promote the temperature, associated with the black hole horizon in the AdS$_5$ to be time dependent. Such metric alone would not satisfy the Einstein's equations of motion, but it could be a valid metric if we coupled black brane metric to some external matter. 

The metric that describes a spatially homogeneous plasma with a time-dependent temperature
is given in Eddington-Finkelstein coordinates by (\ref{ansatz})
%given in eq.~\r{ansatz}, with radial coordinate transformed using time dependent gauge parameter $\xi(v)$ has components
with
\beq
\begin{split}
A &=  \( \(\frac{1}{u} + \xi(t) \)^2 f(u,t) -2 \frac{\partial \xi}{\partial t} \), \\
\Sigma &= \frac{1}{u} + \xi(t), \quad B = F = 0,
\end{split}
\label{AppendixBmetric}
\eeq
with 
\beq
f(u,v) = 1 - \(\frac{u}{(1 + u\xi(t))u_h(t)}\)^4
\eeq
where $u_h(t)=1/(\pi\,T(t))$ and where $\xi(t)$ describes the residual
diffeomorphism  introduced in (\ref{shiftdiff}).  As we discussed there, we
fix $\xi$ by demanding that the apparent horizon be at $u=1$. 
%
%
%
%It is convenient to choose coordinate system such that the apparent horizon is located at constant radial position $u=1$. %The apparent horizon is determined by solving the equation (for position independent metric)
%\beq
%\partial_v \Sigma - u^2 \frac{A}{2} \partial_u \Sigma = 0
%\eeq
%which can be solved to obtain the value of gauge parameter $\xi(v)$ at each time point. 
%So the apparent horizon at $u=1$ is determined by
%\beq
%\partial_v \xi(v) + \frac{1}{2} \((1+\xi)^2 f(u=1,v) - 2\partial_v \xi \) = 0
%\eeq
%and the gauge parameter is uniquely determined to be 
In this simplified setup, 
in coordinates in which $\xi=0$ the apparent horizon is at $u=u_h(t)$
whereas ensuring that it is at $u=1$ corresponds to %in (\ref{ansatz}) with (\ref{AppendixBmetric})
choosing
\beq
%f(u=1,v) = 0 \rightarrow 
\xi(t) = \frac{1}{u_h(t)} - 1\ ,
\eeq
which ensures that $f(u=1,t)=0$.

%In this case, the apparent horizon is located at the same radial coordinate as the horizon associated with the temperature of the black hole. 

\begin{figure}[t]
\includegraphics[scale=0.45]{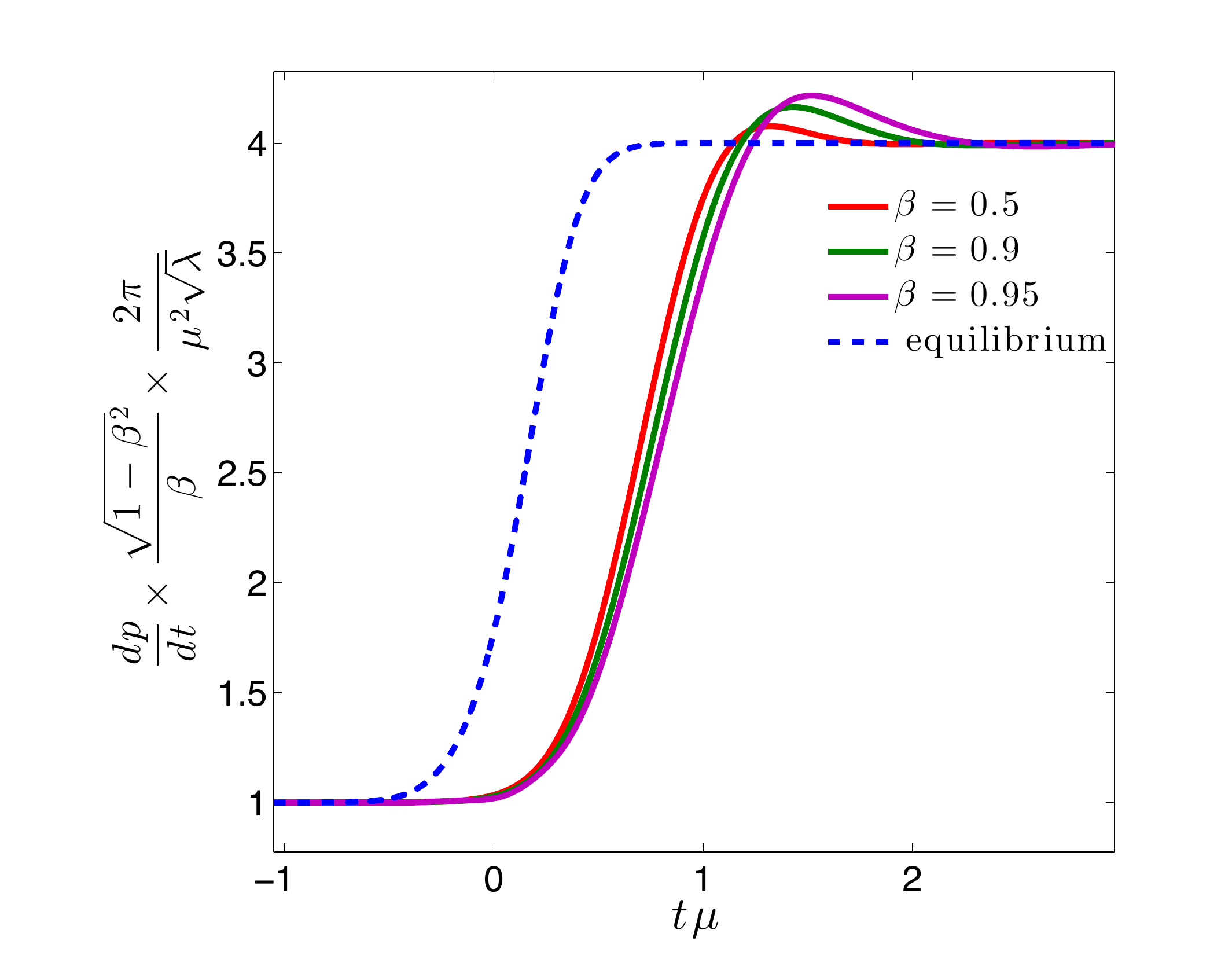}
\caption{Drag force, scaled by $\beta\gamma$,
for a heavy quark with velocity $\beta$ moving through a spatially homogeneous
plasma with a time dependent temperature $T(t)$ 
that increases from $\pi T=\mu$ at early times to $\pi T=2\mu$
at late times and is described by (\ref{uhvar}).
The different solid curves show the scaled drag force on quarks
with different velocities.
The dashed curve is the equilibrium expectation (\ref{EqbmDrag}). Because 
$T_e=T_\perp=T_\parallel=T(t)$ there is only a single
dashed curve.  
% with the black brane temperature following the smoothed step function profile (dashed line). Solid lines represent the normalized force for velocities $\beta = 0.5,0.9,0.95$. 
}
\label{varT2}
\end{figure}

\begin{figure}
\includegraphics[scale=0.48]{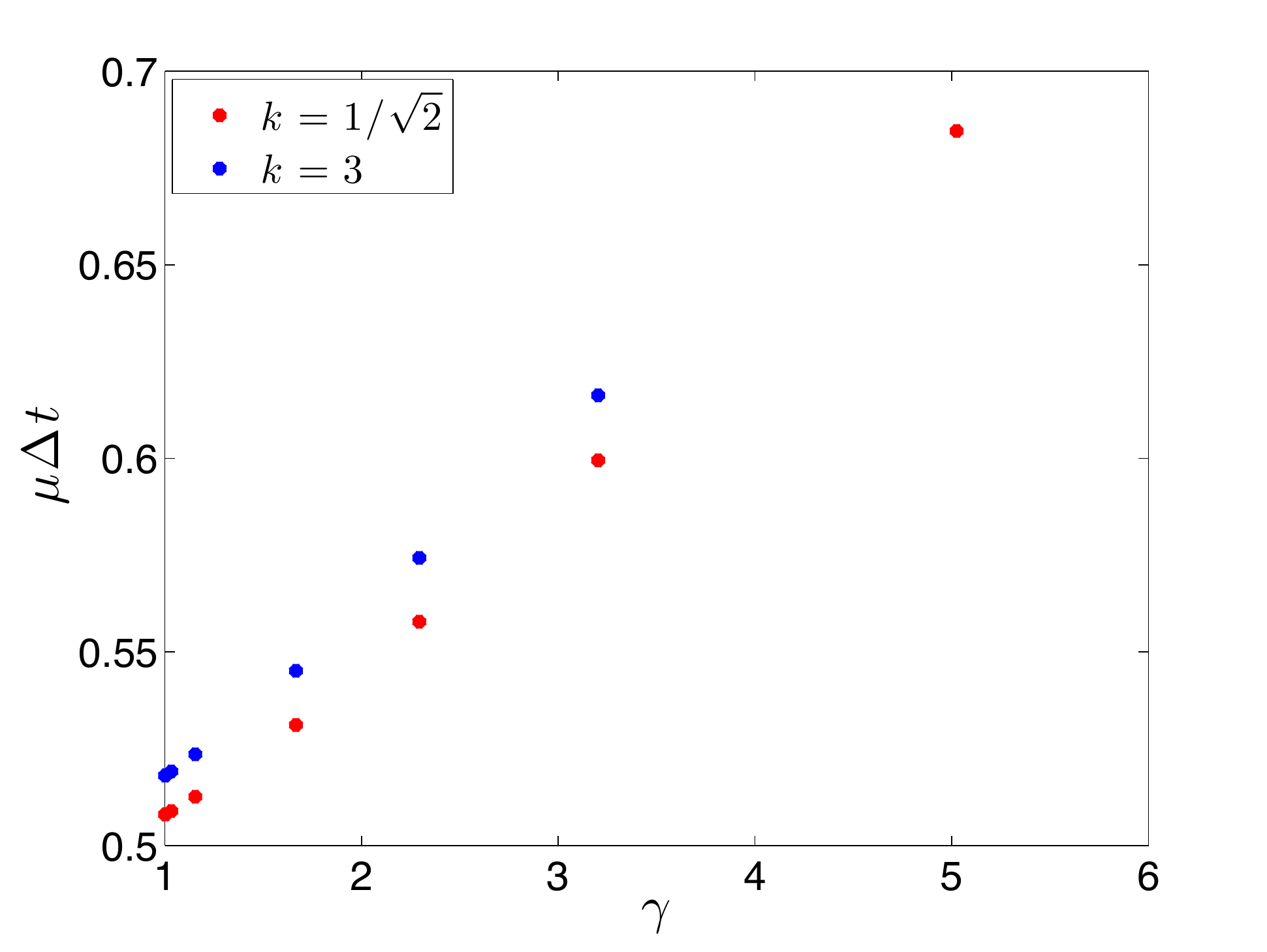}
\caption{The delay time $\Delta t$, defined from results as in Fig.~\ref{varT2} as described in the text,
as a function of $\gamma$ for two values of the parameter $k$ that controls the rapidity
with which the temperature increases.
%In both cases, 
%The linear scaling seems to be a good guide for both horizon profiles just like in the case of the colliding shockwaves.
}
\label{varT_delay}
\end{figure}

%Particularly interesting geometry to consider is when the (apparent) horizon approaches the step function. More exactly, we consider the temperature profile 
We shall choose a time-dependent temperature that starts at $\pi T= \mu$
at early times and ends at $\pi T=2\mu$ at late times, rising smoothly during
a window in time that is $\sim 1/(k\mu)$ wide,
with $k$ a parameter. We choose
\beq
u_h(t) = 1 - \frac{1}{4} \(1 + {\rm erf} (k\, t\, \mu ) \)\ .
\label{uhvar}
\eeq
%where ${\rm erf} (k t)$ is an error function and $k$ is parameter, so that 
For large values of $k$, $u_h(t)$ (and therefore $T(t)$) becomes a step function at $t=0$. 
In Fig.~\ref{varT2}, we choose $k=3$.
% we consider how the drag force is delayed when the relatively smooth step function, $k = 3$ in the temperature profile \r{uhvar}. 
We have plotted the drag force rescaled in such a way that the dashed curve,
which is obtained by substituting $T(t)$ into (\ref{EqbmDrag}) and
shows what the drag force would be in an equilibrium plasma
with the instantaneous temperature $T(t)$, is independent of
the quark velocity $\beta$ and is a plot of $(\pi T(t)/\mu)^2$.
We see that the actual drag force, shown as the solid
curves, is $\beta$-dependent even when rescaled as in the Figure
and shows a significant, and somewhat $\beta$-dependent, time delay.
The temperature of the plasma has risen quite significantly before
the drag force begins to rise; once the temperature reaches
its final plateau the drag force is only about half way up its rise;
and, finally, the drag force over-shoots before approaching its
new equilibrium value from above.

%The force is delayed as compared to equilibrium expectation and the delay time increases with increasing $\beta$. 
%In addition, the drag force slightly overshoots the equilibrium expectation before approaching the constant value after the change in temperature happens and the magnitude of this overshoot increases with increasing the drag velocity.  

There are various ways in which we could choose an operational definition of
the time delay from the results in Fig.~\ref{varT2}.  We shall define the time
delay $\Delta t$ as the delay between the times when the dashed
and solid curves cross the midpoint between the initial drag force and the
final drag force, i.e.~in Fig.~\ref{varT2} when they cross 2.5.  In Fig.~\ref{varT_delay}
we show how $\Delta t$ depends on the quark velocity for two different
values of $k$, the parameter that controls the rapidity with which $T(t)$ changes.
For both values of $k$ we see that at low velocities there is a time 
delay of around $0.5$, which we note is around $1/(\pi T_{\rm final})$ where $T_{\rm final}$ is the
final temperature.  And, for both values of $k$ we see that the time delay increases
with increasing $\beta$ in a way that is close to, but not exactly, linear with $\gamma$.
%
%The behavior of the delay time as a function of relativistic $\gamma = (1-\beta^2)^{-1/2}$ factor is illustrated in the Figure \ref{varT_delay}.
%The delay time $t_{\rm delay} = t_2 - t_1$ is defined so that $t_1$ is the time at which equilibrium expectation takes the average value or when the $u_h^{-2}$ takes the average value $(u_h^{-2})_{\rm av}$ (in the case of Figure \ref{varT2}, $(u_h^{-2})_{\rm av} = 2.5$).  
%As the velocities $\beta \to 0$, the delay time $t_{\rm delay}$ is non-vanishing and approaches the constant value. Nevertheless, when the velocities are increased, the delay is growing (at least approximately) linearly. This seems to be in accord with the delay time $t_p$ growing linearly for the colliding sheets as shown in the Figure \ref{peak_d}. We interpret it as consequence that the change of the horizon radius at $u=u_h$ takes time to be communicated along the string to boundary where the drag force is measured. This information traveling at the speed of light takes time $t_{\rm delay} \sim \mathcal O\( 1/(\pi T_{\rm final}) \)$ to propagate and this is the amount of time that the string gets delayed. 
%

From a gravitational perspective, the fact that there is a time delay can be described
in qualitative terms as if when the horizon at $u=u_h$ moves
the drag force only learns of this
at a time that is delayed by
of order the light-travel-time for information from the horizon travelling through  the bulk 
to reach the near-boundary region, where the drag on the heavy quark is encoded.
If taken literally, this interpretation would suggest that in the setup of this Appendix
the time delay should start out around $1/(\pi T_{\rm initial})$ and then drop to 
around $1/(\pi T_{\rm final})$. This interpretation should not be taken too literally, however,
because making $u_h$ time dependent makes the entire bulk metric at all $u$ time dependent,
not just the metric near the horizon.

\vfill\eject


\begin{thebibliography}{99}

 %\cite{Herzog:2006gh}
\bibitem{Herzog:2006gh} 
  C.~P.~Herzog, A.~Karch, P.~Kovtun, C.~Kozcaz and L.~G.~Yaffe,
  %``Energy loss of a heavy quark moving through N=4 supersymmetric Yang-Mills plasma,''
  JHEP {\bf 0607}, 013 (2006)
  [hep-th/0605158].
  %%CITATION = HEP-TH/0605158;%%
  
  %\cite{Gubser:2006bz}
\bibitem{Gubser:2006bz} 
  S.~S.~Gubser,
  %``Drag force in AdS/CFT,''
  Phys.\ Rev.\ D {\bf 74}, 126005 (2006)
  [hep-th/0605182].
  %%CITATION = HEP-TH/0605182;%%
  
  %\cite{CasalderreySolana:2006rq}
\bibitem{CasalderreySolana:2006rq} 
  J.~Casalderrey-Solana and D.~Teaney,
  %``Heavy quark diffusion in strongly coupled N=4 Yang-Mills,''
  Phys.\ Rev.\ D {\bf 74}, 085012 (2006)
  [hep-ph/0605199].
  %%CITATION = HEP-PH/0605199;%%
  
 %\cite{Maldacena:1997re}
\bibitem{Maldacena:1997re} 
  J.~M.~Maldacena,
  %``The Large N limit of superconformal field theories and supergravity,''
  Adv.\ Theor.\ Math.\ Phys.\  {\bf 2}, 231 (1998)
  [hep-th/9711200].
  %%CITATION = HEP-TH/9711200;%%
  %8838 citations counted in INSPIRE as of 27 Feb 2013
  
  %\cite{Witten:1998qj}
\bibitem{Witten:1998qj} 
  E.~Witten,
  %``Anti-de Sitter space and holography,''
  Adv.\ Theor.\ Math.\ Phys.\  {\bf 2}, 253 (1998)
  [hep-th/9802150].
  %%CITATION = HEP-TH/9802150;%%
  %5922 citations counted in INSPIRE as of 27 Feb 2013
 
   %\cite{Karch:2002sh}
\bibitem{Karch:2002sh} 
  A.~Karch and E.~Katz,
  %``Adding flavor to AdS / CFT,''
  JHEP {\bf 0206}, 043 (2002)
  [hep-th/0205236].
  %%CITATION = HEP-TH/0205236;%%

 %\cite{CasalderreySolana:2011us}
\bibitem{CasalderreySolana:2011us} 
  J.~Casalderrey-Solana, H.~Liu, D.~Mateos, K.~Rajagopal and U.~A.~Wiedemann,
  %``Gauge/String Duality, Hot QCD and Heavy Ion Collisions,''
  arXiv:1101.0618 [hep-th].
  %%CITATION = ARXIV:1101.0618;%%
  %136 citations counted in INSPIRE as of 27 Feb 2013


 %\cite{Chesler:2010bi}
\bibitem{Chesler:2010bi} 
  P.~M.~Chesler and L.~G.~Yaffe,
%  ``Holography and colliding gravitational shock waves in asymptotically AdS_5 spacetime,''
  Phys.\ Rev.\ Lett.\  {\bf 106}, 021601 (2011)
  [arXiv:1011.3562 [hep-th]].
  %%CITATION = ARXIV:1011.3562;%%
  
  %\cite{Casalderrey-Solana:2013aba}
\bibitem{Casalderrey-Solana:2013aba} 
  J.~Casalderrey-Solana, M.~P.~Heller, D.~Mateos and W.~van der Schee,
  %``From full stopping to transparency in a holographic model of heavy ion collisions,''
  arXiv:1305.4919 [hep-th].
  %%CITATION = ARXIV:1305.4919;%%
  
  \bibitem{PaulLarryToAppear}
  P.~M.~Chesler and L.~G.~Yaffe, to appear shortly.
  
 
%\cite{Janik:2005zt}
\bibitem{Janik:2005zt} 
  R.~A.~Janik and R.~B.~Peschanski, 
  %``Asymptotic perfect fluid dynamics as a consequence of Ads/CFT,''
  Phys.\ Rev.\ D {\bf 73}, 045013 (2006)
  [hep-th/0512162].
  %%CITATION = HEP-TH/0512162;%%
  %201 citations counted in INSPIRE as of 25 Mar 2013


 %\cite{Chesler:2008hg}
\bibitem{Chesler:2008hg} 
  P.~M.~Chesler and L.~G.~Yaffe,
  %``Horizon formation and far-from-equilibrium isotropization in supersymmetric Yang-Mills plasma,''
  Phys.\ Rev.\ Lett.\  {\bf 102}, 211601 (2009)
  [arXiv:0812.2053 [hep-th]].
  %%CITATION = ARXIV:0812.2053;%%
  %105 citations counted in INSPIRE as of 12 Apr 2013
  
  %\cite{Chesler:2009cy}
\bibitem{Chesler:2009cy} 
  P.~M.~Chesler and L.~G.~Yaffe,
  %``Boost invariant flow, black hole formation, and far-from-equilibrium dynamics in N = 4 supersymmetric Yang-Mills theory,''
  Phys.\ Rev.\ D {\bf 82}, 026006 (2010)
  [arXiv:0906.4426 [hep-th]].
  %%CITATION = ARXIV:0906.4426;%%
  
  %\cite{Booth:2009ct}
\bibitem{Booth:2009ct} 
  I.~Booth, M.~P.~Heller and M.~Spalinski,
  %``Black brane entropy and hydrodynamics: The Boost-invariant case,''
  Phys.\ Rev.\ D {\bf 80}, 126013 (2009)
  [arXiv:0910.0748 [hep-th]].
  %%CITATION = ARXIV:0910.0748;%%
  
  %\cite{Heller:2011ju}
\bibitem{Heller:2011ju} 
  M.~P.~Heller, R.~A.~Janik and P.~Witaszczyk,
  %``The characteristics of thermalization of boost-invariant plasma from holography,''
  Phys.\ Rev.\ Lett.\  {\bf 108}, 201602 (2012)
  [arXiv:1103.3452 [hep-th]].
  %%CITATION = ARXIV:1103.3452;%%
  
  %\cite{Heller:2012km}
\bibitem{Heller:2012km} 
  M.~P.~Heller, D.~Mateos, W.~van der Schee and D.~Trancanelli,
  %``Strong Coupling Isotropization of Non-Abelian Plasmas Simplified,''
  Phys.\ Rev.\ Lett.\  {\bf 108}, 191601 (2012)
  [arXiv:1202.0981 [hep-th]].
  %%CITATION = ARXIV:1202.0981;%%
  
  %\cite{vanderSchee:2012qj}
\bibitem{vanderSchee:2012qj} 
  W.~van der Schee,
  %``Holographic thermalization with radial flow,''
  Phys.\ Rev.\ D {\bf 87}, 061901 (2013)
  [arXiv:1211.2218 [hep-th]].
  %%CITATION = ARXIV:1211.2218;%%
  
  %\cite{Heller:2013oxa}
\bibitem{Heller:2013oxa} 
  M.~P.~Heller, D.~Mateos, W.~van der Schee and M.~Triana,
  %``Holographic isotropization linearized,''
  arXiv:1304.5172 [hep-th].

%\cite{Shen:2010uy}
\bibitem{Shen:2010uy} 
  C.~Shen, U.~Heinz, P.~Huovinen and H.~Song,
  %``Systematic parameter study of hadron spectra and elliptic flow from viscous hydrodynamic simulations of Au+Au collisions at $\sqrt{s_{NN}}=200$ GeV,''
  Phys.\ Rev.\ C {\bf 82}, 054904 (2010)
  [arXiv:1010.1856 [nucl-th]].
  %%CITATION = ARXIV:1010.1856;%%

%\cite{Kolb:2003dz}
\bibitem{Kolb:2003dz} 
  P.~F.~Kolb and U.~W.~Heinz,
  %``Hydrodynamic description of ultrarelativistic heavy ion collisions,''
  in Hwa, R.C. (ed.) et al.: {\it Quark gluon plasma} 634-714 (2003)
  [nucl-th/0305084].
  %%CITATION = NUCL-TH/0305084;%%

%\cite{Heinz:2004pj}
\bibitem{Heinz:2004pj} 
  U.~W.~Heinz,
  %``Thermalization at RHIC,''
  AIP Conf.\ Proc.\  {\bf 739}, 163 (2005)
  [nucl-th/0407067].
  %%CITATION = NUCL-TH/0407067;%%

%\cite{Carrington:2013tz}
\bibitem{Carrington:2013tz} 
  M.~E.~Carrington, K.~Deja and S.~Mrowczynski,
  %``Parton Energy Loss in the Extremly Prolate Quark-Gluon Plasma,''
  PoS ConfinementX {\bf }, 175 (2012)
  [arXiv:1301.4563 [hep-ph]].
  %%CITATION = ARXIV:1301.4563;%%

  %\cite{Grumiller:2008va}
\bibitem{Grumiller:2008va} 
  D.~Grumiller and P.~Romatschke,
  %``On the collision of two shock waves in AdS(5),''
  JHEP {\bf 0808}, 027 (2008)
  [arXiv:0803.3226 [hep-th]].
  %%CITATION = ARXIV:0803.3226;%%
  %63 citations counted in INSPIRE as of 15 Mar 2013

%\cite{de Haro:2000xn}
\bibitem{de Haro:2000xn} 
  S.~de Haro, S.~N.~Solodukhin and K.~Skenderis,
  %``Holographic reconstruction of space-time and renormalization in the AdS / CFT correspondence,''
  Commun.\ Math.\ Phys.\  {\bf 217}, 595 (2001)
  [hep-th/0002230].
  %%CITATION = HEP-TH/0002230;%%
  %549 citations counted in INSPIRE as of 19 Mar 2013


%\cite{Skenderis:2000in}
\bibitem{Skenderis:2000in} 
  K.~Skenderis,
  %``Asymptotically Anti-de Sitter space-times and their stress energy tensor,''
  Int.\ J.\ Mod.\ Phys.\ A {\bf 16}, 740 (2001)
  [hep-th/0010138].
  %%CITATION = HEP-TH/0010138;%%

  %\cite{Herzog:2006se}
\bibitem{Herzog:2006se} 
  C.~P.~Herzog,
  %``Energy Loss of Heavy Quarks from Asymptotically AdS Geometries,''
  JHEP {\bf 0609}, 032 (2006)
  [hep-th/0605191].
  %%CITATION = HEP-TH/0605191;%%
  %92 citations counted in INSPIRE as of 25 Feb 2013

%\cite{Chesler:2012pw}
\bibitem{Chesler:2012pw} 
  P.~Chesler, M.~Lekaveckas and K.~Rajagopal,
  %``Far-from-equilibrium heavy quark energy loss at strong coupling,''
  Nucl.\ Phys.\ A904-905 {\bf 2013}, 861c (2013)
  [arXiv:1211.2186 [hep-th]].


  %\cite{Giecold:2009wi}
\bibitem{Giecold:2009wi} 
  G.~C.~Giecold,
  %``Heavy quark in an expanding plasma in AdS/CFT,''
  JHEP {\bf 0906}, 002 (2009)
  [arXiv:0904.1874 [hep-th]].
  %%CITATION = ARXIV:0904.1874;%%
  %12 citations counted in INSPIRE as of 20 Mar 2013  
  
  %\cite{Stoffers:2011fx}
\bibitem{Stoffers:2011fx} 
  A.~Stoffers and I.~Zahed,
  %``Holographic Jets in an Expanding Plasma,''
  Phys.\ Rev.\ C {\bf 86}, 054905 (2012)
  [arXiv:1110.2943 [hep-th]].
  %%CITATION = ARXIV:1110.2943;%%
  %5 citations counted in INSPIRE as of 20 Mar 2013

%\cite{Abbasi:2012qz}
\bibitem{Abbasi:2012qz} 
  N.~Abbasi and A.~Davody,
  %``Moving Quark in a Viscous Fluid,''
  JHEP {\bf 1206}, 065 (2012)
  [arXiv:1202.2737 [hep-th]].
  %%CITATION = ARXIV:1202.2737;%%
  %2 citations counted in INSPIRE as of 20 Mar 2013
  
%\cite{Bjorken:1982qr}
\bibitem{Bjorken:1982qr} 
  J.~D.~Bjorken,
  %``Highly Relativistic Nucleus-Nucleus Collisions: The Central Rapidity Region,''
  Phys.\ Rev.\ D {\bf 27}, 140 (1983).
  %%CITATION = PHRVA,D27,140;%%
  %2058 citations counted in INSPIRE as of 15 May 2013

%\cite{Peigne:2005rk}
\bibitem{Peigne:2005rk} 
  S.~Peigne, P.~-B.~Gossiaux and T.~Gousset,
  %``Retardation effect for collisional energy loss of hard partons produced in a QGP,''
  JHEP {\bf 0604}, 011 (2006)
  [hep-ph/0509185].
  %%CITATION = HEP-PH/0509185;%%

%\cite{Guijosa:2011hf}
\bibitem{Guijosa:2011hf} 
  A.~Guijosa and J.~F.~Pedraza,
  %``Early-Time Energy Loss in a Strongly-Coupled SYM Plasma,''
  JHEP {\bf 1105}, 108 (2011)
  [arXiv:1102.4893 [hep-th]].
  %%CITATION = ARXIV:1102.4893;%%



%\cite{Chernicoff:2012iq}
\bibitem{Chernicoff:2012iq} 
  M.~Chernicoff, D.~Fernandez, D.~Mateos and D.~Trancanelli,
  %``Drag force in a strongly coupled anisotropic plasma,''
  JHEP {\bf 1208}, 100 (2012)
  [arXiv:1202.3696 [hep-th]].
  %%CITATION = ARXIV:1202.3696;%%
  
  %\cite{NataAtmaja:2010hd}
\bibitem{NataAtmaja:2010hd} 
  A.~Nata Atmaja and K.~Schalm,
  %``Anisotropic Drag Force from 4D Kerr-AdS Black Holes,''
  JHEP {\bf 1104}, 070 (2011)
  [arXiv:1012.3800 [hep-th]].
  %%CITATION = ARXIV:1012.3800;%%
  
  %\cite{Fadafan:2012qu}
\bibitem{Fadafan:2012qu} 
  K.~B.~Fadafan and H.~Soltanpanahi,
  %``Energy loss in a strongly coupled anisotropic plasma,''
  JHEP {\bf 1210}, 085 (2012)
  [arXiv:1206.2271 [hep-th]].
  %%CITATION = ARXIV:1206.2271;%%
  
  
%\cite{Moore:2004tg}
\bibitem{Moore:2004tg} 
  G.~D.~Moore and D.~Teaney,
  %``How much do heavy quarks thermalize in a heavy ion collision?,''
  Phys.\ Rev.\ C {\bf 71}, 064904 (2005)
  [hep-ph/0412346].
  %%CITATION = HEP-PH/0412346;%%

  %\cite{vanHees:2005wb}
\bibitem{vanHees:2005wb} 
  H.~van Hees, V.~Greco and R.~Rapp,
  %``Heavy-quark probes of the quark-gluon plasma at RHIC,''
  Phys.\ Rev.\ C {\bf 73}, 034913 (2006)
  [nucl-th/0508055].
  %%CITATION = NUCL-TH/0508055;%%

%\cite{vanHees:2007me}
\bibitem{vanHees:2007me} 
  H.~van Hees, M.~Mannarelli, V.~Greco and R.~Rapp,
  %``Nonperturbative heavy-quark diffusion in the quark-gluon plasma,''
  Phys.\ Rev.\ Lett.\  {\bf 100}, 192301 (2008)
  [arXiv:0709.2884 [hep-ph]].
  %%CITATION = ARXIV:0709.2884;%%
  
%\cite{Akamatsu:2008ge}
\bibitem{Akamatsu:2008ge} 
  Y.~Akamatsu, T.~Hatsuda and T.~Hirano,
  %``Heavy Quark Diffusion with Relativistic Langevin Dynamics in the Quark-Gluon Fluid,''
  Phys.\ Rev.\ C {\bf 79}, 054907 (2009)
  [arXiv:0809.1499 [hep-ph]].
  %%CITATION = ARXIV:0809.1499;%%


%\cite{Rapp:2009my}
\bibitem{Rapp:2009my} 
  R.~Rapp and H.~van Hees,
  %``Heavy Quarks in the Quark-Gluon Plasma,''
  in R. C. Hwa, X.-N. Wang (eds.) {\it Quark Gluon Plasma 4}, World Scientific, 111 (2010)
  [arXiv:0903.1096 [hep-ph]].
  %%CITATION = ARXIV:0903.1096;%%
 
  %\cite{Alberico:2011zy}
\bibitem{Alberico:2011zy} 
  W.~M.~Alberico, A.~Beraudo, A.~De Pace, A.~Molinari, M.~Monteno, M.~Nardi and F.~Prino,
  %``Heavy-flavour spectra in high energy nucleus-nucleus collisions,''
  Eur.\ Phys.\ J.\ C {\bf 71}, 1666 (2011)
  [arXiv:1101.6008 [hep-ph]].
  %%CITATION = ARXIV:1101.6008;%%


  
   
 
\end{thebibliography}
\end{document}